\documentclass[preprintnumbers,amsmath,amssymb,superscriptaddress]{revtex4-2}

\usepackage{mathrsfs}
\usepackage{amssymb}
\usepackage{graphicx}
\usepackage{dcolumn}
\usepackage{bm}
\usepackage{textcomp}
\usepackage{amsmath}
\usepackage[titletoc]{appendix}
\usepackage{accents}
\usepackage{ntheorem}
\usepackage{color}
\usepackage{datetime2} 

\newtheorem{proposition}{Proposition}

\newcommand\ket[1]{\ensuremath{|#1\rangle}}
\newcommand\bra[1]{\ensuremath{\langle#1|}}
\newcommand\iprod[2]{\ensuremath{\langle#1|#2\rangle}}
\newcommand\oprod[2]{\ensuremath{|#1\rangle\langle#2|}}
\newcommand\mean[1]{\ensuremath{\langle #1\rangle}}
\newcommand\tr{\mathop{\rm tr}\nolimits}
\newcounter{RomanNumber}

\makeatletter
\def\widebar{\accentset{{\cc@style\underline{\mskip10mu}}}}
\def\Widebar{\accentset{{\cc@style\underline{\mskip8mu}}}}
\makeatother

\begin{document}

\title{The finite key effect of side-channel-secure quantum key distribution beyond post-selection technique}
\author{Cong Jiang}\email{Corresponding author: jiangcong@jiqt.org}
\affiliation{Jinan Institute of Quantum Technology and Jinan branch, Hefei National Laboratory, Jinan, Shandong 250101, China}
\affiliation{State Key Laboratory of Low Dimensional Quantum Physics, Department of Physics, Tsinghua University, and Frontier Science Center for Quantum Information, Beijing 100084, China}
\author{Zong-Wen Yu}
\affiliation{Data Communication Science and Technology Research Institute, Beijing 100191, China}

\author{Xiang-Bin Wang}\email{Corresponding author: xbwang@mail.tsinghua.edu.cn}
\affiliation{Jinan Institute of Quantum Technology and Jinan branch, Hefei National Laboratory, Jinan, Shandong 250101, China}
\affiliation{State Key Laboratory of Low Dimensional Quantum Physics, Department of Physics, Tsinghua University, and Frontier Science Center for Quantum Information, Beijing 100084, China}
\affiliation{ International Quantum Academy, Shenzhen 518048, China.}

\begin{abstract}
By applying the framework of entropic uncertainty relation (EUR) and the Quantum Leftover Hash Lemma (QLHL), we introduce a security-proof method for variable-length side-channel-secure (SCS) quantum key distribution (QKD) against coherent attacks. This method reframes composable security as a statistical fluctuation problem of phase errors, enabling direct proofs against coherent attacks through observables and virtual observables. It yields tight key rates for the SCS protocol and reduces pulse requirements by over two orders of magnitude compared to prior works that employ the post-selection technique. {Utilizing the characteristic of the SCS protocol that it has no untagged-bit bit-flip errors, we prove that the secure key length for the SCS protocol can be determined after Cascade error correction, using its actual parity disclosures. We also explain how the same argument extends to protocols with compatible bit-error-pattern and phase-error measurements.} We further identify sufficient conditions under which the final key length may be determined after error correction in a broader class of QKD protocols. Under the framework of EUR and QLHL, we clarify the applicability of several commonly used concentration bounds to variable-length QKD and the appropriate manner of their implementation. This work enhances the practical value of the SCS protocol and clarifies the security justification of key-rate formulas used in practical variable-length QKD implementations.
\end{abstract}


\maketitle

\section{Introduction}\label{intro}
Quantum key distribution (QKD) leverages quantum mechanics to enable secure key exchange, detecting eavesdroppers through disturbances in quantum states~\cite{bennett1984quantum,gisin2002quantum,xu2020secure,pirandola2020advances,scarani2009security}. Rooted in foundational protocols such as BB84 (1984)~\cite{bennett1984quantum}, QKD offers information-theoretic security against quantum-capable adversaries, unlike classical cryptography, which is vulnerable to algorithms such as Shor’s. The theory and experimental implementation of practically secure, long-distance QKD have advanced rapidly in recent years.

Recent progress in long-distance QKD has been driven by Twin-Field (TF) QKD protocols~\cite{lu2018overcoming,wang2018twin,ma2018phase,lin2018simple,curty2018simple,cui2019twin}, which surpass the repeaterless bound; among these, the sending-or-not-sending (SNS) QKD variant~\cite{wang2018twin} is particularly notable. Using the SNS protocol, QKD has been demonstrated over 1002 km optical fiber~\cite{liu2023experimental,liu20231002}, currently the world’s longest ground-based QKD distance. In addition, numerous long-distance and field experiments based on TF-QKD have been reported~\cite{pittaluga2025long, pittaluga2021600, chen2021twin, liu2021field, wang2022twin, zhou2023twin,zhou2024independent,chen2024twin,chen2022quantum,fang2020implementation}. Furthermore, satellite-based QKD, including the Micius (QUESS) satellite~\cite{liao2017satellite} and the Jinan-1 satellite~\cite{li2025micro}, extends secure links to global scales.

Imperfections in practical implementations are a primary source of security vulnerabilities in QKD systems~\cite{zapatero2025implementation}. In recent years, several theoretical frameworks have been developed to address such real-world imperfections. For instance, the decoy-state method guarantees provable security even with imperfect single-photon sources~\cite{hwang2003quantum,wang2005beating,lo2005decoy}. Measurement-device-independent (MDI) QKD protocols eliminate all side-channel attacks targeting the detection system~\cite{lo2012measurement,braunstein2012side,lu2018overcoming,wang2018twin,ma2018phase,lin2018simple,curty2018simple,cui2019twin}. On the source side, imperfections such as pattern effects, state-preparation flaws, and mode dependencies can be more tightly bounded using the reference-technique framework and the quantum-coin approach, thereby significantly enhancing resilience against side-channel attacks~\cite{pereira2020quantum,zapatero2021security,li2025quantum,curras2025security}. For non-MDI protocols, detector-side imperfections—including detection-efficiency mismatch, dark counts, and dead-time effects—can be effectively addressed through techniques based on entropic uncertainty relations (EUR), squashing maps, and related tools~\cite{tupkary2024phase,nahar2025imperfect}.

Since QKD keys are ultimately used in higher-level cryptographic tasks, the security of QKD protocols must be established within the composable security framework~\cite{renner2005security}. Finite-key effects quantify how composable security impacts the achievable key rate. Computing these effects using EUR and the Quantum Leftover Hashing Lemma (QLHL)~\cite{tomamichel2012tight} is a widely used approach in security proofs for most QKD protocols, including BB84~\cite{lim2014concise}, MDI-QKD~\cite{curty2014finite}, TF-type protocols (such as SNS~\cite{jiang2019unconditional}, CAL~\cite{curras2021tight}, and Phase-Matching~\cite{zeng2020symmetry}), and the Mode-Paring protocol~\cite{zeng2022mode,xie2022breaking}. Ref.~\cite{tomamichel2012tight} provides a rigorous finite-key analysis for the single-photon BB84 protocol secure against coherent attacks, also based on EUR and QLHL. However, that method~\cite{tomamichel2012tight} applies only to protocols producing fixed-length keys. Such protocols rely on predetermined threshold conditions: after a protocol run, if all thresholds are met, a key of predefined length $\ell$ is output; otherwise, no key is produced.


For fixed-length QKD protocols, improving the probability that the protocol does not abort generally comes at the cost of reducing the key length that can be certified upon success~\cite{tupkary2024security}. Variable-length QKD protocols avoid this limitation by determining the final key length adaptively from the statistics actually observed during the protocol. As a result, they can make efficient use of the experimental data without requiring prior knowledge of the expected channel behaviour. Ref.~\cite{hayashi2014security} offers a rigorous finite-key analysis for a variable-length, decoy-state BB84 protocol secure against coherent attacks, using the phase-error approach. Ref.~\cite{tupkary2024security} presents a proof framework for finite-key effects in variable-length QKD protocols secure against collective attacks, based on Rényi entropy and QLHL, while Ref.~\cite{kanitschar2025composable} extends it to high-dimensional cases. In principle, the methods in Refs.~\cite{tupkary2024security,kanitschar2025composable} can be further extended—through post-selection techniques~\cite{christandl2009postselection, nahar2024postselection}—to achieve security against coherent attacks. Recently, based on the EUR and QLHL, a direct security framework for variable-length QKD protocols against coherent attacks was proposed in Ref.~\cite{tupkary2025phase}. In these frameworks, however, the final key length or an upper bound on the error-correction leakage must be fixed before error correction; therefore, the common engineering practice of inserting the actual number of disclosed reconciliation bits into the key-length formula after error correction is not automatically justified by those results.  

In this work, we extend the security framework in Ref.~\cite{tupkary2025phase} to the case of the side-channel-secure (SCS) protocol~\cite{wang2019practical,jiang2024side}, in which the observed values can only be obtained after error correction. Notably, this approach allows us to derive the coherent-attack–resistant key-rate formula for the SCS protocol directly, without employing post-selection as in Ref.~\cite{jiang2024side}. {In the proof, we use the complete window-type sequence to specify the bit-error pattern before the untagged key bits are measured in the virtual protocol. For Cascade~\cite{tupkary2023using}, the actual number of Alice's parity disclosures is fixed by this pattern and the independent algorithmic randomness. This allows the final key length to be calculated after reconciliation. We also discuss an extension to other protocols based on a virtual joint $ZZ$-parity measurement.} The security framework for variable-length QKD also clarifies which concentration bounds are applicable to QKD and how they should be applied.

The remainder of this paper is organized as follows. In Sec.~\ref{main_scs}, we give the key-rate formula obtained by applying the EUR and QLHL framework to the SCS protocol. Sec.~\ref{core} provides a detailed security proof for the SCS protocol, and Sec.~\ref{numerical} presents numerical results. We then discuss the implications of our findings and provide a brief summary.

\section{Results}

The core of the security proof for the SCS protocol consists in establishing a relation between the phase-error rate and the counting rates of three distinct states. Of these three states, two appear in the real protocol, whereas the third state—commonly denoted as $\ket{\phi_2}$—is absent from both the real protocol and the equivalent entanglement-based protocol constructed in Refs.~\cite{wang2019practical,jiang2024side}. In the asymptotic case, the issue admits a simple treatment: the counting rate of any state is bounded above by 1. However, in the finite-key regime against coherent attacks, the analysis becomes considerably more involved. Standard approaches result in a divergent (infinitely large) term. In Ref.~\cite{jiang2024side}, the finite-key problem was addressed by first deriving the key rate secure against collective attacks, followed by the application of the post-selection technique~\cite{christandl2009postselection, nahar2024postselection} to obtain the key rate secure against coherent attacks. While effective, this method yields a key rate that is excessively sensitive to the total number of pulses.

In the present work, we introduce a modest modification to the SCS protocol: all bits are independently and randomly assigned to one of two subsets with equal probability, which are subsequently subjected to independent error correction and privacy amplification. For this modified protocol, we construct the corresponding equivalent entanglement-based version. Because the bits are divided into two subsets, the security analysis of one subset can exploit a virtual measurement performed on the other subset—namely, a virtual observable corresponding to the $\ket{\phi_2}$ state. Although the actual counting rate of the $  |\phi_2\rangle  $ state is unavailable in the real protocol, a tight upper bound on this rate can be rigorously estimated. By incorporating this virtual measurement, we directly derive the finite-key secure key rate of the protocol against coherent attacks beyond the post-selection technique.

In the methods presented in Ref.~\cite{tupkary2024security,kamin2025improved,tupkary2025phase}, the amount of key leakage \(\lambda\) during the error correction process must be determined prior to error correction. Ref.~\cite{tupkary2024security,tupkary2025phase} demonstrates that this requirement can be slightly relaxed to predetermining a maximum allowable leakage \(\lambda_{\max}\) before error correction. Still, this precludes the use of the actual leakage from the error correction process in the key rate calculation. {In this work, we prove that Alice and Bob can first perform error correction using the Cascade protocol described in Ref.~\cite{tupkary2023using} and subsequently compute the final key length from Alice's actual parity disclosures. The proof retains the complete bit-error pattern and all public randomness; it does not assume that a general adaptive communication length is independent of the raw key. Appendix~\ref{framework} extends this Cascade argument to qubit virtual protocols with compatible $ZZ$-parity and phase-error measurements.} {Appendix~\ref{actual_leakage_general} proves that, under a conditional Kraft bound on the complete reconciliation transcript, the realized disclosure cost can be subtracted in the key-length formula even when it depends on the raw strings. This provides a sufficient condition for using actual leakage after adaptive reconciliation and recovers the Cascade parity count used here as a special case.}

\subsection{The real SCS protocol and the main results}\label{main_scs}
Based on the fact that there is no side channel in the vacuum state, we propose the side-channel-secure (SCS) QKD protocol that is secure against to source side channels~\cite{wang2019practical}. There are only two states in the SCS protocol, and only the upper bound of the intensity needs to be characterized without characterizing the state in the whole space. 


To keep the topic of this article focused, we assume that there is no bit-dependent correlation between the emitted pulses and that Eve cannot enter the laboratory. Combined with the method of Ref.~\cite{jiang2025side}, the conclusions of this article can be naturally extended to the case where these two assumptions are removed. 

In the SCS protocol~\cite{wang2019practical,jiang2023side,jiang2024side}, the two senders Alice and Bob respectively send quantum states to an untrusted measurement party Charlie. In Alice's side, there are two sources named "$o_A$" and "$x_A$" which produce two states with different intensities. We denote the states produced by sources $o_A$ and $x_A$ in the $k$-th time window by $\rho_{o_{Ak}}$ and $\rho_{x_{Ak}}$ respectively. In general, source $o_A$ is designed to produce states whose intensity is as close to $0$ as possible, i.e., source $o_A$ is a vacuum source or an almost-vacuum source. Similarly, in Bob side, there are two sources named "$o_B$" and "$x_B$" which respectively produce quantum states $\rho_{o_{Bk}}$ and $\rho_{x_{Bk}}$ in the $k$-th time window. We suppose the following equation holds for all time windows
\begin{equation}\label{character}
\begin{split}
&\bra{0}\rho_{o_{Ak}}\ket{0}\ge{a_{v0}}\ge 0.5,\quad \bra{0}\rho_{x_{Ak}}\ket{0}\ge{a_{0}}\ge 0.5,\\
&\bra{0}\rho_{o_{Bk}}\ket{0}\ge{b_{v0}}\ge 0.5,\quad \bra{0}\rho_{x_{Bk}}\ket{0}\ge{b_{0}}\ge 0.5,
\end{split}
\end{equation}

The SCS protocol~\cite{wang2019practical,jiang2023side,jiang2024side} contains the following main steps.

\textbf{Step 1.} In the \textit{k}-th time window, Alice (Bob) randomly prepare a pulse from source $o_A$ or $x_A$ ($o_B$ or $x_B$) with probabilities $p_o$ and ${p_x=1-p_o}$. If source $o_A$ ($x_A$) is chosen in Alice's side, Alice takes the corresponding bit of this time window as bit $0$ ($1$). If source $o_B$ ($x_B$) is chosen in Bob's side, Bob takes the corresponding bit of this time window as bit $1$ ($0$). 

\textbf{Step 2.} After receiving the signal pulses, Charlie first performs phase compensation and then interferometry measurement at his measurement station. The measurement results would be announced to Alice and Bob. If the right-side detector clicks, it is regarded as an effective time window.

Remark: In this work, we assume that when two identical coherent lights arrive at Charlie's detection station, most of the energy reaches the left-side detector. And we assume
\begin{equation}
p_o\ge p_x,
\end{equation}
which is the case of the optimized parameters of SCS protocol.

\textbf{Step 3.} After preparing $N$ time windows and received all measurement results, Alice and Bob jointly and randomly divide all time windows into two disjoint parts, such that each time window is assigned to either part 1 or part 2 {independently with probability $1/2$, using shared randomness}. For each part, Alice and Bob obtains two raw key strings $Z_A^\beta$ and $Z_B^\beta$ corresponding to the effective time windows, where $\beta=1,2$.

\begingroup
\textbf{Step 4.} Alice and Bob perform Cascade error correction~\cite{tupkary2023using} separately on the two pairs of strings $Z_A^\beta,Z_B^\beta$. Once the partition and string lengths are known, Cascade's other inputs, including the number of passes, block sizes, and independent public random seeds $R$ for its internal permutations, can be fixed before measuring $Z_A^\beta,Z_B^\beta$ in the equivalent entanglement protocol.

Bob obtains an estimate $\hat Z_A^\beta$ and retains his original $Z_B^\beta$ for Step 5. For a prescribed correctness parameter $\varepsilon_{\mathrm{cor}}$, set $v=\lceil\log_2(2/\varepsilon_{\mathrm{cor}})\rceil$. Alice announces the $v$-bit value of a fresh independent two-universal verification hash. Agreement of the hashes is denoted by $\Omega_{EV}$; a part that fails verification outputs the empty key. This choice bounds the probability of accepting unequal reconciled strings by $\varepsilon_{\mathrm{cor}}$ for each part. The two parts use independent randomness, and each part is accepted or aborted separately.
\endgroup

\textbf{Step 5.} By comparing $\hat{Z}_A^\beta$ with $Z_B^\beta$, Bob obtain a guess of the positions of all types of effective windows:
\\ for the bits that are identical in \(\hat{Z}_A^\beta\) and \(Z_B^\beta\), Bob treats them as ${\mathcal{Z}}$-windows;
\\ for the bits that is $1$ in \(\hat{Z}_A^\beta\) but $0$ in \(Z_B^\beta\), Bob treats them as ${\mathcal{B}}$-windows;
\\ for the bits that is $0$ in \(\hat{Z}_A^\beta\) but $1$ in \(Z_B^\beta\), Bob treats them as ${\mathcal{O}}$-windows.

If $Z_A^\beta=\hat{Z}_A^\beta$, Bob will know the the correct number of each kinds of effective windows: windows of Alice and Bob choosing sources $o_Ao_B$ (named $\mathcal{O}$ windows), windows of Alice and Bob choosing sources $x_Ax_B$ (named $\mathcal{B}$ windows), and windows of Alice and Bob choosing sources $o_Ax_B$ or $x_Ao_B$ (named $\mathcal{Z}$ windows)~\cite{jiang2024side}. In the SCS protocol, the untagged bits are the bits in $\mathcal{Z}$ windows.

\begingroup
Based on this classification information, Bob obtains a tentative window-type sequence $\hat\omega=(\hat t_1,\ldots,\hat t_{m_\beta})$, where $m_\beta$ is the number of clicking windows in part $\beta$ and each $\hat t_k$ belongs to $\{\mathcal O,\mathcal B,\mathcal Z\}$. We write $\hat\Omega_i$ for the event $\hat\omega=\omega_i$, where $\{\omega_i\}$ enumerates all possible complete window-type sequences. The counts $n_{\mathcal O,i},n_{\mathcal B,i},n_{\mathcal Z,i}$ are functions of $\omega_i$. Bob announces these counts; the full sequence is retained as additional side information in the proof and need not be announced in the real protocol. Alice and Bob determine the key length by
\endgroup

\begin{equation} \label{keyrate1}
{\ell_{i,h}=\max\left\{0,\left\lfloor n_{\mathcal Z,i}[1-H(e_{\mathrm{ph},i})]-\lambda_h-v-\log_2\frac{1}{4\varepsilon_{PA}^2}\right\rfloor\right\}.}
\end{equation}
Here,  
\begin{equation}\label{ephi}
e_{\text{ph},i}=\text{KU}\left(\frac{p_op_x}{4}\left[\sqrt{\frac{2}{p_o^2}\text{IKU}({n_{\mathcal{O},i}})}+\sqrt{\frac{2}{p_x^2}\text{IKU}({n_{\mathcal{B},i}})}+\sqrt{\frac{8}{p_x^2}\text{IKU}\left(\text{CU}(\frac{p_x^2}{8}c_2^2N)\right)}\right]^2 \right)/n_{{\mathcal{Z}},i},
\end{equation}
and $\text{IKU}(\cdot)$ is defined in Eq.~\eqref{estimator1}, {$\text{KU}(\cdot)$ is defined in Eq.~\eqref{estimator3}}, $\text{CU}(\cdot)$ is defined in Eq.~\eqref{estimator_c} and 
\begin{equation}
{c_2=2\sqrt{(1-e^{-\mu_A/2})(1-e^{-\mu_B/2})}.}
\end{equation}
\begingroup
Here $\lambda_h$ is the number of parity bits announced by Alice during Cascade; the verification hash is accounted for separately by $v=\lceil\log_2(2/\varepsilon_{\mathrm{cor}})\rceil$. If $n_{\mathcal Z,i}=0$ or $e_{\mathrm{ph},i}>1/2$, no key is extracted from that part. The entropy cost of all other Cascade messages is treated explicitly in the proof below.
\endgroup 

\textbf{Step 6.} For each part, if the event $\hat{\Omega}_i$ and $\Omega_{EV}$ happened and the length of the cost bit in the error correction is $\lambda_h$, Alice and Bob apply privacy amplification to $Z_A$ and $\hat{Z}_A$ via a two-universal hash function to produce final keys of length $\ell_{i,h}$, $S_A$ and $S_B$.

While Alice and Bob extract the secure final keys for part $\beta$ according to Eq.~\eqref{keyrate1}, the security of the final keys meets the standards of composable security, i.e., $\varepsilon_{com}$-secure against coherent attacks, where
\begin{equation}
\varepsilon_{com}=\varepsilon_{\mathrm{cor}}+\varepsilon_{\mathrm{sec}}=2\varepsilon_{\mathrm{cor}}+\varepsilon_{PA}+2\sqrt{5\varepsilon_p}.
\end{equation}
Here $\varepsilon_{\mathrm{cor}}$ is the prescribed upper bound on the probability of accepting unequal reconciled strings, $\varepsilon_{PA}$ is the privacy-amplification security parameter, and $\varepsilon_p$ is the failure budget for each concentration bound.

\begingroup
As proved in Sec.~\ref{core} and Appendix~\ref{scs_delta}, each key remains secret when the other key and its complete public transcript are included in Eve's side information. The two keys can therefore be used simultaneously with $2\varepsilon_{com}$ composable security, without assuming independence between the parts.
\endgroup

\subsection{The security of SCS protocol against coherent attacks}\label{core}
To avoid confusion, we reiterate the meanings of the following terms:
\\ \textbf{The real protocol}: the protocol described in Sec.~\ref{main_scs}.
\\ \textbf{The perfect protocol} or \textbf{the actual perfect protocol}: the protocol only differs from the real protocol in the state preparation process. In the perfect protocol, the perfect vacuum and perfect WCS sources are assumed.
\\ \textbf{The entanglement version of the perfect protocol} or \textbf{the entanglement protocol}: the protocol described above and if we aim to prove the security of part 1 (which is defined in step 3 of Sec.~\ref{main_scs}), we regard part 1 as the signal part defined in Eq.~\eqref{entangle}, and part 2 as the estimation part. If we aim to prove the security of part 2, we regard part 2 as the signal part and part 1 as the estimation part.

According to Ref.~\cite{jiang2024side}, we can construct the perfect protocol with following perfect states in Alice and Bob's sides respectively: $s_A=\{\ket{0},\ket{\sqrt{\mu_A}}\}$ and 
$s_B=\{\ket{0},\ket{\sqrt{\mu_B}}\}$, where
\begin{equation}\label{muamub}
\begin{split}
e^{-\mu_A} = \left|\sqrt{ a_{0} \cdot a_{v0}}-\sqrt{(1-{a_{0}})(1-{a_{v0}})}\right|^2,\\
e^{-\mu_B} = \left|\sqrt{ b_{0} \cdot b_{v0}}-\sqrt{(1-{b_{0}})(1-{b_{v0}})}\right|^2,
\end{split}
\end{equation}
and $\ket{\sqrt{\mu_A}},\ket{\sqrt{\mu_B}}$ are perfect coherent states with intensities $\mu_A,\mu_B$ respectively. \textbf{The perfect protocol only differs from the real protocol in the state preparation process}: in the perfect protocol, the states produced by sources $o_A$ and $x_A$ ( $o_B$ and $x_B$) in the all time windows are respectively $\ket{0}$ and $\ket{\sqrt{\mu_A}}$ ($\ket{0}$ and $\ket{\sqrt{\mu_B}}$). Such a perfect protocol can be mapped to the real protocol through attenuation and unitary transformation, thus in the security proof, we can regard that Alice and Bob are actually execute the perfect protocol~\cite{jiang2024side}.

To proof the security of the SCS protocol described in Sec.~\ref{main_scs}, we consider the following equivalent entanglement version of the perfect protocol. In each time window, Alice and Bob prepared the following state:
\begin{equation}\label{entangle}
\ket{\Phi}=\frac{1}{\sqrt{2}}(\ket{\text{signal}}_{L2}\otimes\ket{\phi_{\text{sig}}}+\ket{\text{est}}_{L2}\otimes\ket{\phi_{\text{est}}}),
\end{equation}
where
\begin{equation}
\begin{split}
&\ket{\phi_{\text{sig}}}=\ket{\text{un}}_{L1}\otimes\{p_o\ket{01}_I\otimes \ket{00}_S+p_x\ket{10}_I\otimes\ket{\sqrt{\mu_A}\sqrt{\mu_B}}_S \\
&+\sqrt{2p_op_x}[1/\sqrt{2}(\ket{00}_I\otimes\ket{0\sqrt{\mu_B}}_S+\ket{11}_I\otimes\ket{\sqrt{\mu_A}0}_S]\},\\
&\ket{\phi_{\text{est}}}=2p_x\ket{\text{balance}}_{L1}\otimes \ket{\phi_b}+\sqrt{1-4p_x^2}\ket{\text{rest}}_{L1}\otimes \ket{\phi_r},\\
&\ket{\phi_b}=\frac{1}{2}(\ket{01}_I\otimes \ket{00}_S+\ket{10}_I\otimes\ket{\sqrt{\mu_A}\sqrt{\mu_B}}_S+\ket{00}_I\otimes\ket{0\sqrt{\mu_B}}_S+\ket{11}_I\otimes\ket{\sqrt{\mu_A}0}_S),\\
&\ket{\phi_r}=\frac{1}{\sqrt{1-4p_x^2}}[\sqrt{(p_o^2-p_x^2)}\ket{01}_I\otimes \ket{00}_S+\sqrt{p_x(p_o-p_x)}(\ket{00}_I\otimes\ket{0\sqrt{\mu_B}}_S+\ket{11}_I\otimes\ket{\sqrt{\mu_A}0}_S)].
\end{split}
\end{equation}
Here, the subsystem $I,L1,L2$ is the local memories in Alice's and Bob's laboratory, where $I$ stores the qubit information of each time window, $L1$ stores the classical information of which kinds of parts it belongs to: the balance part, the rest part or the undefined part, $L2$ stores the classical information of whether it is a signal window used to extract the final keys or an estimation window used to perform the phase-error estimation. The subsystem $S$ is send out to Charlie.

It is easy to check that
\begin{equation}\label{equal}
\begin{split}
&\tr_{I,L1}(\oprod{\phi_{\text{sig}}}{\phi_{\text{sig}}})=\tr_{I,L1}(\oprod{\phi_{\text{est}}}{\phi_{\text{est}}}),\\
&=p_o^2\oprod{00}{00}_S+p_x^2\oprod{\sqrt{\mu_A}\sqrt{\mu_B}}{\sqrt{\mu_A}\sqrt{\mu_B}}_S+p_op_x(\oprod{0\sqrt{\mu_B}}{0\sqrt{\mu_B}}_S+\oprod{\sqrt{\mu_A}0}{\sqrt{\mu_A}0}_S).
\end{split}
\end{equation}
Eq.~\eqref{equal} shows that the states sent out to Charlie from the states $\ket{\phi_{\text{sig}}}$ and $\ket{\phi_{\text{est}}}$ are the same, and are the same with those in the the perfect protocol. 

A virtual entanglement protocol that is completely equivalent to the actual perfect protocol is constructed as follows:

\textbf{Step 1.} Alice and Bob prepare the state $\ket{\Phi}^{\otimes N}$ and send out all the subsystems $S$ to Charlie (Eve). In what follows, we no longer distinguish between Charlie and Eve.  

\begingroup
\textbf{Step 2.} Charlie announces the detection results after interacting with the optical systems and his ancillary system $E$. Denote the announced outcome by $\xi_j$. The corresponding conditional state is
\begin{equation}\label{initial}
\ket{\Psi_{\xi_j}}=\frac{1}{\sqrt{\Pr(\xi_j)}}\hat M_{\xi_j}\hat U_{ES}
(\ket\Phi^{\otimes N}\otimes\ket e_E).
\end{equation}
Eve's remaining information is retained in $E$ and need not be measured.

\textbf{Step 3.} Alice and Bob perform the following measurements:
\\ (i) They measure $L2$ to learn the signal/estimation partition. Together with the clicking positions, this fixes both raw-key lengths. They then sample and publicly share all Cascade randomness $R$, independently of the bit values conditional on this record. Its prescribed permutations can thus be fixed before the bit measurement; subsequent adaptive queries still depend on parity disagreements. Independent verification seeds may also be sampled at this stage but are not used by Cascade.
\\ (ii) For clicking signal windows, they measure $I$ with projectors
$P_{\mathcal O}=\oprod{01}{01}$, $P_{\mathcal B}=\oprod{10}{10}$, and $P_{\mathcal Z}=\oprod{00}{00}+\oprod{11}{11}$, obtaining the complete sequence $\omega_i$. Its outcome is denoted by $\Omega_i$. In particular, this measurement identifies the untagged positions without resolving their bit values.
\\ (iii) They measure the untagged signal pairs in the $Z$ basis to obtain the remaining bits of $Z_A,Z_B$. Operations on the estimation part can be performed before or after these signal measurements. Equivalently, that entire part can be retained as a quantum system $Q$ and processed later.

The sequence $\omega_i$ fixes the full bit-error pattern $D=Z_A\oplus Z_B$:
\begin{equation}\label{pattern_main}
D=d(\omega_i),\qquad d_k(\omega_i)=
\begin{cases}1,&t_k\in\{\mathcal O,\mathcal B\},\\0,&t_k=\mathcal Z.\end{cases}
\end{equation}
It also fixes the tagged bit values and specifies the error positions used in Cascade. The type measurement followed by the bit measurement has the same outcome statistics and Eve states as the direct bit measurement in the real protocol. Its record can therefore be supplied to Eve in the proof without modifying the real protocol.

\textbf{Step 4--6.} Error correction, verification, and privacy amplification are as in Sec.~\ref{main_scs}. For the proof of one part, we divide the classical information relevant to reconciliation into three classes:
\\ $C_1$: the first-class record consisting of $\xi_j$, the partition, algorithmic parameters $\Theta$ (the pass schedule, block-size rule, and deterministic tie-breaking and termination rules), independent public seeds including $R$, and the virtual type record $\omega_i$. The other part is retained in $Q$ and processed later, so its actual transcript is not included in this record.
\\ $C_2$: the ordered parity bits announced by Alice during signal-part Cascade.
\\ $C_3$: the other Cascade messages, including Bob's parities, query and correction indices, and control and termination messages.

Verification values, denoted by $V$, are counted separately, and the privacy-amplification seed is fresh and independent. Let $\tilde C$ denote all public information at the end of the signal-part protocol, including the seeds, $V$, and its acceptance flag. These records may be correlated with each other and with the data from the other part.

Let $c$ denote a value of $C_1$, which fixes the partition, $R=r_c$, and $\Theta=\theta_c$. With $A,B$ the ordered untagged signal registers and $Q$ all complementary local systems, the state after Steps 3(i)--(ii), before the bit measurement, is
\begin{equation}\label{state_extension_main}
\rho_{ABQEC_1|\Omega_i,\xi_j}
=\sum_c\Pr(C_1=c|\Omega_i,\xi_j)
\rho_{ABQE|\Omega_i,\xi_j,C_1=c}\otimes\oprod c c_{C_1}.
\end{equation}
The sum runs over compatible records of positive probability, with each quantum block given by the corresponding conditional state. We apply the entropy bounds to these blocks with the full $Q$ as side information, and then average over $c$, as detailed in Appendix~\ref{scs_delta}.

The alternative phase measurement is the joint $X$-parity measurement on each untagged pair. In the even-$Z$-parity subspace it has eigenstates $\ket{X_\pm}=(\ket{00}\pm\ket{11})/\sqrt2$. Following the detector convention below, $\ket{X_+}$ counts as a phase error. Accordingly, in the EUR we use Alice's local $X$ outcome $X_A$ and the complemented local $X$ outcome $X_B$ of Bob, so that $N_{\mathrm{ph}}$ is their Hamming distance. The type projectors commute with this phase-error measurement. The raw-bit and phase measurements are alternative measurements on the same conditional state, not simultaneous measurements of individual $Z$ and $X$ values.

We first state the general variable-length secrecy requirement, retaining the actual leakage as a variable. Let $\Omega_g$ be the event $\hat Z_A=Z_A$ and $\hat\Omega_i$ the tentative type-sequence event in the real protocol. With $\rho_{S_A}^{\mathrm{unif},\ell_{i,h}}$ the uniform state on $\ell_{i,h}$ bits, a sufficient secrecy condition is
\begin{equation}\label{sec_def2}
\begin{split}
\mathcal D_{\mathrm{sec}}:=\sum_{i,j,h}&\frac{\Pr(\hat\Omega_i,\xi_j,\lambda_h,\Omega_{EV})}{2}
\left\|\rho_{S_A\tilde C QE|\hat\Omega_i,\xi_j,\lambda_h,\Omega_{EV}}\right.\\[-2pt]
&\left.\quad-\rho_{S_A}^{\mathrm{unif},\ell_{i,h}}\otimes
\rho_{\tilde C QE|\hat\Omega_i,\xi_j,\lambda_h,\Omega_{EV}}\right\|_1
\le\varepsilon_{\mathrm{sec}}.
\end{split}
\end{equation}
Equivalently, each summand is the trace distance of the corresponding subnormalized states, with event restrictions denoted by $\land$. A bound retaining virtual classical labels implies a bound without them by contractivity of trace distance. Empty-key outputs contribute zero.

The sequences $\omega$ and $\hat\omega$ are obtained by applying Step 5 to $(Z_A,Z_B)$ and $(\hat Z_A,Z_B)$, respectively, where $Z_B$ is Bob's original string. On $\Omega_g$, they agree exactly and verification necessarily passes. Therefore
\[
\hat\Omega_i\land\Omega_{EV}\land\Omega_g
=\Omega_i\land\Omega_g.
\]
For each fixed-length block in Eq.~\eqref{sec_def2}, split the accepted subnormalized state into its $\Omega_g$ and $\Omega_g^c$ contributions. The triangle inequality bounds its secrecy distance by the sum of the two distances; the incorrect contribution is at most its trace. Relabeling only the correct contribution by the above event identity gives~\cite{metger2023security}
\begin{equation}\label{rid_equal}
\mathcal D_{\mathrm{sec}}\le\sum_{i,j,h}\Pr(\Omega_i,\xi_j,\lambda_h)\Delta_{i,j,h}
+\Pr(\Omega_g^c,\Omega_{EV}).
\end{equation}
Here $\Delta_{i,j,h}$ retains the subnormalized restriction to $\Omega_g$ inside the $(\Omega_i,\xi_j,\lambda_h)$ branch. The verification length chosen in Step 4 gives $\Pr(\Omega_g^c,\Omega_{EV})\le\varepsilon_{\mathrm{cor}}$. The reduction is derived in Appendix~\ref{scs_delta}.

To bound Eq.~\eqref{rid_equal}, refine each $(i,j,h)$ branch by $C_1=c$. Let $\sigma^{i,j,c,h}_{Z_AC_2C_3VQE}$ be its normalized state before verification postselection and privacy amplification. The QLHL requires the smooth min-entropy conditioned on all these messages. The leakage chain rule reduces this to their conditional alphabet and the raw-key marginal
\begin{equation}\label{length_general}
\rho^{i,j,c,h}_{Z_AQE}
=\sum_{z_A,z_B}\Pr(z_A,z_B|\Omega_i,\xi_j,C_1=c,\lambda_h)
\oprod{z_A}{z_A}\otimes\rho_{QE}^{z_A,z_B,i,j,c},
\end{equation}
where $\rho^{i,j,c,h}=\tr_{C_2C_3V}\sigma^{i,j,c,h}$. The probabilities in this expression are the weights of the state whose entropy is needed. For general reconciliation, $\lambda_h=L_{r_c,\theta_c}(z_A,z_B)$ may vary among strings with the same error pattern, even with the seeds fixed. Length selection can then change this state.

For Cascade, the complete error pattern $d(\omega_i)$ and the fixed seeds and parameters determine the path, stopping point, and residual error pattern. Appendix~\ref{scs_delta} proves that
\begin{equation}\label{cascade_length_main}
\lambda_h=m_{r_c,\theta_c}(\omega_i)=:m(i,c),
\qquad C_3=f(C_1,C_2).
\end{equation}
Different seeds may give different lengths for the same $\omega_i$. Once $C_1=c$ is fixed, however, a compatible length imposes no further selection:
\begin{equation}\label{length_independence_main}
\Pr(Z_A,Z_B|\Omega_i,\xi_j,C_1=c,\lambda_h)
=\Pr(Z_A,Z_B|\Omega_i,\xi_j,C_1=c).
\end{equation}
The fixed-input $QE$ blocks are unchanged by recording the classical messages. Hence
\begin{equation}\label{rid_h}
\rho^{i,j,c,h}_{Z_AQE}
=\tr_{C_2C_3V}\sigma^{i,j,c,h}_{Z_AC_2C_3VQE}
=\rho_{Z_AQE|\Omega_i,\xi_j,C_1=c}
=:\rho^{i,j,c}_{Z_AQE}.
\end{equation}
This identifies the raw-key state required by the QLHL with the state controlled by the EUR on the same $(i,j,c)$ branch. Its smoothing parameter need not be bounded separately on each branch.

Since $C_2$ has $\lambda_h$ bits, $C_3$ is reconstructible, and $V$ has $v$ bits, their conditional alphabet has at most $2^{\lambda_h+v}$ elements. The smooth leakage chain rule~\cite{tomamichel2015quantum} gives
\begin{equation}\label{leak_main}
H_{\min}^{\delta}(Z_A|C_2C_3VQE)_{i,j,c}
\ge H_{\min}^{\delta}(Z_A|QE)_{i,j,c}-\lambda_h-v.
\end{equation}
Define $p_{i,j,c}=\Pr(\Omega_i,\xi_j,C_1=c)$ and
$q_{i,j,c}=\Pr(N_{\mathrm{ph}}\ge N_{\mathrm{ph},i}^{\mathrm{est}}|\Omega_i,\xi_j,C_1=c)$ in the virtual phase-measurement experiment. For a positive-key branch with $q_{i,j,c}<1$, the smooth EUR and the phase-error entropy bound~\cite{tomamichel2011uncertainty,tupkary2025phase} yield
\begin{equation}\label{eur_main}
H_{\min}^{\sqrt{q_{i,j,c}}}(Z_A|QE)_{i,j,c}
\ge n_{\mathcal Z,i}[1-H(e_{\mathrm{ph},i})].
\end{equation}
Applying the QLHL and averaging over $c$ gives
\begin{equation}\label{ep11a}
\sum_{i,j,h}\Pr(\Omega_i,\xi_j,\lambda_h)\Delta_{i,j,h}
\le\varepsilon_{PA}+2\sqrt{\Delta},
\end{equation}
where, after summing all first-class refinements,
\begin{equation}\label{ep11}
\begin{split}
\Delta&=\sum_{i,j,c}p_{i,j,c}q_{i,j,c}\\
&=\sum_j\Pr(\xi_j)\sum_i
\Pr(N_{\mathrm{ph}}\ge N_{\mathrm{ph},i}^{\mathrm{est}},\Omega_i|\xi_j)
\le\varepsilon.
\end{split}
\end{equation}
The second equality follows by summing over $c$, as shown in Appendix~\ref{scs_delta}. The concentration analysis in Sec.~\ref{rigorous} bounds this averaged probability by $\Delta\le5\varepsilon_p$. Thus Eqs.~\eqref{rid_equal}--\eqref{ep11} give $\varepsilon_{\mathrm{sec}}=\varepsilon_{\mathrm{cor}}+\varepsilon_{PA}+2\sqrt\varepsilon$. The $\varepsilon_{\mathrm{cor}}$ contribution arises here because the SCS phase-error estimate uses window counts inferred from Bob\textquotesingle s corrected string, whose agreement with the true counts is guaranteed on $\Omega_g$. When the estimation inputs are available independently of successful reconciliation, the general arguments in Appendices~\ref{framework} and \ref{actual_leakage_general} give secrecy $\varepsilon_{PA}+2\sqrt\varepsilon$ without this additional term.

Other reconciliation algorithms can be used by fixing a maximum leakage $\lambda_{\max}$ before error correction and aborting before exceeding this budget~\cite{tupkary2024security,tupkary2025phase}. Under the complete-transcript bound specified in Appendix~\ref{max_leakage}, Eq.~\eqref{keyrate1} applies with $\lambda_h$ replaced by $\lambda_{\max}$, while the phase-error estimator and the security parameter remain unchanged. Appendix~\ref{scs_delta} gives the Cascade leakage argument and the weighted QLHL proof. 

Finally, consider both keys together. Write $S_\beta=S_A^\beta$, let $L_\beta$ be the public random variable specifying its actual output length, and let $\boldsymbol\ell=(\ell_1,\ell_2)$ be a value of $\boldsymbol L=(L_1,L_2)$. Length zero denotes the empty key. The proof for part $\beta$ retains the entire complementary system $Q=Q_{3-\beta}$ as side information. Its real processing is a trace-preserving channel whose outputs include the other key, its random length, its transcript, and its acceptance flag. Applying this complete channel and then decomposing the output into classical length blocks gives the following averaged bound, where $F$ contains all public records, including $L_1,L_2$:
\begin{equation}\label{joint_marginal_main}
\begin{split}
\sum_{\boldsymbol\ell}\frac{\Pr(\boldsymbol L=\boldsymbol\ell)}2
&\Bigl\|\rho_{S_\beta S_{3-\beta}FE|\boldsymbol L=\boldsymbol\ell}\\[-2pt]
&\quad-\rho^{\mathrm{unif}}_{S_\beta|\ell_\beta}\otimes
\rho_{S_{3-\beta}FE|\boldsymbol L=\boldsymbol\ell}\Bigr\|_1
\le\varepsilon_{\mathrm{sec}},\qquad\beta=1,2.
\end{split}
\end{equation}
Appendix~\ref{scs_delta} derives this bound with all output lengths retained and combines the two bounds to obtain joint secrecy $2\varepsilon_{\mathrm{sec}}$. The complementary processing preserves the averaged signal phase-error probability; conditional states at individual lengths may differ. The virtual estimation measurement and the real processing remain alternative operations on $Q$.
\endgroup

\subsection{Upper bounded $\Delta$ by introducing virtual observable}\label{rigorous}

\begingroup
We bound $\Delta$ in Eq.~\eqref{ep11} by conditioning on $\xi_j$ and averaging over the random partition and virtual measurement outcomes.
\endgroup

The method proposed in this paper ultimately reduces the computation of the security parameter in composable security to a statistical fluctuation problem, namely, deriving an upper bound on \(\Delta\). In the process of deriving this upper bound on \(\Delta\), we may incorporate hypothetical measurement procedures, {provided that, after their auxiliary outcomes are summed, they preserve the joint statistics of the signal-part type record and phase-error event}. To facilitate our proof, we consider the following virtual steps in lieu of Step 3 in Sec.~\ref{core}:

\textbf{(Virtual) Step 3.} Alice and Bob (they) performs the following measurement operations step by step:
\\ (i) For all windows, Alice and Bob first measure subsystems $L1,L2$ to learn which kinds of parts it belong to.
\\ (ii) They measure the subsystems $C$ to learn the information whether the $i$-th time window cause a click or not. 
\\(iii a) For the clicking windows, if it belongs to the signal part, they measure the subsystems $I$ in the basis $\{\oprod{01}{01}_I,\oprod{10}{10}_I,\oprod{00}{00}_I+\oprod{11}{11}_I\}$ to learn the information that which kind of window it belongs, $\mathcal{O}$, $\mathcal{B}$, or $\mathcal{Z}$. If it belongs to the balance part of the estimation part, they measure the subsystems $I$ in the basis $\{\oprod{--}{--}_I, \mathcal{I}-\oprod{--}{--}_I\}$, where $\ket{-}=\frac{1}{\sqrt{2}}(\ket{0}_I-\ket{1}_I)$ and $\mathcal{I}$ is the identity operator.     
\\(iii b) For the not-clicking windows, if it belongs to the balance part of the estimation part, they measure the subsystems $I$ in the basis $\{\oprod{--}{--}_I, \mathcal{I}-\oprod{--}{--}_I\}$.
\\ (iv) For the clicking $\mathcal{Z}$ windows, they measure the subsystems $I$ in the basis $\{\ket{X_+}_I=\frac{1}{\sqrt{2}}(\ket{00}_I+\ket{11}_I), \ket{X_-}_I=\frac{1}{\sqrt{2}}(\ket{00}_I-\ket{11}_I)\}$. A phase error occurs when Alice and Bob measure the state $\ket{X_+}_I$. 

After Charlie announces all the measurement outcome and before Alice and Bob measure their local systems, Alice, Bob and Eve share the state $\ket{\Psi_{\xi_j}}$. To simplify the expression, we denote $\hat{M}_{\text{eve}}=1/\sqrt{\Pr(\xi_j)}\hat{M}_{{\xi_j}}\hat{U}_{ES}\ket{e}_E$. Let $M_s$ be the number of total clicking windows, and we rearrange the state $\ket{\Phi}^{\otimes N}$ by putting all clicking windows at first which results that
\begin{equation}\label{initial22}
\ket{\Psi_{\xi_j}}=\hat{M}_{\text{eve}}(\ket{\Phi}^{\otimes M_s} \otimes \ket{\Phi}^{\otimes N-M_s}).
\end{equation}

Alice and Bob measure the state in Eq.~\eqref{initial22} round by round. Let $F_u$ be the measurement result in the $u$-th round, and let $\vec{F}_{u-1}=F_1,\cdots,F_{u-1}$ be the measurement results of the former $u-1$ round.

For clarify, we define
\\$n_\mathcal{O}$: the observed values of the number of $\ket{01}_I$ detected in the clicking signal windows, i.e., the number of the clicking events caused by the sent out state $\ket{00}_S$ in the signal windows.
\\$n_\mathcal{B}$: the observed values of the number of $\ket{10}_I$ detected in the clicking signal windows, i.e., the number of the clicking events caused by the sent out state $\ket{\sqrt{\mu_A}\sqrt{\mu_B}}_S$ in the signal windows.
\\$n_\mathcal{Z}$: the observed values of the number of $\oprod{00}{00}_I+\oprod{11}{11}_I$ detected in the clicking signal windows, i.e., the number of the clicking events caused by the sent out state $\oprod{0\sqrt{\mu_B}}{0\sqrt{\mu_B}}_S+\oprod{\sqrt{\mu_A}0}{\sqrt{\mu_A}0}_S$ in the signal windows.  
\\$n_\mathcal{D}$: the virtual observed values of the number of $\ket{--}_I$ detected in the clicking estimation windows, i.e., the number of the clicking events caused by the sent out state $\ket{\phi_2}$ in the estimation windows. Note $n_\mathcal{D}$ can only be observed in the virtual step 3, and can not be observed in the actual protocol.
\\$n_\mathcal{D}^{\text{all}}$: the virtual observed values of the number of $\ket{--}_I$ detected in all estimation windows, i.e., the number of the sent out state $\ket{\phi_2}$ in the estimation windows. Note $n_\mathcal{D}^{\text{all}}$ can only be observed in the virtual step 3, and can not be observed in the actual protocol.
\\$N_{\text{ph}}$: the observed values of the number of phase errors in the clicking signal windows. Note $N_{\text{ph}}$ can only be observed in the virtual step 3, and can not be observed in the actual protocol.
\\$\mean{n_\mathcal{O}}$: the expected values of $n_\mathcal{O}$. Mathematically, $\mean{n_\mathcal{O}}=\sum_{u=1}^{M_s}\Pr(F_u=\mathcal{O}|\vec{F}_{u-1})$.
\\$\mean{n_\mathcal{B}}$: the expected values of $n_\mathcal{B}$. Mathematically, $\mean{n_\mathcal{B}}=\sum_{u=1}^{M_s}\Pr(F_u=\mathcal{B}|\vec{F}_{u-1})$.
\\$\mean{n_\mathcal{D}}$: the expected values of $n_\mathcal{D}$. Mathematically, $\mean{n_\mathcal{D}}=\sum_{u=1}^{M_s}\Pr(F_u=\mathcal{D}|\vec{F}_{u-1})$.
\\$\mean{N_{\text{ph}}}$: the expected values of $N_{\text{ph}}$. Mathematically, $\mean{N_{\text{ph}}}=\sum_{u=1}^{M_s}\Pr(F_u=X_+|\vec{F}_{u-1})$.

Remark: Keep in mind that all the observed values and mean values are defined in the condition that Charlie has observed $\xi$.

After the virtual step 3 (iii) defined in beginning of this section, Alice and Bob obtains a group of measurement results $\vec{F}_N$. Let $\{\tilde{\Omega}_s\}$ be the set that contains all possible different $\vec{F}_N$ and each $\tilde{\Omega}_s$ corresponds to a unique $\vec{F}_N$. Let $n_{\mathcal{O},s},n_{\mathcal{B},s},n_{\mathcal{Z},s}, N_{\text{ph},s}, n_{\mathcal{D},s}$ be the corresponding number of the corresponding measurement results in $\tilde{\Omega}_s$. Based on the values of $n_{\mathcal{O},s},n_{\mathcal{B},s}$ and Eq.~\eqref{ephi}, Alice and Bob obtained $N_{\text{ph},s}^\text{est}$, and if $N_{\text{ph},s}\ge N_{\text{ph},s}^\text{est}$, the event $N_{\text{ph}} \ge N_{\text{ph},i}^{\text{est}},\Omega_i | \xi_j$ happened, which results
\begin{equation}\label{eq43}
\Delta=\sum_j\Pr(\xi_j)\sum_s \Pr(\tilde{\Omega}_s|\xi_j)\Pr(N_{\text{ph},s}\ge N_{\text{ph},s}^\text{est}|\tilde{\Omega}_s,\xi_j).
\end{equation}
Note conditioned on the event $\tilde{\Omega}_s$, $\Pr(N_{\text{ph},s}\ge N_{\text{ph},s}^\text{est}|\tilde\Omega_s,\xi_j)$ equals $0$ or $1$. 

Next, we will demonstrate that the phase-error rate estimator constructed in Sec.~\ref{intuition} leads to 
\begin{equation}\label{main22}
\Delta\le 5\varepsilon_p=\varepsilon,
\end{equation}

We first calculate $\Delta_j=\sum_s \Pr(\tilde{\Omega}_s|\xi_j)\Pr(N_{\text{ph},s}\ge N_{\text{ph},s}^\text{est}|\tilde{\Omega}_s)$. We can divide events $\tilde{\Omega}_{s}$ into three categories
\begin{equation}
\begin{split}
&\text{case 1: } N_{\text{ph},s}\ge N_{\text{ph},s}^\text{est} \text{ and } \mean{N_{\text{ph},s}^{\text{est}}} \ge \mean{N_{\text{ph},s}}\\
&\text{case 2: } N_{\text{ph},s}\ge N_{\text{ph},s}^\text{est} \text{ and } \mean{N_{\text{ph},s}^{\text{est}}} < \mean{N_{\text{ph},s}},\\
&\text{case 3: } N_{\text{ph},s}< N_{\text{ph},s}^\text{est}.
\end{split}
\end{equation} 
Note unlike the Chernoff bound, the exceptions defined in the Kato's inequality is determined by $\xi_j$ and $\tilde{\Omega}_{s}$.

Then we have
\begin{equation}\label{eq1}
\Delta_j=\Pr(N_{\text{ph}}\ge N_{\text{ph}}^\text{est},\ \mean{N_{\text{ph}}^{\text{est}}} \ge \mean{N_{\text{ph}}}|\xi_j)+\Pr(N_{\text{ph}}\ge N_{\text{ph}}^\text{est}, \mean{N_{\text{ph}}^{\text{est}}} < \mean{N_{\text{ph}}}|\xi_j), 
\end{equation}
where
\begin{align}
&\Pr(N_{\text{ph}}\ge N_{\text{ph}}^\text{est},\ \mean{N_{\text{ph}}^{\text{est}}} \ge \mean{N_{\text{ph}}}|\xi_j)\equiv \sum_{s} \Pr(\tilde{\Omega}_{s}|\xi_j)\cdot\Pr(N_{\text{ph},s}\ge N_{\text{ph},s}^\text{est}, \mean{N_{\text{ph},s}^{\text{est}}} \ge \mean{N_{\text{ph},s}}|\xi_j),\\
&\Pr(N_{\text{ph}}\ge N_{\text{ph}}^\text{est},\ \mean{N_{\text{ph}}^{\text{est}}} < \mean{N_{\text{ph}}}|\xi_j)\equiv \sum_{s} \Pr(\tilde{\Omega}_{s}|\xi_j)\cdot\Pr(N_{\text{ph},s}\ge N_{\text{ph},s}^\text{est}, \mean{N_{\text{ph},s}^{\text{est}}} <\mean{N_{\text{ph},s}}|\xi_j). 
\end{align}
Hereafter, quantities with the subscript $_{s}$ refer to the actual observed values, while those without subscripts denote the corresponding random variables.

For the first term in \eqref{eq1}, we have
\begin{equation}\label{eq1.1}
\begin{split}
&\Pr(N_{\text{ph}}\ge N_{\text{ph}}^\text{est},\ \mean{N_{\text{ph}}^{\text{est}}} \ge \mean{N_{\text{ph}}}|\xi_j)\\
&=\Pr(N_{\text{ph}}\ge \mean{N_{\text{ph}}^{\text{est}}}+\sqrt{-0.5M_s{\ln\varepsilon_p} },\mean{N_{\text{ph}}^{\text{est}}} \ge \mean{N_{\text{ph}}}|\xi_j)\\
&\le\Pr(N_{\text{ph}}\ge \mean{N_{\text{ph}}}+\sqrt{-0.5M_s{\ln\varepsilon_p}},\mean{N_{\text{ph}}^{\text{est}}} \ge \mean{N_{\text{ph}}}|\xi_j)\\
&\le\Pr(N_{\text{ph}}\ge \mean{N_{\text{ph}}}+\sqrt{-0.5M_s{\ln\varepsilon_p}}|\xi_j)\\
&\le \varepsilon_p.
\end{split}
\end{equation}
We use $N_{\text{ph}}^\text{est}=\text{KU}(\mean{N_{\text{ph}}^{\text{est}}})$ and the definition of $\text{KU}(\cdot)$ in Eq.~\eqref{estimator3} in the first equality. In the first inequality, we use the fact that if the event ''$N_{\text{ph}}\ge \mean{N_{\text{ph}}^{\text{est}}}+\sqrt{-0.5M_s{\ln\varepsilon_p}},\mean{N_{\text{ph}}^{\text{est}}} \ge \mean{N_{\text{ph}}}|\xi_j$'' occurs, the event ''$N_{\text{ph}}\ge \mean{N_{\text{ph}}}+\sqrt{-0.5M_s{\ln\varepsilon_p}},\mean{N_{\text{ph}}^{\text{est}}} \ge \mean{N_{\text{ph}}}|\xi_j$'' must occur. In the third inequality, we use the definition of $\mean{N_{\text{ph}}}=\sum_{u=1}^{M_s}\Pr(F_u=X_+|\vec{F}_{u-1})$ and Eq.~\eqref{kato3}. 

Note $\Pr(N_{\text{ph}}\ge \mean{N_{\text{ph}}}+\sqrt{-0.5M_s{\ln\varepsilon_p}}|\xi_j)$ means if Alice and Bob measure the local memories of the clicking windows in state $\ket{\Psi_{\xi_j}}$ defined in Eqs.~(\ref{initial},\ref{initial22}) according to the virtual step 3, the probability that the event $N_{\text{ph}}\ge \mean{N_{\text{ph}}}+\sqrt{-0.5M_s{\ln\varepsilon_p}}$ occurs. Thus the condition $\xi_j$ here only represents the state corresponding to the stochastic process considered here, and does not affect the application of Eq.~\eqref{kato3}. 

For the second term in \eqref{eq1}, we have
\begin{equation}\label{eq1.2}
\Pr(N_{\text{ph}}\ge N_{\text{ph}}^\text{est}, \mean{N_{\text{ph}}^{\text{est}}} < \mean{N_{\text{ph}}}|\xi_j)\le \Pr(\mean{N_{\text{ph}}^{\text{est}}} < \mean{N_{\text{ph}}}|\xi_j)
\end{equation} 

According to Eqs.~(\ref{eq24},\ref{eq29}), if 
\begin{equation}
\mean{n_\mathcal{O}^{\text{est}}}\ge \mean{n_\mathcal{O}},\mean{n_\mathcal{B}^{\text{est}}}\ge \mean{n_\mathcal{B}},\mean{n_\mathcal{D}^{\text{est}}}\ge \mean{n_\mathcal{D}},
\end{equation}
then $\mean{N_{\text{ph}}^{\text{est}}} \ge \mean{N_{\text{ph}}}$. Thus
\begin{equation}\label{term2}
\begin{split}
&\Pr(\mean{N_{\text{ph}}^{\text{est}}} < \mean{N_{\text{ph}}}|\xi_j)\\
&\le \Pr[\neg(\mean{n_\mathcal{O}^{\text{est}}}\ge \mean{n_\mathcal{O}}\land \mean{n_\mathcal{B}^{\text{est}}}\ge \mean{n_\mathcal{B}}\land \mean{n_\mathcal{D}^{\text{est}}}\ge \mean{n_\mathcal{D}})|\xi_j]\\
&=\Pr(\mean{n_\mathcal{O}^{\text{est}}}< \mean{n_\mathcal{O}}\lor \mean{n_\mathcal{B}^{\text{est}}}< \mean{n_\mathcal{B}}\lor \mean{n_\mathcal{D}^{\text{est}}}< \mean{n_\mathcal{D}}|\xi_j)\\
&\le \Pr(\mean{n_\mathcal{O}^{\text{est}}}< \mean{n_\mathcal{O}}|\xi_j)+\Pr(\mean{n_\mathcal{B}^{\text{est}}}< \mean{n_\mathcal{B}}|\xi_j)+\Pr(\mean{n_\mathcal{D}^{\text{est}}}< \mean{n_\mathcal{D}}|\xi_j).
\end{split}
\end{equation}

For the first term in Eq.~\eqref{term2}, we have

\begin{equation}\label{term2.1}
\begin{split}
&\Pr(\mean{n_\mathcal{O}^{\text{est}}}< \mean{n_\mathcal{O}}|\xi_j)\\
&=\Pr(n_\mathcal{O}+\left[b+a \left(\frac{2 n_\mathcal{O}}{M_s} - 1 \right) \right] \sqrt{M_s}< \mean{n_\mathcal{O}}|\xi_j)\\
&\le {\exp\left[\frac{-2(b^{ 2}-a^{ 2})}{(1+\frac{4a}{3\sqrt{M_s}})^2}\right]}\\
&=\varepsilon_p,
\end{split}
\end{equation}
where we use $\mean{n_\mathcal{O}^{\text{est}}}=\text{IKU}(n_{\mathcal{O}})$ and the definition of $\text{IKU}(\cdot)$ in Eq.~\eqref{estimator1} in the first equality, and apply Eq.~\eqref{kato1} in the first inequality, and apply Eq.~\eqref{abp} in the second equality. For the second term in Eq.~\eqref{term2}, with the same argument, we have
\begin{equation}\label{term2.2}
\begin{split}
&\Pr(\mean{n_\mathcal{B}^{\text{est}}}< \mean{n_\mathcal{B}}|\xi_j)\le \varepsilon_p,
\end{split}
\end{equation}

For the third term in Eq.~\eqref{term2}, we have
\begin{equation}\label{term2.3}
\Pr(\mean{n_\mathcal{D}^{\text{est}}}< \mean{n_\mathcal{D}}|\xi_j)=\Pr(\mean{n_\mathcal{D}^{\text{est}}}< \mean{n_\mathcal{D}},n_\mathcal{D}^{\text{all}}< n_\mathcal{D}^{\text{est,all}}|\xi_j)+\Pr(\mean{n_\mathcal{D}^{\text{est}}}< \mean{n_\mathcal{D}},n_\mathcal{D}^{\text{all}}\ge n_\mathcal{D}^{\text{est,all}}|\xi_j)
\end{equation}
{For the first term in Eq.~\eqref{term2.3}, we use that $\text{IKU}(x)$ is nondecreasing for the chosen parameters $a\ge-\sqrt{M_s}/2$; the parameter convention is specified in Appendix~\ref{tightbound}. Thus}
\begin{equation}\label{term2.3.1}
\begin{split}
&\Pr(\mean{n_\mathcal{D}^{\text{est}}}< \mean{n_\mathcal{D}},n_\mathcal{D}^{\text{all}}< n_\mathcal{D}^{\text{est,all}}|\xi_j)\\
&=\Pr(n_\mathcal{D}^{\text{est,all}}+\left[b+a \left(\frac{2 n_\mathcal{D}^{\text{est,all}}}{M_s} - 1 \right) \right] \sqrt{M_s}< \mean{n_\mathcal{D}},n_\mathcal{D}^{\text{all}}< n_\mathcal{D}^{\text{est,all}}|\xi_j)\\
&\le \Pr(n_\mathcal{D}^{\text{all}}+\left[b+a \left(\frac{2 n_\mathcal{D}^{\text{all}}}{M_s} - 1 \right) \right] \sqrt{M_s}< \mean{n_\mathcal{D}},n_\mathcal{D}^{\text{all}}< n_\mathcal{D}^{\text{est,all}}|\xi_j),\\
&\le \Pr(n_\mathcal{D}^{\text{all}}+\left[b+a \left(\frac{2 n_\mathcal{D}^{\text{all}}}{M_s} - 1 \right) \right] \sqrt{M_s}< \mean{n_\mathcal{D}}|\xi_j),\\
&\le \Pr(n_\mathcal{D}+\left[b+a \left(\frac{2 n_\mathcal{D}}{M_s} - 1 \right) \right] \sqrt{M_s}< \mean{n_\mathcal{D}}|\xi_j)\\
&\le \varepsilon_p.
\end{split}
\end{equation}
For the second term in Eq.~\eqref{term2.3}, we have
\begin{equation}
\Pr(\mean{n_\mathcal{D}^{\text{est}}}< \mean{n_\mathcal{D}},n_\mathcal{D}^{\text{all}}\ge n_\mathcal{D}^{\text{est,all}}|\xi_j)\le\Pr(n_\mathcal{D}^{\text{all}}\ge n_\mathcal{D}^{\text{est,all}}|\xi_j) 
\end{equation}

Combining Eqs.~(\ref{eq1},\ref{eq1.1},\ref{eq1.2},\ref{term2},\ref{term2.1},\ref{term2.2},\ref{term2.3},\ref{term2.3.1}), we have
\begin{equation}
\begin{split}
\Delta_j \le 4\varepsilon_p+\Pr(n_\mathcal{D}^{\text{all}}\ge n_\mathcal{D}^{\text{est,all}}|\xi_j).
\end{split}
\end{equation}
Then
\begin{equation}
\Delta=\sum_{j}\Pr(\xi_j) \Delta_j\le \sum_{j}\Pr(\xi_j)[4\varepsilon_p+\Pr(n_\mathcal{D}^{\text{all}}\ge n_\mathcal{D}^{\text{est,all}}|\xi_j)]=4\varepsilon_p+\Pr(n_\mathcal{D}^{\text{all}}\ge n_\mathcal{D}^{\text{est,all}}).
\end{equation}

$\Pr(n_\mathcal{D}^{\text{all}}\ge n_\mathcal{D}^{\text{est,all}})$ is the probability that the event $n_\mathcal{D}^{\text{all}}\ge n_\mathcal{D}^{\text{est,all}}$ happened while Alice and Bob measure their local systems in state $\hat{U}_{ES} (\ket{\Phi}^{\otimes N} \otimes \ket{e}_E)$ one by one in the basis $\{\oprod{\text{est}}{\text{est}}_{L2}\otimes \oprod{\text{balance}}{\text{balance}}_{L1} \otimes \oprod{--}{--}_I, \mathcal{I}-\oprod{\text{est}}{\text{est}}_{L2}\otimes \oprod{\text{balance}}{\text{balance}}_{L1} \otimes \oprod{--}{--}_I\}$. For each time windows, the probability that $\oprod{--}{--}_I$ occurs is independent and identically distributed, and 
\begin{equation}
p_{\mathcal{D}}=\tr(\oprod{\text{est}}{\text{est}}_{L2}\otimes \oprod{\text{balance}}{\text{balance}}_{L1} \otimes \oprod{--}{--}_I\ket{\Phi}\bra{\Phi})=\frac{p_x^2}{8}c_2^2.
\end{equation}
Thus
\begin{equation}
\Pr(n_\mathcal{D}^{\text{all}}\ge n_\mathcal{D}^{\text{est,all}})=\Pr\left(n_\mathcal{D}^{\text{all}}\ge \text{CU}(\frac{p_x^2}{8}c_2^2N)\right)\le \varepsilon_p,
\end{equation}
where we have applied $n_\mathcal{D}^{\text{est,all}}=\text{CU}(\frac{p_x^2}{8}c_2^2N)$ and Eqs.~(\ref{B1}-\ref{B3}).

Finally, by setting $5\varepsilon_p=\varepsilon$, we obtains Eq.~\eqref{main22}.

\subsection{Numerical simulation}\label{numerical}

To evaluate the performance of our improved security-proof method for the SCS protocol, we conducted numerical simulations using a linear channel model. This model assumes symmetric channels between Alice-Charlie and Bob-Charlie, with loss characterized by the fiber attenuation coefficient and detector efficiency. The misalignment error and dark count rates are also incorporated to simulate realistic conditions.

The experimental parameters used in the simulations are listed in Table~\ref{exproperty}. These values are chosen to reflect typical settings in practical QKD systems, including a low dark count rate, moderate misalignment, and high detector efficiency.

\begin{table}[htbp]
\centering
\begin{tabular}{cccccc}
\hline
$p_d$& $e_d$ &$\eta_d$ & $f$ & $\alpha_f $ & $2\varepsilon_{\text{com}}$ \\
\hline
$1.0\times10^{-9}$& {$3\%$}  & {$60.0\%$} & $1.16$ & $0.2$ & $10^{-10}$\\ 
\hline
\end{tabular}
\caption{Experimental parameters for numerical simulations. Here $p_d$ is the dark counting rate per pulse of Charlie's detectors; $e_d$ is the misalignment-error probability; $\eta_d$ is the detection efficiency of Charlie's detectors; $f$ is the error correction inefficiency; $\alpha_f$ is the fiber loss coefficient (dB/km).}\label{exproperty}
\end{table}

The key rates were calculated using the formula given in Eq.~\eqref{keyrate1}, incorporating the phase error rate from Eq.~\eqref{ephi}. We optimized the probabilities $p_o$ and $p_x$, as well as the intensities $\mu_A$ and $\mu_B$, for each distance to maximize the key rate. We set $\varepsilon_{\mathrm{cor}}=\varepsilon_{\text{PA}}=\sqrt{5\varepsilon_{p}}.$

\begin{figure}
\centering
\includegraphics[width=9cm]{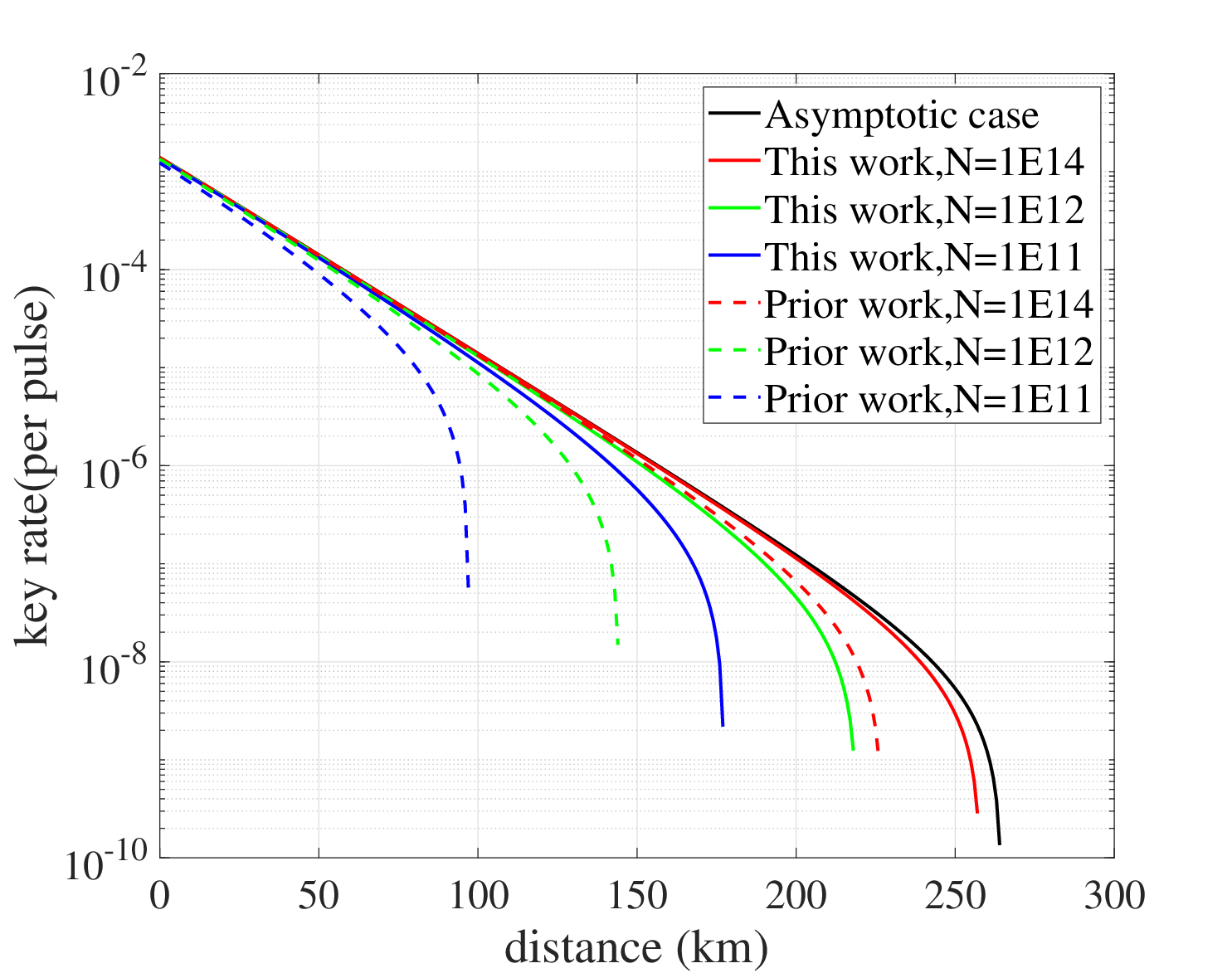}
\caption{Comparison of key rates with different $N$. The curves show the secret key rate (bits per pulse) versus fiber distance (km) for total pulse numbers $N = 10^{11}$, $10^{12}$, and $10^{14}$. Our method outperforms prior works, enabling secure QKD at longer distances with fewer pulses.}
\label{core1}
\end{figure}

Figure~\ref{core1} compares the secret key rates as a function of distance for different total pulse numbers $N = 10^{11}, 10^{12}, 10^{14}$. The results demonstrate that our method achieves higher key rates and longer distances compared to previous approaches using post-selection techniques. Notably, for $N = 10^{12}$, secure keys can be distributed over distances exceeding 200 km. {The reduction in required pulses by over two orders of magnitude demonstrates the smaller finite-key penalty of our direct security analysis compared with the post-selection approach.}

\begin{figure}
\centering
\includegraphics[width=9cm]{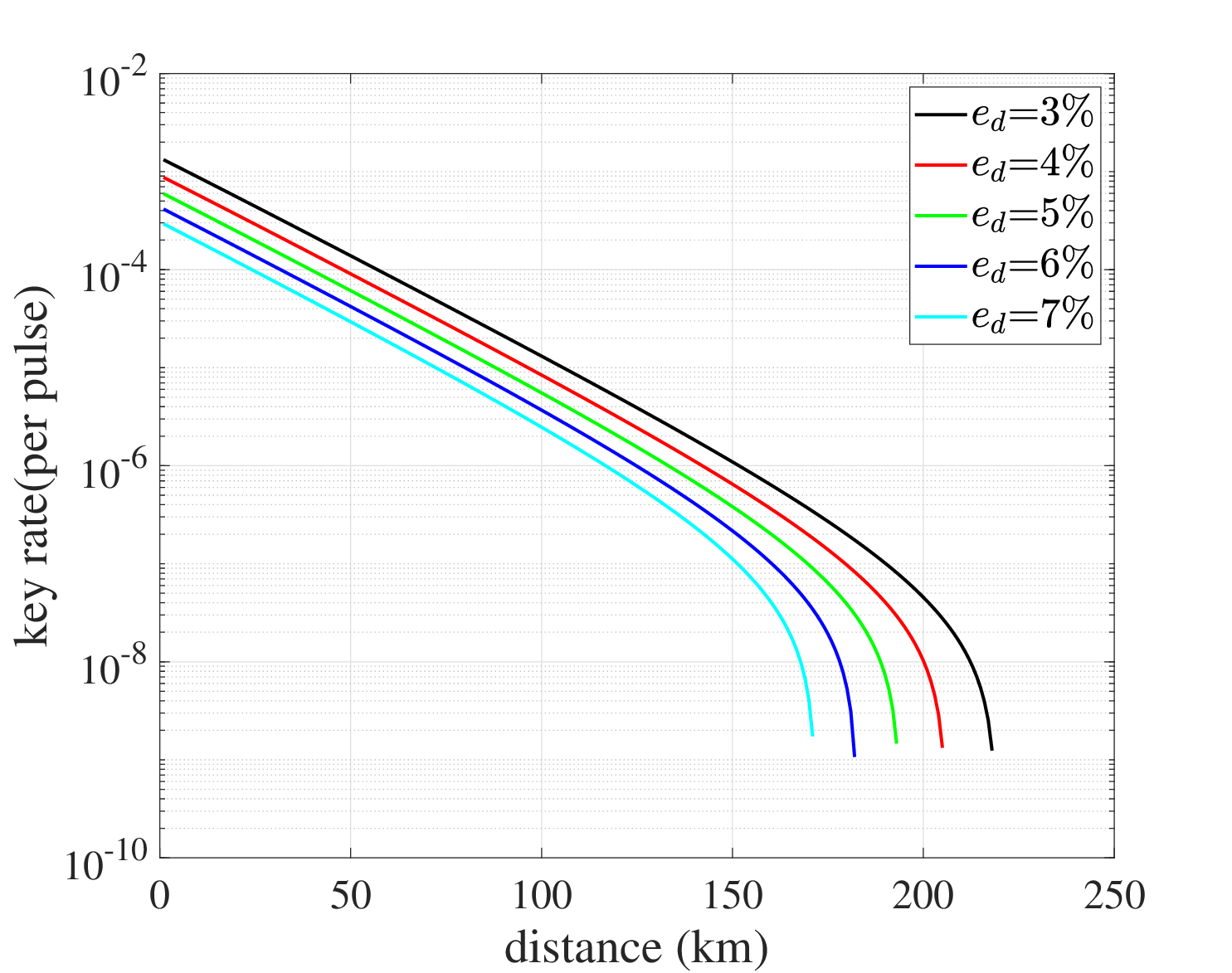}
\caption{Comparison of key rates with different $e_d$. The curves show the secret key rate (bits per pulse) versus fiber distance (km) for the misalignment-error probability $e_d = 3\%,4\%,5\%,6\%$ and $7\%$. High $e_d$ is commonly observed in both deployed field fiber links and free-space channels. The results shows that the maximum transmission distance of the SCS protocol decreases only mildly with increasing $e_d$.}
\label{core2}
\end{figure}

Figure~\ref{core2} compares the secret key rates as a function of distance for the misalignment-error probability $e_d = 3\%,4\%,5\%,6\%$ and $7\%$. High $e_d$ is commonly observed in both deployed field fiber links and free-space channels~{\cite{chen2021twin,li2026free}}. The results shows that the maximum transmission distance of the SCS protocol decreases only mildly with increasing $e_d$.It is worth noting that even with $  e_d  $ as high as 5\%, the maximum transmission distance of the SCS protocol remains close to 200 km.

\section{Discussion}\label{discussion}
In deriving an upper bound on $\Delta$, our primary technique involves constructing a phase-error estimator. Specifically, we first define appropriate expected values; second, establish the relationship between the expected value of the phase-flip error and that of the corresponding observables; third, employ the inverse form of a concentration inequality to derive an upper bound on the expected phase-error based on the observed values of the observables; and finally, apply a concentration inequality to obtain an upper bound on the phase-error rate itself, leveraging the aforementioned bound on its expected value. Subsequently, based on this constructed phase-error estimator, we utilize techniques analogous to those in Sec.~\ref{rigorous} to prove the associated failure probability. Employing this approach, we can clarify whether a given concentration bound, together with its inverse use, provides a valid protocol-level failure-probability bound in variable-length QKD.

For a generic random process, when the expectation $\mathcal{E}$ and a suitable concentration inequality are known, one typically bounds the random variable $\mathcal{X}$ from above or below with a prescribed failure probability. QKD, however, almost always poses the inverse problem: given a realized observation $\mathcal{X}$, infer an upper or lower bound on the expectation $\mathcal{E}$, again with controlled failure probability.

For this inverse problem—bounding the expectation from a realized observation—the relevant probability is
\[
\Pr(\mathcal{E} \ge \operatorname{Xup}(\mathcal{X}) \mid \mathcal{X}),
\]
where $\operatorname{Xup}(\cdot)$ denotes the inverse bound. {When $\mathcal{E}$ is fixed by the preceding conditioning record, as in the inverse Chernoff argument below, this conditional probability is either 0 or 1. For a correlated process, the sum of conditional expectations can itself be random and must be treated within the corresponding concentration argument.}

The prior applications of the inverse bound in QKD are, in essence, attempts to compute the following average failure probability, namely
\begin{equation}\label{DD1}
\sum_i \Pr(\mathcal{X}_i) \cdot \Pr\bigl(\mathcal{E} \ge \operatorname{Xup}(\mathcal{X}_i) \mid \mathcal{X}_i\bigr),
\end{equation}
or, more generally,
\begin{equation}\label{DD2}
\sum_i \Pr\bigl(\mathcal{E} \ge \operatorname{Xup}(\mathcal{X}_i) , \mathcal{X}_i\bigr).
\end{equation}
{This average is over all possible observations in a single execution of the protocol. Different observations can lead to different output key lengths, and Eq.~\eqref{DD1} is the resulting protocol-level failure probability used in the variable-length security proof.}

{In Appendix~\ref{confirm}, we examine Kato's inequality~\cite{kato2020concentration,curras2021tight}, the Chernoff bound~\cite{chernoff1952measure,zhang2017improved} and their inverse forms, as well as Serfling's random-sampling inequality~\cite{serfling1974probability}. These bounds provide protocol-level parameter-estimation guarantees under the stated conditioning assumptions. In the Kato argument used here, each observable class has a pair $(a,b)$ fixed before its observations. Choosing these parameters from the same data requires an additional uniform bound or an allocation of failure probabilities over the possible choices, as discussed in Appendix~\ref{kato_no}.}

{In summary, we establish a composable finite-key security proof for variable-length SCS-QKD against coherent attacks without resorting to the post-selection technique. By introducing a virtual observable in an equivalent entanglement-based protocol, we obtain a finite-key phase-error bound that substantially improves the achievable key rates and reduces the required pulse numbers. The proof also justifies determining the final key length after Cascade error correction using the actual parity disclosures and post-reconciliation window statistics, while ensuring the joint security of the keys extracted from both subsets. A supplementary Kraft-bound argument provides a sufficient condition for using actual leakage in more general reconciliation procedures. Furthermore, we clarify the conditioning and parameter-selection requirements for applying concentration bounds in variable-length QKD. These results strengthen the practical performance of SCS-QKD and place its adaptive key-length determination on a rigorous composable-security footing.}

\section*{Data availability} The data that support the findings of this study are available from the corresponding author upon reasonable request.
\section*{Code availability}
The code used to generate the numerical results is available from the corresponding author upon reasonable request.

\section*{Acknowledgements} This work was supported by National Natural Science Foundation of China Grant Nos. 12374473, 12174215, 12104184; Innovation Program for Quantum Science and Technology No. 2021ZD0300705; the Taishan Scholars Program.

\section*{Author contributions} X.-B.W. conceived the original idea and supervised the project. C.J. carried out the theoretical derivations and main calculations. C.J. and Z.-W.Y. jointly wrote the manuscript. All authors contributed to the scientific discussions and finalized the manuscript.

\section*{Competing interests} The authors declare no competing interests.

\bibliography{refs-jiang.bib}
\color{black}

\appendix

\section{{Determine the secure final key length after error correction for variable-length QKD against coherent attacks}}\label{framework}
Here we extend the Cascade argument to QKD protocols admitting an entanglement-based qubit description with the conventional phase-error observable, using the Cascade rules specified in Sec.~\ref{main_scs}. We first consider estimation and tagging inputs that are available independently of successful reconciliation, so that the implemented entropy bound remains valid on every accepted run. Estimation inputs reconstructed only after error correction, as in SCS, are treated separately at the end.

For the protocol considered in this appendix, let $\omega_i$ denote the parameter-estimation and tagging data, and let $n_{\mathrm{un},i}$ be the number of untagged raw-key pairs. The detection record is denoted by $\xi_j$, and $C_1=c$ specifies the pre-bit-measurement data, including the partition, independent public seeds (also the verification seed), and algorithmic parameters. With the complementary systems retained in $Q$, the conditional state before the raw-key measurement can be written as
\begin{equation}\label{initial1}
\rho_{ABQE|\Omega_i,\xi_j,C_1=c}.
\end{equation}
No product decomposition between the classical records and the quantum systems is assumed.

\textbf{Virtual parity measurement.} Before reading the raw bits, measure each pair's $ZZ$ parity with
\begin{equation}\label{zz_projectors}
P_{d_k}^{(k)}=\frac{\mathcal I+(-1)^{d_k}Z_{A_k}Z_{B_k}}2,
\qquad P_d=\bigotimes_k P_{d_k}^{(k)}.
\end{equation}
The outcome $D=d$ is the complete bit-error pattern, not merely its Hamming weight. The conditional state is
\begin{equation}\label{zz_state}
\rho_{ABQE|\Omega_i,\xi_j,C_1=c,D=d}
=\frac{P_d\rho_{ABQE|\Omega_i,\xi_j,C_1=c}P_d}
{\tr(P_d\rho_{ABQE|\Omega_i,\xi_j,C_1=c})}.
\end{equation}
One may give $D$ to Eve in the proof. This is a virtual nonlocal measurement; it is not an extra operation required of the real protocol. Each $P_d$ commutes with the subsequent local $Z$-basis measurement, so summing $d$ recovers the real raw-key state and its Eve correlations.

For the conventional phase error, the per-pair observable is the $XX$ parity, with the sign adjusted to the protocol's correlation convention. The elementary commutation relations are
\begin{equation}\label{zz_xx_commute}
[Z_{A_k}Z_{B_k},X_{A_k}X_{B_k}]=0,\qquad
[P_d,Z_{A_k}]=[P_d,Z_{B_k}]=0.
\end{equation}
Thus the virtual parity measurement can precede either the raw-bit measurement or the phase-error measurement. Individual local $X$ outcomes generally do not commute with $P_d$, but the phase-error pattern does. In particular, this does not assert that the individual $Z$ bits and the phase-error observable commute with each other.

Suppose the parameter-estimation and tagging procedure leaves the retained qubits with this description and that its phase-error failure effect $B_i$ commutes with $P_d$. This holds, for example, when the relevant estimation data have already been generated on separate systems and $B_i$ is the spectral projector for a threshold on the retained pairs' $XX$ parities. Then
\begin{equation}\label{zz_marginal}
\sum_d\tr\left(B_iP_d\rho_{ABQE|\Omega_i,\xi_j,C_1=c}P_d\right)
=\tr\left(B_i\rho_{ABQE|\Omega_i,\xi_j,C_1=c}\right).
\end{equation}
Assume also that summing the first-class record preserves the phase-failure probability of the protocol's parameter-estimation experiment, as established for SCS in Appendix~\ref{scs_delta}. Summing over $c$ and $d$ then gives the averaged phase-error probability to which the parameter-estimation bound applies. The conditional failure probabilities on individual branches may differ. Protocols with different effective observables require their own compatibility check.

\textbf{Cascade and privacy amplification.} Fix the full error pattern of every string entering Cascade, including any tagged positions, and all algorithmic seeds and parameters. The induction in Appendix~\ref{scs_delta} gives $\lambda_h=m(d,r_c,\theta_c)$, and reconstructs all other Cascade messages from Alice's parity string and the first-class record. If only the untagged subset is used for the EUR, the remaining raw-key systems must also be retained as side information or accounted for by the tagging argument. In the SCS protocol their bit values are fixed by $\omega_i$.

For $q_{i,j,c,d}=\Pr(N_{\mathrm{ph}}\ge N_{\mathrm{ph},i}^{\mathrm{est}}|\Omega_i,\xi_j,C_1=c,D=d)$, on positive-key branches with $q_{i,j,c,d}<1$, the conditional smooth EUR and the phase-error entropy bound used in Appendix~\ref{scs_delta} give
\begin{equation}\label{general_cascade_entropy}
H_{\min}^{\sqrt{q_{i,j,c,d}}}(Z_A|C_2C_3VQE)_{i,j,c,d}
\ge n_{\mathrm{un},i}\left[1-H\left(\frac{N_{\mathrm{ph},i}^{\mathrm{est}}}{n_{\mathrm{un},i}}\right)\right]-m(d,r_c,\theta_c)-v.
\end{equation}
Although $d$ is not known in the real protocol, the actual parity count $\lambda_h$ is known after Cascade. The realizable key-length rule is therefore
\begin{equation}\label{keyrate2}
\ell_{i,h}=\max\left\{0,\left\lfloor n_{\mathrm{un},i}
\left[1-H\left(N_{\mathrm{ph},i}^{\mathrm{est}}/n_{\mathrm{un},i}\right)\right]
-\lambda_h-v-\log_2\frac1{4\varepsilon_{PA}^2}\right\rfloor\right\},
\end{equation}
with the same integer and empty-key conventions as Eq.~\eqref{keyrate1}. The phase-error estimator must be computable from the real observed data; no unobserved $d$ is needed to evaluate this rule.

For this first case, apply privacy amplification to all accepted runs, including runs with an incorrect reconciled string. The phase-error and tagging inputs, and hence the entropy term in Eq.~\eqref{keyrate2}, remain valid on these runs. Any further public announcement must be determined by the retained record and the counted messages, or its leakage must be included explicitly.

To handle verification without assuming successful reconciliation, write $\alpha=(i,j,c,d)$, $q_\alpha=q_{i,j,c,d}$, $g_i=n_{\mathrm{un},i}[1-H(N_{\mathrm{ph},i}^{\mathrm{est}}/n_{\mathrm{un},i})]$, and $m_\alpha=m(d,r_c,\theta_c)$. Let $\sigma^\alpha_{Z_AC_2C_3VQE}$ denote the normalized state before selecting the acceptance event. Equation~\eqref{general_cascade_entropy} bounds its smooth min-entropy by $g_i-m_\alpha-v$. Keep the accepted state subnormalized. A single smoothing of $\sigma^\alpha$, extended to the classical acceptance register, gives a nearby accepted substate whose min-entropy obeys the same lower bound: the accepted substate is bounded above by the smoothed full state. The extension and restriction argument is detailed in Appendix~\ref{actual_leakage_general}. The subnormalized QLHL~\cite{tomamichel2017largely} then gives the accepted-branch secrecy distance
\begin{equation}\label{general_accepted_qlhl}
\mathcal D^{\mathrm{acc}}_\alpha
\le 2\sqrt{q_\alpha}
+\frac12\sqrt{2^{\ell_{i,h}-g_i+m_\alpha+v}}
\le 2\sqrt{q_\alpha}+\varepsilon_{PA}.
\end{equation}
This inequality is used for positive-key branches with $q_\alpha<1$; empty-key branches contribute zero, and $q_\alpha=1$ is covered by the trivial bound $\mathcal D^{\mathrm{acc}}_\alpha\le1$. The acceptance flag is constant within this substate, and aborted runs output the empty key. No conditioning on $\Omega_g$ or division by an acceptance probability is needed.

Averaging with the actual branch probabilities gives
\begin{equation}\label{sec_11}
\mathcal D_{\mathrm{sec}}
\le\varepsilon_{PA}+2\sqrt{\Delta}
\le\varepsilon_{PA}+2\sqrt\varepsilon,
\end{equation}
where
\begin{equation}\label{sec_22}
\begin{split}
\Delta&=\sum_{i,j,c,d}\Pr(\Omega_i,\xi_j,C_1=c,D=d)q_{i,j,c,d}\\
&=\sum_{i,j}\Pr(N_{\mathrm{ph}}\ge N_{\mathrm{ph},i}^{\mathrm{est}},\Omega_i,\xi_j)\le\varepsilon.
\end{split}
\end{equation}
With $v=\lceil\log_2(2/\varepsilon_{\mathrm{cor}})\rceil$, correctness is bounded separately by $\varepsilon_{\mathrm{cor}}$, yielding $\varepsilon_{com}=\varepsilon_{\mathrm{cor}}+\varepsilon_{PA}+2\sqrt\varepsilon$. Using the actual Cascade leakage after error correction does not itself add a correctness term to secrecy.

If the estimation inputs are instead inferred from the corrected string and their agreement with the true inputs is guaranteed only when $\Omega_g$ holds, the preceding entropy argument controls only the correctly reconciled contribution. Splitting off incorrectly accepted runs as in Eq.~\eqref{rid_equal} then gives
\begin{equation}\label{general_postec_secrecy}
\mathcal D_{\mathrm{sec}}
\le\varepsilon_{\mathrm{cor}}+\varepsilon_{PA}+2\sqrt\varepsilon.
\end{equation}
This is the SCS case, where the window counts are reconstructed after error correction. The additional term accounts for their possible misidentification; it is not a general cost of verification or of postponing the key-length calculation.

\section{{The proof of Eqs.~\eqref{ep11a} and \eqref{ep11}}}\label{scs_delta}
We prove the conditional Cascade leakage bound, the weighted smooth-entropy estimate, and the joint security of the two parts. All classical records are retained with their actual joint distribution. Write $(i,j,c)$ for a fixed value of $(\Omega_i,\xi_j,C_1=c)$, and let $p_{i,j,c}=\Pr(\Omega_i,\xi_j,C_1=c)$.

\subsection{From inferred to true window types}\label{verification_reduction}
We first derive Eq.~\eqref{rid_equal} from Eq.~\eqref{sec_def2}, independently of the subsequent Cascade-specific entropy argument. Bob applies the same deterministic classification to $(\hat Z_A,Z_B)$ that defines the true sequence from $(Z_A,Z_B)$. Here $Z_B$ is his original raw string. Thus $\Omega_g=\{\hat Z_A=Z_A\}$ implies $\hat\omega=\omega$. Identical strings also have identical verification hashes for every seed, so
\begin{equation}\label{verification_event_identity}
\hat\Omega_i\land\Omega_{EV}\land\Omega_g
=\Omega_i\land\Omega_g.
\end{equation}
On this event all inferred counts and the selected key length $\ell_{i,h}$ equal their true-record values.

For any fixed input length and unequal pair $Z_A,\hat Z_A$, independent two-universal verification with the prescribed length $v=\lceil\log_2(2/\varepsilon_{\mathrm{cor}})\rceil$ bounds false acceptance by $\varepsilon_{\mathrm{cor}}$. Its seed is independent of the inputs and the reconciliation path; Cascade does not use it to choose its queries or stopping point. Averaging over the input pairs and all previous records therefore gives
\begin{equation}\label{verification_failure_mass}
\Pr(\Omega_g^c,\Omega_{EV})\le\varepsilon_{\mathrm{cor}}.
\end{equation}
This bounds the unconditional probability of incorrect acceptance. In the present SCS reduction it also bounds the contribution of accepted runs on which the inferred window counts may be wrong.

Fix $(i,j,h)$ as indexed by the inferred event in Eq.~\eqref{sec_def2}, and abbreviate $u_{i,h}=\rho_{S_A}^{\mathrm{unif},\ell_{i,h}}$. Define positive subnormalized operators on $S_A\tilde C QE$ by
\begin{equation}\label{verification_state_split}
\begin{split}
\tau_{i,j,h}&=\rho_{S_A\tilde C QE\land\hat\Omega_i\land\xi_j\land\lambda_h\land\Omega_{EV}},\\
\tau^a_{i,j,h}&=\rho_{S_A\tilde C QE\land\hat\Omega_i\land\xi_j\land\lambda_h\land\Omega_{EV}\land\Omega_a},
\quad a\in\{g,b\},
\end{split}
\end{equation}
where $\Omega_b=\Omega_g^c$ and $\tau_{i,j,h}=\tau^g_{i,j,h}+\tau^b_{i,j,h}$. For any such operator $\tau$, write
\[
\mathfrak d_{i,h}(\tau)
=\tfrac12\|\tau-u_{i,h}\otimes\tr_{S_A}\tau\|_1.
\]
The replacement by $u_{i,h}$ is linear within this fixed-length block. The triangle inequality and positivity give
\begin{equation}\label{verification_distance_split}
\begin{split}
\mathfrak d_{i,h}(\tau_{i,j,h})
&\le\mathfrak d_{i,h}(\tau^g_{i,j,h})
+\mathfrak d_{i,h}(\tau^b_{i,j,h}),\\
\mathfrak d_{i,h}(\tau^b_{i,j,h})
&\le\tfrac12\bigl(\|\tau^b_{i,j,h}\|_1
+\|u_{i,h}\otimes\tr_{S_A}\tau^b_{i,j,h}\|_1\bigr)
=\tr\tau^b_{i,j,h}.
\end{split}
\end{equation}
The real and ideal operators use the same good and bad marginals, so the bad contribution is bounded by its probability mass.

Equation~\eqref{verification_event_identity} identifies the good operator exactly:
\[
\tau^g_{i,j,h}
=\rho_{S_A\tilde C QE\land\Omega_i\land\xi_j\land\lambda_h\land\Omega_g}.
\]
Its output length is still $\ell_{i,h}$, so the ideal state in this block is unchanged. Consequently, with the coarse $\Delta_{i,j,h}$ defined below,
\[
\mathfrak d_{i,h}(\tau^g_{i,j,h})
=\Pr(\Omega_i,\xi_j,\lambda_h)\Delta_{i,j,h}.
\]
In this expression $\Delta_{i,j,h}$ is normalized only with respect to $(\Omega_i,\xi_j,\lambda_h)$; its restriction to $\Omega_g$ remains subnormalized. Since the inferred labels partition all accepted runs,
\[
\sum_{i,j,h}\tr\tau^b_{i,j,h}=\Pr(\Omega_g^c,\Omega_{EV}).
\]
Summing Eq.~\eqref{verification_distance_split} gives Eq.~\eqref{rid_equal}, with the actual joint distribution of the inferred sequence, public transcript, and $QE$ retained throughout.

\subsection{Length conditioning and the public random seed}
Conditioning on $C_1=c$ fixes the complete reconciliation randomness $R=r_c$ and parameters $\Theta=\theta_c$. Here $R$ specifies every randomized choice, including later random choices, which can equivalently be sampled before the raw-bit measurement once the input length is known. Fixing $(r_c,\theta_c)$ makes any stipulated reconciliation procedure deterministic on its two inputs. Its realized length is therefore $L_{r_c,\theta_c}(z_A,z_B)$, even when its path and termination are adaptive. Determinism given both inputs does not imply determinism given their error pattern.

For any compatible branch of positive probability, Bayes' rule gives
\[
\Pr(z_A,z_B|\Omega_i,\xi_j,C_1=c,\lambda_h)
=\frac{\mathbf 1_{\{L_{r_c,\theta_c}(z_A,z_B)=\lambda_h\}}}
{\Pr(\lambda_h|\Omega_i,\xi_j,C_1=c)}\Pr(z_A,z_B|\Omega_i,\xi_j,C_1=c).
\]
Here $\mathbf 1_{\{\cdot\}}$ denotes the indicator function: it equals $1$ when $L_{r_c,\theta_c}(z_A,z_B)=\lambda_h$ and $0$ otherwise.
For a general algorithm the indicator can exclude some inputs allowed by $(i,j,c)$, even though $r_c$ is fixed. Tracing out the transcript does not undo this selection. This is why Eq.~\eqref{length_general} uses the length-conditioned distribution and cannot initially be replaced by the raw-key state obtained from the pre-bit-measurement state,
\begin{equation}\label{error_1}
\rho^{i,j,c}_{Z_AQE}
=\sum_{z_A,z_B}\Pr(z_A,z_B|\Omega_i,\xi_j,C_1=c)
\oprod{z_A}{z_A}\otimes\rho_{QE}^{z_A,z_B,i,j,c}.
\end{equation}
The fixed-input $QE$ blocks in this equation and Eq.~\eqref{length_general} are the same: classical reconciliation acts on the input strings and independent randomness, not on $QE$. These blocks need not be independent of either string.

The QLHL on a length branch requires the entropy of $Z_A$ conditioned on the complete messages. With $\sigma^{i,j,c,h}$ as defined in the main text and $\mathcal T_{i,j,c,h}$ the conditional alphabet of $(C_2,C_3,V)$, the leakage chain rule gives
\[
H_{\min}^{\delta}(Z_A|C_2C_3VQE)_{\sigma^{i,j,c,h}}
\ge H_{\min}^{\delta}(Z_A|QE)_{\rho^{i,j,c,h}}
-\log_2|\mathcal T_{i,j,c,h}|.
\]
Thus one must both identify the appropriate raw-key marginal and bound the transcript alphabet. For Cascade, the induction below proves
\[
L_{r_c,\theta_c}(z_A,z_B)=m_{r_c,\theta_c}(\omega_i)
\quad\text{whenever }z_A\oplus z_B=d(\omega_i).
\]
On every supported $(i,j,c,h)$ branch the indicator and denominator in Bayes' rule then both equal one. This proves Eq.~\eqref{length_independence_main} and, with the unchanged $QE$ blocks, Eq.~\eqref{rid_h}. The identity holds for each fixed seed value. The final secrecy bound averages these values through $c$ with their actual probabilities.

\subsection{The communication path for a fixed error pattern}
Consider two input pairs $(z,z\oplus d)$ and $(z',z'\oplus d)$ with the same full error pattern. Fix every random seed, all pass parameters, and deterministic tie-breaking and termination rules. Describe the randomly permuted blocks of every Cascade pass as subsets of the original position labels. At any stage let $b$ be Bob's correction mask relative to his initial string. His current difference from Alice is $d^{\mathrm{cur}}=d\oplus b$. Suppose both runs have followed the same path up to this stage. The next queried set $I$ is the same. Alice's parity $a_I$ and Bob's current parity $b_I$ obey
\begin{equation}\label{cascade_parity_main}
a_I\oplus b_I=\bigoplus_{k\in I}d_k^{\mathrm{cur}}
\end{equation}
in both runs, although their absolute parity values may differ. Consequently both runs choose the same half in BINARY, locate the same bit to flip, update the same earlier blocks in look-back, and select the same next block. Updating the common correction mask preserves the induction hypothesis. A finite pass schedule and all stipulated stopping rules therefore terminate on the same path.

This induction does not assume successful reconciliation. It proves that the query sets, number of Alice parity announcements, final correction mask, and residual error pattern are functions of $(d,R,\Theta)$ alone, where $\Theta$ denotes the fixed algorithmic parameters. In particular,
\begin{equation}\label{cascade_invariance}
\lambda(z,z\oplus d;R,\Theta)=\lambda(z',z'\oplus d;R,\Theta)=m(d,R,\Theta)
\end{equation}
for every pair $z,z'$. The event $\Omega_g$ is also a function of these same data: it holds exactly when the residual pattern is zero. Absolute parity values may differ between the two runs.

The public randomness is sampled after Step 3(i), when the clicking set, partition, and hence both input lengths are known, but before the raw-key measurement. Its distribution may depend on these public sizes; conditional on this pre-existing record it is sampled independently of the bit values and quantum systems. The prescribed pass partitions can then be fixed, while BINARY and look-back remain adaptive to parity disagreements. All these seeds and parameters remain in the conditioning record. If a block-size rule uses other prior data, those data must likewise be included in $C_1$. Any extra input-dependent stopping or adaptation outside the stated parity-disagreement rules would require a new argument.

For each query, Bob's announced parity is Alice's announced parity XOR the current error parity. Query indices and correction/termination messages are reconstructible by the same induction. Hence, conditional on $C_1$, $C_3$ is a deterministic function of $C_2$. Equivalently, the actual joint state obeys the Markov relation
\begin{equation}\label{cascade_markov}
Z_AQE\ \longleftrightarrow\ C_1C_2\ \longleftrightarrow\ C_3,
\qquad C_3=f(C_1,C_2).
\end{equation}
The full conditional state before verification can be written as
\begin{equation}\label{cascade_cq}
\rho_{Z_AC_2C_3QE|i,j,c}
=\sum_{z,a}p(z,a|i,j,c)\oprod z z\otimes\oprod a a
\otimes\oprod{f(c,a)}{f(c,a)}\otimes\rho_{QE}^{z,a,i,j,c}.
\end{equation}
Appending $C_3$ from $C_2$ and then discarding it are channels acting only on the conditioning system. The smooth min-entropy is therefore unchanged by this addition. At fixed $(i,j,c)$, $C_2$ is an $m(i,c)$-bit string, including repeated parity announcements if any, so its alphabet has size at most $2^{m(i,c)}$. This conditional alphabet bound justifies subtracting the realized $\lambda_h=m(i,c)$.

\subsection{Phase-error smoothing and the leakage chain rule}

Define the refined good-event secrecy distance by
\begin{equation}\label{ab1}
\Delta_{i,j,c,h}=\frac12\left\|
\rho_{(S_A\tilde C QE|\Omega_i,\xi_j,C_1=c,\lambda_h)\land\Omega_g}
-\rho_{S_A}^{\mathrm{unif},\ell_{i,h}}\otimes
\rho_{(\tilde C QE|\Omega_i,\xi_j,C_1=c,\lambda_h)\land\Omega_g}
\right\|_1.
\end{equation}
We first express the two operators in this equation in terms of the state before privacy amplification, then bound its smooth min-entropy, and finally average the resulting distances with the real branch probabilities.

Fix a supported $(i,j,c,h)$ branch. Equation~\eqref{cascade_invariance} gives $\lambda_h=m(i,c)$, and the final residual error pattern also depends only on $(i,c)$. Thus $\Omega_g$ is either true for every input in this branch or false for every input. In the latter case both operators in Eq.~\eqref{ab1} are zero. An empty-key branch also has $\Delta_{i,j,c,h}=0$. We therefore first consider a branch on which $\Omega_g$ holds and $\ell_{i,h}>0$. Verification always accepts on this branch, the inferred sequence equals $\omega_i$, and the selected length is $\ell_{i,h}$. The restriction to $\Omega_g$ in Eq.~\eqref{ab1} then leaves a normalized state; it introduces no additional division by a success probability.

Let $U$ denote the fresh independent privacy-amplification seed, with state $\rho_U$. The normalized state before privacy amplification is the $\sigma^{i,j,c,h}_{Z_AC_2C_3VQE}$ defined in the main text. Let $\mathcal P_{i,h}$ apply the two-universal hash specified by $U$ to $Z_A$, output $\ell_{i,h}$ bits, retain $U$ and all side-information registers, and discard $Z_A$. Its output obeys
\[
\begin{split}
\rho^{i,j,c,h}_{S_AUC_2C_3VQE}
&=\mathcal P_{i,h}\bigl(\sigma^{i,j,c,h}_{Z_AC_2C_3VQE}\otimes\rho_U\bigr),\\
\rho^{i,j,c,h}_{UC_2C_3VQE}
&=\rho_U\otimes\sigma^{i,j,c,h}_{C_2C_3VQE}.
\end{split}
\]
On the fixed good branch, $\tilde C$ consists of $U,C_2,C_3,V$, the fixed public components of $c$, and their deterministic copies, including the announced counts and the acceptance flag. Appending or removing these copies does not change the trace distance. Hence, writing $u_{i,h}=\rho_{S_A}^{\mathrm{unif},\ell_{i,h}}$, Eq.~\eqref{ab1} becomes
\[
\Delta_{i,j,c,h}
=\frac12\left\|
\mathcal P_{i,h}\bigl(\sigma^{i,j,c,h}\otimes\rho_U\bigr)
-u_{i,h}\otimes\rho_U\otimes\sigma^{i,j,c,h}_{C_2C_3VQE}
\right\|_1.
\]
The QLHL~\cite{tomamichel2012tight} therefore gives, for any $0\le\delta<1$,
\[
\Delta_{i,j,c,h}
\le 2\delta+\frac12
\sqrt{2^{\ell_{i,h}-H_{\min}^{\delta}(Z_A|C_2C_3VQE)_{\sigma^{i,j,c,h}}}}.
\]
In particular, the classical messages are retained in the entropy conditioning, and the public seed $U$ is retained in the output state. The bound concerns their joint state, with all message values included.

We next reduce the entropy in this bound to the raw-key marginal. At fixed $c$, $C_3=f(c,C_2)$ by Eq.~\eqref{cascade_markov}; appending $C_3$ from $C_2$ and discarding it are channels on the conditioning registers, so this register does not change the smooth min-entropy. Moreover, $C_2$ has $\lambda_h=m(i,c)$ bits and $V$ has $v$ bits, giving at most $2^{\lambda_h+v}$ possible pairs $(C_2,V)$. The relevant marginal is
\[
\rho^{i,j,c,h}_{Z_AQE}
=\tr_{C_2C_3V}\sigma^{i,j,c,h}_{Z_AC_2C_3VQE}
=\rho^{i,j,c}_{Z_AQE},
\]
The last equality follows from the Bayes formula preceding Eq.~\eqref{error_1}: on a supported Cascade branch, both its indicator and its denominator equal one. Conditioning on $\lambda_h$ therefore leaves the input probabilities unchanged, and the fixed-input $QE$ blocks are the same, giving Eq.~\eqref{rid_h}. Thus the classical leakage chain rule~\cite{tomamichel2015quantum} gives
\begin{equation}\label{ab3}
\begin{split}
H_{\min}^{\delta}(Z_A|C_2C_3VQE)_{\sigma^{i,j,c,h}}
&=H_{\min}^{\delta}(Z_A|C_2VQE)_{\sigma^{i,j,c,h}}\\
&\ge H_{\min}^{\delta}(Z_A|QE)_{\rho^{i,j,c,h}}-\lambda_h-v\\
&=H_{\min}^{\delta}(Z_A|QE)_{\rho^{i,j,c}}-\lambda_h-v.
\end{split}
\end{equation}
The last equality explicitly removes the length condition from the state at the same smoothing parameter $\delta$. The numerical leakage $\lambda_h$ remains in the subtraction. Substitution into the QLHL reduces the distance in Eq.~\eqref{ab1} to
\[
\Delta_{i,j,c,h}
\le 2\delta+\frac12
\sqrt{2^{\ell_{i,h}-H_{\min}^{\delta}(Z_A|QE)_{\rho^{i,j,c}}+\lambda_h+v}}.
\]

It remains to bound this raw-key entropy and choose $\delta$. Write $\rho^{i,j,c}_{ABQE}=\rho_{ABQE|\Omega_i,\xi_j,C_1=c}$ for the state before the raw-bit measurement, and define
\[
q_{i,j,c}
=\Pr(N_{\mathrm{ph}}\ge N_{\mathrm{ph},i}^{\mathrm{est}}|\Omega_i,\xi_j,C_1=c).
\]
This is the conditional failure probability in the alternative phase-measurement experiment on the same state; it determines the smoothing parameter used below. Let $X_A,X_B$ denote the $n_{\mathcal Z,i}$-bit outcomes of the untagged registers, with Bob's outcome complemented as in the main text. For $q_{i,j,c}<1$, the standard phase-error bound on the smooth max-entropy (Lemma~7 of Ref.~\cite{tupkary2025phase}) gives
\begin{equation}\label{phase_max_entropy}
H_{\max}^{\sqrt{q_{i,j,c}}}(X_A|X_B)_{\rho^{i,j,c}}
\le n_{\mathcal Z,i}H(e_{\mathrm{ph},i}),
\qquad e_{\mathrm{ph},i}=N_{\mathrm{ph},i}^{\mathrm{est}}/n_{\mathcal Z,i}.
\end{equation}
Here $H$ is the binary Shannon entropy, and $e_{\mathrm{ph},i}\le1/2$ on the branches under consideration; at $e_{\mathrm{ph},i}=1/2$ the bound follows from $H_{\max}(X_A|X_B)\le n_{\mathcal Z,i}$. All measurement outcomes refer to the same conditional pre-measurement state. Since $\omega_i$ fixes the tagged positions and their bit values, inserting or removing those values leaves the smooth min-entropy unchanged. The smooth EUR~\cite{tomamichel2011uncertainty,tomamichel2012tight}, followed by data processing under Bob's $X$ measurement, therefore yields
\begin{equation}\label{phase_min_entropy}
\begin{split}
H_{\min}^{\sqrt{q_{i,j,c}}}(Z_A|QE)_{\rho^{i,j,c}}
&\ge n_{\mathcal Z,i}-H_{\max}^{\sqrt{q_{i,j,c}}}(X_A|B)_{\rho^{i,j,c}}\\
&\ge n_{\mathcal Z,i}-H_{\max}^{\sqrt{q_{i,j,c}}}(X_A|X_B)_{\rho^{i,j,c}}\\
&\ge n_{\mathcal Z,i}[1-H(e_{\mathrm{ph},i})]=:g_i.
\end{split}
\end{equation}
Choosing $\delta=\sqrt{q_{i,j,c}}$ in Eq.~\eqref{ab3} consequently gives
\[
H_{\min}^{\sqrt{q_{i,j,c}}}(Z_A|C_2C_3VQE)_{\sigma^{i,j,c,h}}
\ge g_i-\lambda_h-v.
\]
Substituting this into the smooth QLHL above gives
\begin{equation}\label{ab2}
\begin{split}
\Delta_{i,j,c,h}
&\le2\sqrt{q_{i,j,c}}+\frac12\sqrt{2^{\ell_{i,h}-g_i+\lambda_h+v}}\\
&\le2\sqrt{q_{i,j,c}}+\varepsilon_{PA}.
\end{split}
\end{equation}
The last inequality follows from Eq.~\eqref{keyrate1}, which gives
$\ell_{i,h}-g_i+\lambda_h+v\le-\log_2[1/(4\varepsilon_{PA}^2)]$
on a positive-key branch. If $q_{i,j,c}=1$, the same final bound follows directly from $\Delta_{i,j,c,h}\le1\le2\sqrt{q_{i,j,c}}+\varepsilon_{PA}$. Together with the zero contributions from empty-key branches and branches outside $\Omega_g$, Eq.~\eqref{ab2} thus bounds every supported branch.

We now restore the actual branch probabilities. The coarse distance $\Delta_{i,j,h}$ in Eq.~\eqref{rid_equal} is defined by Eq.~\eqref{ab1} without conditioning on $C_1=c$. The unfixed components of $C_1$ are classical public records already in $\tilde C$; its virtual type component is fixed by $\Omega_i$. Both the real and ideal states therefore decompose into the same orthogonal $c$ blocks, giving
\[
\Delta_{i,j,h}
=\sum_c\Pr(C_1=c|\Omega_i,\xi_j,\lambda_h)\Delta_{i,j,c,h}.
\]
These weights are the probabilities of the original protocol. The same deterministic-length property of Cascade gives $\Pr(\lambda_h|\Omega_i,\xi_j,C_1=c)=\mathbf1_{\{\lambda_h=m(i,c)\}}$, and hence
\[
\Pr(\Omega_i,\xi_j,C_1=c,\lambda_h)
=p_{i,j,c}\,\mathbf1_{\{\lambda_h=m(i,c)\}},
\qquad p_{i,j,c}=\Pr(\Omega_i,\xi_j,C_1=c).
\]
There is thus one supported value of $h$ for each $(i,j,c)$. Multiplying the coarse distance by its branch probability and summing gives
\begin{equation}\label{weighted_scs}
\begin{split}
\sum_{i,j,h}\Pr(\Omega_i,\xi_j,\lambda_h)\Delta_{i,j,h}
&=\sum_{i,j,c,h}\Pr(\Omega_i,\xi_j,C_1=c,\lambda_h)\Delta_{i,j,c,h}\\
&\le\sum_{i,j,c}p_{i,j,c}\bigl(\varepsilon_{PA}+2\sqrt{q_{i,j,c}}\bigr)\\
&=\varepsilon_{PA}\sum_{i,j,c}p_{i,j,c}
+2\sum_{i,j,c}p_{i,j,c}\sqrt{q_{i,j,c}}\\
&\le\varepsilon_{PA}+2\sqrt{\sum_{i,j,c}p_{i,j,c}q_{i,j,c}}.
\end{split}
\end{equation}
The last step uses $\sum_{i,j,c}p_{i,j,c}\le1$ and
\[
\left(\sum_{i,j,c}p_{i,j,c}\sqrt{q_{i,j,c}}\right)^2
\le\left(\sum_{i,j,c}p_{i,j,c}\right)
\left(\sum_{i,j,c}p_{i,j,c}q_{i,j,c}\right).
\]
Thus the security distance has been reduced to the weighted phase-error failure probability.

In the virtual experiment, the commuting window-type and phase-error measurements, together with the preceding classical records, define a joint probability distribution for $N_{\mathrm{ph}},\Omega_i,\xi_j,C_1$. The events $\{C_1=c\}$ form a mutually exclusive and exhaustive partition of this probability space. By the definitions of $p_{i,j,c}$ and $q_{i,j,c}$ and the law of total probability,
\begin{equation}\label{phase_record_average}
\begin{split}
\sum_c p_{i,j,c}q_{i,j,c}
&=\sum_c\Pr(N_{\mathrm{ph}}\ge N_{\mathrm{ph},i}^{\mathrm{est}},\Omega_i,\xi_j,C_1=c)\\
&=\Pr(N_{\mathrm{ph}}\ge N_{\mathrm{ph},i}^{\mathrm{est}},\Omega_i,\xi_j)\\
&=\Pr(\xi_j)\Pr(N_{\mathrm{ph}}\ge N_{\mathrm{ph},i}^{\mathrm{est}},\Omega_i|\xi_j).
\end{split}
\end{equation}
This identity does not require statistical independence between $C_1$ and the phase-error outcome. Summing over $i,j$ gives the averaged failure probability in Eq.~\eqref{ep11}. Since the estimator depends only on the counts specified by $\omega_i$, grouping the type sequences with the same counts and applying the concentration bound in Sec.~\ref{rigorous} yields
\[
\Delta=\sum_{i,j,c}p_{i,j,c}q_{i,j,c}
=\sum_j\Pr(\xi_j)\sum_i
\Pr(N_{\mathrm{ph}}\ge N_{\mathrm{ph},i}^{\mathrm{est}},\Omega_i|\xi_j)
\le5\varepsilon_p=\varepsilon.
\]
Combining this averaged bound with Eq.~\eqref{weighted_scs} and the incorrect-acceptance contribution in Eq.~\eqref{rid_equal} completes the reduction from Eq.~\eqref{ab1}:
\[
\begin{split}
\mathcal D_{\mathrm{sec}}
&\le\Pr(\Omega_g^c,\Omega_{EV})
+\sum_{i,j,h}\Pr(\Omega_i,\xi_j,\lambda_h)\Delta_{i,j,h}\\
&\le\varepsilon_{\mathrm{cor}}+\varepsilon_{PA}+2\sqrt{\Delta}\\
&\le\varepsilon_{\mathrm{cor}}+\varepsilon_{PA}+2\sqrt{\varepsilon}
=\varepsilon_{\mathrm{sec}}.
\end{split}
\]
The correctness contribution enters here through the inferred-to-true-count reduction in Eq.~\eqref{rid_equal}. This proves Eqs.~\eqref{ep11a} and \eqref{ep11}.

\subsection{Complementary processing and joint key extraction}
Fix the partition and focus first on part 1. Let $Q$ contain the full complementary local system, including part 2 before any of its measurements. For fixed public inputs, its entire real processing, with the independent seeds averaged according to the protocol, defines a quantum instrument $\{\mathcal N_{\ell_2,f_2}\}$. Here $\ell_2$ is the output length and $f_2$ is the remaining part-2 public transcript, including its acceptance flag. The complete channel is
\[
\mathcal N(\tau_Q)
=\sum_{\ell_2,f_2}\mathcal N_{\ell_2,f_2}(\tau_Q)_{S_2}
\otimes\oprod{\ell_2,f_2}{\ell_2,f_2}_{L_2F_2}.
\]
Each branch map is completely positive and trace non-increasing, and the sum of their output traces equals $\tr\tau_Q$. The key register is the direct sum of the spaces of bit strings of all allowed lengths; the $\ell_2=0$ sector contains the empty string. Thus this single trace-preserving channel includes every possible key length and abort outcome. Dependence on the common public inputs is implemented by classical control, with those inputs retained.

Writing $y=(\ell_2,f_2)$, and extending the maps by the identity on $E$, trace preservation gives, for the signal-part type-and-phase failure effect $B_i$,
\begin{equation}\label{complement_marginal}
\sum_y\tr\left[(B_i\otimes\mathcal I)(\operatorname{id}_{AB}\otimes\mathcal N_y)(\rho)\right]
=\tr[(B_i\otimes\mathcal I)\rho].
\end{equation}
The operations act on distinct systems and commute. In particular, this equality sums all lengths and outcomes; conditioning on a particular $y$ may change the signal state.

The virtual $\ket{--}$ measurement used to estimate the $\ket{\phi_2}$ contribution in Sec.~\ref{rigorous} and the actual part-2 processing are alternative operations on $Q$. Averaging over all outcomes of either operation leaves the reduced signal state unchanged, and hence preserves the phase-error failure probability in Eq.~\eqref{complement_marginal}. The concentration argument conditions on the detection-stage record $\xi_j$ and averages the additional records according to Eq.~\eqref{phase_record_average}, bounding the $\Delta$ in Eq.~\eqref{weighted_scs}.

We now apply the channel to the secrecy bound, keeping its random output length explicitly. For either $\beta=1,2$, let $F_\beta$ contain the public records available after processing part $\beta$, including the common initial data, with $L_\beta$ displayed separately. Let $\sigma^{\ell_\beta}_{S_\beta F_\beta QE}$ be the subnormalized state before processing the complementary system, restricted to $L_\beta=\ell_\beta$. Its trace is $\Pr(L_\beta=\ell_\beta)$, and no restriction on the future length $L_{3-\beta}$ has been made. Set $u_\beta^{\ell_\beta}=\rho^{\mathrm{unif}}_{S_\beta|\ell_\beta}$. Grouping the events in Eq.~\eqref{sec_def2} by their output length and discarding the virtual labels gives
\[
\sum_{\ell_\beta}\frac12\left\|
\sigma^{\ell_\beta}_{S_\beta F_\beta QE}
-u_\beta^{\ell_\beta}\otimes\sigma^{\ell_\beta}_{F_\beta QE}
\right\|_1\le\varepsilon_{\mathrm{sec}}.
\]
For each $\ell_\beta$, apply the complete complementary channel to the conditioning systems. This channel retains $L_\beta,F_\beta$, outputs $S_{3-\beta},L_{3-\beta},F_{3-\beta}$, and does not act on $S_\beta$. It therefore maps the ideal product in this expression to $u_\beta^{\ell_\beta}$ tensored with the actual output marginal. The two processing orders give the same real joint output, since the parts use separate local systems and independent randomness.

Let $F=(F_1,F_2,L_1,L_2)$, with duplicate common records identified, and define the subnormalized blocks of that joint output by
\[
\tau^{\boldsymbol\ell}_{S_1S_2FE}
:=\Pr(\boldsymbol L=\boldsymbol\ell)
\rho_{S_1S_2FE|\boldsymbol L=\boldsymbol\ell},
\qquad \boldsymbol\ell=(\ell_1,\ell_2).
\]
Zero-probability blocks are zero operators. The output length records are classical and mutually orthogonal. Additivity of the trace norm on these blocks, followed by contractivity under the complete channel, yields
\[
\begin{split}
&\sum_{\boldsymbol\ell}\frac12\left\|
\tau^{\boldsymbol\ell}_{S_\beta S_{3-\beta}FE}
-u_\beta^{\ell_\beta}\otimes\tau^{\boldsymbol\ell}_{S_{3-\beta}FE}
\right\|_1\\
&\quad\le\sum_{\ell_\beta}\frac12\left\|
\sigma^{\ell_\beta}_{S_\beta F_\beta QE}
-u_\beta^{\ell_\beta}\otimes\sigma^{\ell_\beta}_{F_\beta QE}
\right\|_1\\
&\quad\le\varepsilon_{\mathrm{sec}},\qquad\beta=1,2.
\end{split}
\]
This proves Eq.~\eqref{joint_marginal_main}. The inequality is an average over the actual joint length distribution, which may be correlated with both keys. It does not assert the same secrecy bound for each normalized conditional length block.

For each pair $\boldsymbol\ell$, the triangle inequality gives
\begin{equation}\label{joint_triangle}
\begin{split}
&\frac12\|\tau^{\boldsymbol\ell}_{S_1S_2FE}
-u_1^{\ell_1}\otimes u_2^{\ell_2}\otimes\tau^{\boldsymbol\ell}_{FE}\|_1\\
&\quad\le\frac12\|\tau^{\boldsymbol\ell}_{S_1S_2FE}
-u_1^{\ell_1}\otimes\tau^{\boldsymbol\ell}_{S_2FE}\|_1\\
&\qquad+\frac12\|\tau^{\boldsymbol\ell}_{S_2FE}
-u_2^{\ell_2}\otimes\tau^{\boldsymbol\ell}_{FE}\|_1.
\end{split}
\end{equation}
The second term is at most the corresponding part-2 distance retaining $S_1$, by partial trace. Summing over all $\boldsymbol\ell$ and using the two averaged bounds above therefore gives
\begin{equation}\label{joint_main}
\begin{split}
\sum_{\boldsymbol\ell}\frac{\Pr(\boldsymbol L=\boldsymbol\ell)}2
&\Bigl\|\rho_{S_1S_2FE|\boldsymbol L=\boldsymbol\ell}\\[-2pt]
&\quad-\rho^{\mathrm{unif}}_{S_1|\ell_1}\otimes\rho^{\mathrm{unif}}_{S_2|\ell_2}
\otimes\rho_{FE|\boldsymbol L=\boldsymbol\ell}\Bigr\|_1
\le2\varepsilon_{\mathrm{sec}}.
\end{split}
\end{equation}
Here $\varepsilon_{\mathrm{sec}}=\varepsilon_{\mathrm{cor}}+\varepsilon_{PA}+2\sqrt{5\varepsilon_p}$. The sum includes all length pairs, including either or both lengths being zero. For length zero, the uniform state is the fixed empty-key state. Finally, the probability that either accepted Alice--Bob key pair disagrees is at most $2\varepsilon_{\mathrm{cor}}$ by the union bound, giving the stated $2\varepsilon_{com}$ composable parameter.

\section{{Security with a predetermined maximum error-correction leakage}}\label{max_leakage}
The security proof also applies when Cascade is replaced by another reconciliation algorithm, provided that a maximum leakage $\lambda_{\max}$ is fixed before error correction and the protocol aborts before exceeding it~\cite{tupkary2024security,tupkary2025phase}. For example, Alice may send the syndrome of a linear code with a predetermined parity-check matrix, such as an LDPC code, and Bob may decode it before the same verification step. More generally, the algorithm may be interactive and adaptive. Under the transcript bound specified below, its leakage contribution to the entropy bound and key-length formula is obtained by replacing $\lambda_h$ with $\lambda_{\max}$; the phase-error estimator and the final security parameter remain unchanged.

Let $T$ contain the complete error-correction transcript, including messages from both parties and any data-dependent control messages specifying the communication path, termination, or abort. The independent public randomness remains in $C_1$, and the $v$ verification bits are accounted for separately. At fixed $C_1=c$, represent the communication by a binary tree whose terminal paths have length at most $\lambda_{\max}$. The terminal paths, including aborted paths, are prefix-free. Thus their alphabet $\mathcal T_c$ satisfies
\begin{equation}\label{budget_kraft}
|\mathcal T_c|2^{-\lambda_{\max}}
\le\sum_{t\in\mathcal T_c}2^{-|t|}\le1.
\end{equation}
Equivalently, it suffices to establish $|\mathcal T_c|\le2^{\lambda_{\max}}$ directly. The budget therefore covers the complete reconciliation leakage. For a syndrome protocol with additional control messages, those messages must also be included in this budget.

Fix $(\Omega_i,\xi_j,C_1=c)$ without conditioning on the realized communication length. Let $\sigma^{i,j,c}_{Z_ATVQE}$ be the normalized state after reconciliation and verification but before restricting to accepted runs. Tracing out the messages gives the same raw-key marginal $\rho^{i,j,c}_{Z_AQE}$ as in Eq.~\eqref{rid_h}. The realized length $\lambda$ is a function of $T$ and remains part of the public record. The smooth leakage chain rule~\cite{tomamichel2015quantum} gives
\begin{equation}\label{budget_entropy}
\begin{split}
H_{\min}^{\delta}(Z_A|T\lambda VQE)_{\sigma^{i,j,c}}
&=H_{\min}^{\delta}(Z_A|TVQE)_{\sigma^{i,j,c}}\\
&\ge H_{\min}^{\delta}(Z_A|QE)_{\rho^{i,j,c}}-\lambda_{\max}-v.
\end{split}
\end{equation}
Thus no independence between the realized length and the raw key is required. Substituting the unchanged phase-error entropy bound in Eq.~\eqref{phase_min_entropy} gives, for $q_{i,j,c}<1$,
\[
H_{\min}^{\sqrt{q_{i,j,c}}}(Z_A|T\lambda VQE)_{\sigma^{i,j,c}}
\ge g_i-\lambda_{\max}-v.
\]
On accepted runs, Alice and Bob use their inferred counts to choose the key length. On the correct-reconciliation event these counts equal those specified by $\omega_i$, and the length is
\begin{equation}\label{budget_keylength}
\ell_i^{\max}=\max\left\{0,\left\lfloor
n_{\mathcal Z,i}[1-H(e_{\mathrm{ph},i})]
-\lambda_{\max}-v-\log_2\frac1{4\varepsilon_{PA}^2}
\right\rfloor\right\},
\end{equation}
with the same empty-key conventions as Eq.~\eqref{keyrate1}.

For completeness, let $G$ be the event that reconciliation completes within the budget and $\Omega_g$ holds, and set $s_{i,j,c}=\Pr(G|\Omega_i,\xi_j,C_1=c)$. Keep the restriction to $G$ subnormalized, as in Appendix~\ref{scs_delta}. If $q_{i,j,c}<s_{i,j,c}$, the standard smooth-entropy bound for restriction to a classical event (Lemma~10 of Ref.~\cite{tomamichel2017largely}) gives
\[
H_{\min}^{\sqrt{q_{i,j,c}}}(Z_A|T\lambda VQE)_{\sigma^{i,j,c}\land G}
\ge H_{\min}^{\sqrt{q_{i,j,c}}}(Z_A|T\lambda VQE)_{\sigma^{i,j,c}}
\ge g_i-\lambda_{\max}-v.
\]
Let $\Delta^{\max}_{i,j,c}$ be the good-event secrecy distance defined as in Eq.~\eqref{ab1}, with no conditioning on $\lambda_h$, output length $\ell_i^{\max}$, and restriction to $G$. Applying the smooth QLHL to the subnormalized state yields
\[
\Delta^{\max}_{i,j,c}
\le2\sqrt{q_{i,j,c}}+\frac12\sqrt{2^{\ell_i^{\max}-g_i+\lambda_{\max}+v}}
\le2\sqrt{q_{i,j,c}}+\varepsilon_{PA}.
\]
If $s_{i,j,c}\le q_{i,j,c}$, the same final bound follows from $\Delta^{\max}_{i,j,c}\le s_{i,j,c}\le q_{i,j,c}\le\sqrt{q_{i,j,c}}$. Empty-key branches contribute zero. This includes arbitrary input-dependent correction success and stopping decisions.

The reduction from inferred to true window types, the averaged phase-error bound, and the joint-key argument in Appendix~\ref{scs_delta} then apply as before. In particular, the same independent verification hash bounds incorrect acceptance by $\varepsilon_{\mathrm{cor}}$, while Eqs.~\eqref{weighted_scs} and \eqref{phase_record_average} give
\[
\begin{split}
\mathcal D_{\mathrm{sec}}
&\le\varepsilon_{\mathrm{cor}}+\sum_{i,j,c}p_{i,j,c}\Delta^{\max}_{i,j,c}\\
&\le\varepsilon_{\mathrm{cor}}+\varepsilon_{PA}+2\sqrt{\sum_{i,j,c}p_{i,j,c}q_{i,j,c}}\\
&\le\varepsilon_{\mathrm{cor}}+\varepsilon_{PA}+2\sqrt{\varepsilon}
=\varepsilon_{\mathrm{sec}}.
\end{split}
\]
The extra $\varepsilon_{\mathrm{cor}}$ in secrecy is retained in this SCS reduction because its counts are inferred after reconciliation. If the estimation inputs are valid independently of reconciliation success, restricting instead to all accepted runs gives the bound without this term, as in Appendix~\ref{framework}. Consequently, replacing $\lambda_h$ by $\lambda_{\max}$ in the leakage subtraction and the key-length formula is sufficient for this bounded-transcript version of the protocol. The remaining phase-error estimation and composable-security arguments, including simultaneous use of the two keys, are unchanged.

\section{An phase-error estimator against coherent attacks for SCS protocol}\label{intuition}
In this part, we demonstrate our phase-error rate estimation process based on virtual step 3 defined in Sec.~\ref{rigorous}. Most of the techniques in this part are from Ref.~\cite{curras2021tight}.
  
Alice and Bob measure the state in Eq.~\eqref{initial22} round by round. Before Alice and Bob conduct the $u$-th round of measurement, where $1\le u\le M_s$, the unnormalized state is 
\begin{equation}\label{before_uround}
\ket{\psi_u}=\hat{M}_{\text{eve}}\left[\otimes_{l=1}^{u-1}\left(\hat{M}_{F_l}\ket{\Phi}\right) \otimes \ket{\Phi}_u\otimes \ket{\Phi}^{\otimes M_s-u}\ket{\Phi}^{\otimes N-M_s}\right],
\end{equation}  
where $\hat{M}_{F_l}$ is the measurement operator of the local subsystems corresponding to measurement outcome $F_l$ in the $l$-th round. Let $\sigma_u$ be 
\begin{equation}\label{sigu1}
\sigma_u=\tr_{\bar{u}}(\oprod{\psi_u}{\psi_u})=\sum_{\vec{\bar{u}}}\iprod{\vec{\bar{u}}}{\psi_u}\iprod{\psi_u}{\vec{\bar{u}}},
\end{equation} 
where $\bar{u}$ represent all the rounds except the $u$-th round, and the states $\{\ket{\vec{\bar{u}}}\}$ represents a basis for all the subsystems $L1,L2,I,S$ of all the rounds except the $u$-th round. Let $\hat{M}_{\vec{\bar{u}}}$ be
\begin{equation}
\hat{M}_{\vec{\bar{u}}}=\bra{\vec{\bar{u}}}\hat{M}_{\text{eve}}\left[\otimes_{l=1}^{u-1}\left(\hat{M}_{F_l}\ket{\Phi}\right) \otimes \ket{\Phi}^{\otimes M_s-u}\ket{\Phi}^{\otimes N-M_s}\right].
\end{equation}
We can rewrite Eq.~\eqref{sigu1} as 
\begin{equation}
\sigma_u=\sum_{\vec{\bar{u}}} \hat{M}_{\vec{\bar{u}}} \oprod{\Phi}{\Phi}_u \hat{M}_{\vec{\bar{u}}}^\dagger.
\end{equation}
$\tr (\sigma_u)$ is the probability of all the measurement results before the $u$-th round, including $F_1,\cdots,F_{u-1}$. Let $\vec{F}_{u-1}=F_1,\cdots,F_{u-1}$, we have
\begin{equation}
\Pr( \vec{F}_{u-1})=\tr (\sigma_u).
\end{equation}

In the $u$-th round, the probability that a phase-error of the signal window occurs conditioned on all the previous measurement results is
\begin{equation}\label{start1}
\begin{split}
\Pr(F_u=X_+| \vec{F}_{u-1})&=\frac{\tr\left[\oprod{\text{signal}}{\text{signal}}_{L2}\otimes \oprod{\text{un}}{\text{un}}_{L1} \otimes \oprod{X_+}{X_+}_I \sigma_u \right]}{\tr (\sigma_u)},\\
&=\frac{\frac{p_op_x}{4}\tr\left[\sum_{\vec{\bar{u}}} \hat{M}_{\vec{\bar{u}}} \left( \ket{0\sqrt{\mu_B}}_S + \ket{\sqrt{\mu_A}0}_S\right)\left( \bra{0\sqrt{\mu_B}}_S + \bra{\sqrt{\mu_A}0}_S\right) \hat{M}_{\vec{\bar{u}}}^\dagger \right]}{\Pr( \vec{F}_{u-1})}\\
&=\frac{\frac{p_op_x}{4} \sum_{\vec{\bar{u}}} \parallel \hat{M}_{\vec{\bar{u}}} \left( \ket{0\sqrt{\mu_B}}_S + \ket{\sqrt{\mu_A}0}_S\right) \parallel^2}{\Pr( \vec{F}_{u-1})},\\
&=\frac{\frac{p_op_x}{4} \parallel \sqrt{\hat{E}_u} \left( \ket{0\sqrt{\mu_B}}_S + \ket{\sqrt{\mu_A}0}_S\right) \parallel^2}{\Pr( \vec{F}_{u-1})},
\end{split}
\end{equation}  
where $\hat{E}_u=\sum_{\vec{\bar{u}}}\hat{M}_{\vec{\bar{u}}}^\dagger \hat{M}_{\vec{\bar{u}}}$. And since $\hat{E}_u$ is positive semi-definite, we can decompose it as $\hat{E}_u=\sqrt{\hat{E}_u}\sqrt{\hat{E}_u}$. 

Hereafter, we omit the subscript $S$ where no confusion arises.

As shown in Ref.~\cite{wang2019practical,jiang2024side}, we have
\begin{equation}
\ket{0\sqrt{\mu_B}} + \ket{\sqrt{\mu_A}0}=\ket{00}+\ket{\sqrt{\mu_A}\sqrt{\mu_B}}+c_2\ket{\phi_2},
\end{equation}
where
\begin{equation}
\ket{\phi_2}=\frac{1}{c_2}(\ket{0\sqrt{\mu_B}} + \ket{\sqrt{\mu_A}0}-\ket{00}-\ket{\sqrt{\mu_A}\sqrt{\mu_B}}),
\end{equation}
and 
\begin{equation}
c_2=2\sqrt{(1-e^{-\mu_A/2})(1-e^{-\mu_B/2})}.
\end{equation}

Applying the Cauchy-Schwartz inequality, we have
\begin{equation}
\begin{split}
&\parallel \sqrt{\hat{E}_u} \left( \ket{0\sqrt{\mu_B}}_S + \ket{\sqrt{\mu_A}0}_S\right) \parallel^2\\
&=\parallel \sqrt{\hat{E}_u} \left(\ket{00}+\ket{\sqrt{\mu_A}\sqrt{\mu_B}}+c_2\ket{\phi_2}\right) \parallel^2\\
&\le \left[\parallel \sqrt{\hat{E}_u} \ket{00} \parallel+\parallel \sqrt{\hat{E}_u} \ket{\sqrt{\mu_A}\sqrt{\mu_B}} \parallel + c_2\parallel \sqrt{\hat{E}_u} \ket{\phi_2} \parallel \right]^2.
\end{split}
\end{equation} 

In the $u$-th round, the probability that the measurement result is $\ket{01}_I$ of the signal window conditioned on all the previous measurement results is
\begin{equation}
\begin{split}
\Pr(F_u=\mathcal{O}| \vec{F}_{u-1})&=\frac{\tr\left[\oprod{\text{signal}}{\text{signal}}_{L2}\otimes \oprod{\text{un}}{\text{un}}_{L1} \otimes \oprod{01}{01}_I \sigma_u \right]}{\tr (\sigma_u)},\\
&=\frac{\frac{p_op_o}{2} \parallel \sqrt{\hat{E}_u} \ket{00}\parallel^2}{\Pr( \vec{F}_{u-1})}.
\end{split}
\end{equation}  

In the $u$-th round, the probability that the measurement result is $\ket{10}_I$ of the signal window conditioned on all the previous measurement results is
\begin{equation}
\begin{split}
\Pr(F_u=\mathcal{B}| \vec{F}_{u-1})&=\frac{\tr\left[\oprod{\text{signal}}{\text{signal}}_{L2}\otimes \oprod{\text{un}}{\text{un}}_{L1} \otimes \oprod{10}{10}_I \sigma_u \right]}{\tr (\sigma_u)},\\
&=\frac{\frac{p_xp_x}{2} \parallel \sqrt{\hat{E}_u} \ket{\sqrt{\mu_A}\sqrt{\mu_B}}\parallel^2}{\Pr( \vec{F}_{u-1})}.
\end{split}
\end{equation}  

In the $u$-th round, the probability that the measurement result is $\ket{--}_I$ of the estimation window conditioned on all the previous measurement results is
\begin{equation}\label{end1}
\begin{split}
\Pr(F_u=\mathcal{D}| \vec{F}_{u-1})&=\frac{\tr\left[\oprod{\text{est}}{\text{est}}_{L2}\otimes \oprod{\text{balance}}{\text{balance}}_{L1} \otimes \oprod{--}{--}_I \sigma_u \right]}{\tr (\sigma_u)},\\
&=\frac{\frac{p_xp_x}{8} \parallel \sqrt{\hat{E}_u} \left(\ket{0\sqrt{\mu_B}} + \ket{\sqrt{\mu_A}0}-\ket{00}-\ket{\sqrt{\mu_A}\sqrt{\mu_B}}\right)\parallel^2}{\Pr( \vec{F}_{u-1})}.
\end{split}
\end{equation}  

Combining Eqs.~(\ref{start1}-\ref{end1}), we have
\begin{equation}\label{main_formula2}
\Pr(F_u=X_+| \vec{F}_{u-1})\le \frac{p_op_x}{4}\left(\sqrt{\frac{2}{p_o^2}\Pr(F_u=\mathcal{O}| \vec{F}_{u-1})}+\sqrt{\frac{2}{p_x^2}\Pr(F_u=\mathcal{B}| \vec{F}_{u-1})}+\sqrt{\frac{8}{p_x^2}\Pr(F_u=\mathcal{D}| \vec{F}_{u-1})} \right)^2.
\end{equation}

For clarify, we define
\\$n_\mathcal{O}$: the observed values of the number of $\ket{01}_I$ detected in the clicking signal windows, i.e., the number of the clicking events caused by the sent out state $\ket{00}_S$ in the signal windows.
\\$n_\mathcal{B}$: the observed values of the number of $\ket{10}_I$ detected in the clicking signal windows, i.e., the number of the clicking events caused by the sent out state $\ket{\sqrt{\mu_A}\sqrt{\mu_B}}_S$ in the signal windows.
\\$n_\mathcal{Z}$: the observed values of the number of $\oprod{00}{00}_I+\oprod{11}{11}_I$ detected in the clicking signal windows, i.e., the number of the clicking events caused by the sent out state $\oprod{0\sqrt{\mu_B}}{0\sqrt{\mu_B}}_S+\oprod{\sqrt{\mu_A}0}{\sqrt{\mu_A}0}_S$ in the signal windows.  
\\$n_\mathcal{D}$: the virtual observed values of the number of $\ket{--}_I$ detected in the clicking estimation windows, i.e., the number of the clicking events caused by the sent out state $\ket{\phi_2}$ in the estimation windows. Note $n_\mathcal{D}$ can only be observed in the virtual step 3, and can not be observed in the actual protocol.
\\$n_\mathcal{D}^{\text{all}}$: the virtual observed values of the number of $\ket{--}_I$ detected in all estimation windows, i.e., the number of the sent out state $\ket{\phi_2}$ in the estimation windows. Note $n_\mathcal{D}^{\text{all}}$ can only be observed in the virtual step 3, and can not be observed in the actual protocol.
\\$N_{\text{ph}}$: the observed values of the number of phase errors in the clicking signal windows. Note $N_{\text{ph}}$ can only be observed in the virtual step 3, and can not be observed in the actual protocol.
\\$\mean{n_\mathcal{O}}$: the expected values of $n_\mathcal{O}$. Mathematically, $\mean{n_\mathcal{O}}=\sum_{u=1}^{M_s}\Pr(F_u=\mathcal{O}|\vec{F}_{u-1})$.
\\$\mean{n_\mathcal{B}}$: the expected values of $n_\mathcal{B}$. Mathematically, $\mean{n_\mathcal{B}}=\sum_{u=1}^{M_s}\Pr(F_u=\mathcal{B}|\vec{F}_{u-1})$.
\\$\mean{n_\mathcal{D}}$: the expected values of $n_\mathcal{D}$. Mathematically, $\mean{n_\mathcal{D}}=\sum_{u=1}^{M_s}\Pr(F_u=\mathcal{D}|\vec{F}_{u-1})$.
\\$\mean{N_{\text{ph}}}$: the expected values of $N_{\text{ph}}$. Mathematically, $\mean{N_{\text{ph}}}=\sum_{u=1}^{M_s}\Pr(F_u=X_+|\vec{F}_{u-1})$.

Remark: Keep in mind that all the observed values and mean values are defined in the condition that Charlie has observed $\xi$.

With Eq.~\eqref{main_formula2}, we have
\begin{equation}\label{eq24}
\begin{split}
\mean{N_{\text{ph}}}&=\sum_{u=1}^{M_s}\Pr(F_u=X_+|\vec{F}_{u-1})\\
&\le \frac{p_op_x}{4}\sum_{u=1}^{M_s}\left(\sqrt{\frac{2}{p_o^2}\Pr(F_u=\mathcal{O}|\vec{F}_{u-1})}+\sqrt{\frac{2}{p_x^2}\Pr(F_u=\mathcal{B}|\vec{F}_{u-1})}+\sqrt{\frac{8}{p_x^2}\Pr(F_u=\mathcal{D}|\vec{F}_{u-1})} \right)^2\\
&\le \frac{p_op_x}{4}\left(\sqrt{\frac{2}{p_o^2}\sum_{u=1}^{M_s}\Pr(F_u=\mathcal{O}|\vec{F}_{u-1})}+\sqrt{\frac{2}{p_x^2}\sum_{u=1}^{M_s}\Pr(F_u=\mathcal{B}|\vec{F}_{u-1})}+\sqrt{\frac{8}{p_x^2}\sum_{u=1}^{M_s}\Pr(F_u=\mathcal{D}|\vec{F}_{u-1})} \right)^2\\
&=\frac{p_op_x}{4}\left(\sqrt{\frac{2}{p_o^2}\mean{n_\mathcal{O}}}+\sqrt{\frac{2}{p_x^2}\mean{n_\mathcal{B}}}+\sqrt{\frac{8}{p_x^2}\mean{n_\mathcal{D}}} \right)^2.
\end{split}
\end{equation}
We use the Jensen’s inequality~\cite{jensen1906fonctions,curras2021tight} in the second inequality. Applying Kato's inequality which was first introduced in Ref.~\cite{kato2020concentration} and further developed in Ref.~\cite{curras2021tight}, we can estimate the upper bounds of $\mean{n_\mathcal{O}},\mean{n_\mathcal{B}}$ according to $n_\mathcal{O},n_\mathcal{B}$. We denote the estimated upper bound by $\mean{n_\mathcal{O}^{\text{est}}},\mean{n_\mathcal{B}^{\text{est}}}$, which are
\begin{equation}
\mean{n_\mathcal{O}^{\text{est}}}=\text{IKU}(n_\mathcal{O}),\quad \mean{n_\mathcal{B}^{\text{est}}}=\text{IKU}(n_\mathcal{B}),
\end{equation}
where the function $\text{IKU}(\cdot)$ is defined in Eq.~\eqref{estimator1}.

To estimated the upper bound of $\mean{n_\mathcal{D}}$, we need its observed values $n_\mathcal{D}$. However, $n_\mathcal{D}$ can not be observed in the actual protocol, but we know that the following inequality must be held
\begin{equation}
n_\mathcal{D}\le n_\mathcal{D}^{\text{all}}.
\end{equation}
Still, $n_\mathcal{D}^{\text{all}}$ can not be observed in the actual protocol, but we can used the following method to get an upper bound $n_\mathcal{D}^{\text{est,all}}$ on $n_\mathcal{D}^{\text{all}}$ to replace $n_\mathcal{D}$ in the estimation of $\mean{N_{\text{ph}}}$, which is 
\begin{equation}
n_\mathcal{D}^{\text{est,all}}=\text{CU}(\frac{p_x^2}{8}c_2^2N).
\end{equation}
And we define 
\begin{equation}
\mean{n_\mathcal{D}^{\text{est}}}=\text{IKU}(n_\mathcal{D}^{\text{est,all}}).
\end{equation}

A rigorous proof of why we can use $n_\mathcal{D}^{\text{est,all}}$ to replace $n_\mathcal{D}$ to estimate $\mean{N_{\text{ph}}}$ is placed in Sec.~\ref{rigorous}.

Finally, we get an estimated upper bound on $\mean{N_{\text{ph}}}$
\begin{equation}\label{eq29}
\mean{N_{\text{ph}}^{\text{est}}}=\frac{p_op_x}{4}\left(\sqrt{\frac{2}{p_o^2}\mean{n_\mathcal{O}^{\text{est}}}}+\sqrt{\frac{2}{p_x^2}\mean{n_\mathcal{B}^{\text{est}}}}+\sqrt{\frac{8}{p_x^2}\mean{n_\mathcal{D}^{\text{est}}}}\right)^2
\end{equation}  
  
Again by applying Kato's inequality, we can get the upper bound of the real value of number of phase errors 
\begin{equation}
N_{\text{ph}}^\text{est}=\text{KU}(\mean{N_{\text{ph}}^{\text{est}}}).
\end{equation}
Finally we can get the upper bound of the phase-flip error rate
\begin{equation}
e_{\text{ph}}=\frac{N_{\text{ph}}^\text{est}}{n_{\mathcal{Z}}}.
\end{equation}

\section{The tight concentration inequality for correlated random variables}\label{tightbound}
In this work, we use the method first introduced in Ref.~\cite{kato2020concentration} and further developed in Ref.~\cite{curras2021tight} to handle statistical fluctuations.

Let $R_1,..., R_{M_s}$ be a sequence of random variables taking values $0$ or $1$, and let ${O_l = \sum_{u=1}^{l} R_u}$. Let $\{\mathcal F_l\}_{l=0}^{M_s}$ be a filtration to which this sequence is adapted; it may contain more information than the binary variables alone~\cite{kato2020concentration}. In the virtual experiment of Sec.~\ref{rigorous}, we work at fixed detection record $\xi_j$ and use $\mathcal F_l=\sigma(F_1,\ldots,F_l)$, where $F_l$ is the complete outcome of round $l$. The conditional probabilities below are evaluated in that same conditional experiment. For any fixed $M_s$ and any fixed $a,b$ such that $b\ge |a|$,

\begin{equation}\label{kato1}
\Pr\left(\sum_{u=1}^{M_s} \Pr \left(R_u = 1 \vert \mathcal{F}_{u-1} \right)-O_{M_s}\ge \left[b+a \left(\frac{2 O_{M_s}}{M_s} - 1 \right) \right] \sqrt{M_s}\right)\le \exp\left[\frac{-2(b^2-a^2)}{(1+\frac{4a}{3\sqrt{M_s}})^2}\right]
\end{equation}

With Eq.~\eqref{kato1}, we can construct the following estimator to estimate the upper bound of the expected value from its observed value
\begin{equation}\label{estimator1}
\mean{{O}_{M_s}^{\text{est}}}=\text{IKU}({O}_{M_s})={O}_{M_s}+\left[b+a \left(\frac{2 {O}_{M_s}}{M_s} - 1 \right) \right] \sqrt{M_s},
\end{equation}

In Eq.~\eqref{estimator1}, we can set $a=0$ and $b=\sqrt{-0.5{\ln\varepsilon_p}}$ or determine $a,b$ by 
\begin{equation}\label{abcal}
\begin{split}
&\resizebox{0.95\linewidth}{!}{$\displaystyle a = \frac{3 \left\{72 \sqrt{M_s} \tilde{O}_{M_s}  (M_s-\tilde{O}_{M_s})\ln \varepsilon_p-16 M_s^{3/2} \ln^2\varepsilon_p+9 \sqrt{2} (M_s-2 \tilde{O}_{M_s}) \sqrt{-M_s^2 \ln \varepsilon_p  [9 \tilde{O}_{M_s} (M_s-\tilde{O}_{M_s})-2 M_s \ln \varepsilon_p]}\right\}}{4 (9M_s-8\ln\varepsilon_p) [9 \tilde{O}_{M_s} (M_s-\tilde{O}_{M_s})-2 {M_s} \ln\varepsilon_p]},$} \\
&b = \frac{\sqrt{18 a^{2} M_s-\left(16 a^{2}+24 a \sqrt{M_s}+9 M_s\right) \ln\varepsilon_p}}{3 \sqrt{2M_s}},
\end{split}
\end{equation}
where $\tilde{O}_{M_s}$ is an empirical value of ${O}_{M_s}$ based on prior knowledge. {We require $a\ge-\sqrt{M_s}/2$ in the uses of $\text{IKU}$ in this work. If the candidate in Eq.~\eqref{abcal} violates this condition, we use $a=0$ and $b=\sqrt{-\tfrac12\ln\varepsilon_p}$ instead. This choice is made from the prior estimate, before observing $O_{M_s}$. In particular,
\[
\text{IKU}(x)=\left(1+\frac{2a}{\sqrt{M_s}}\right)x+(b-a)\sqrt{M_s}
\]
is nondecreasing, which justifies the count replacements in Eq.~\eqref{term2.3.1}.} In both cases, $a,b$ are selected from prior information before observing the estimation data and remain fixed throughout the conditional experiment. In particular, post-reconciliation counts from the same run cannot be used to retune them without a separate uniform or union-bound argument, as discussed in Appendix~\ref{kato_no}. The selected parameters satisfy 
\begin{equation}\label{abp}
\exp\left[\frac{-2(b^{ 2}-a^{2})}{(1+\frac{4a}{3\sqrt{M_s}})^2}\right]=\varepsilon_p,
\end{equation}

Replacing $R_l$ by $1-R_l$ and $a$ by $-a$, we can get
\begin{equation}\label{kato2}
\Pr\left(O_{M_s}-\sum_{i=1}^{M_s} \Pr \left(R_i = 1 \vert \mathcal{F}_{i-1} \right)\ge \left[b+a \left(\frac{2 O_{M_s}}{M_s} - 1 \right) \right] \sqrt{M_s}\right)\le \exp\left[\frac{-2(b^2-a^2)}{(1-\frac{4a}{3\sqrt{M_s}})^2}\right],
\end{equation}
By setting $a=0$ and $b=\sqrt{-0.5{\ln\varepsilon_p}}$ in Eq.~\eqref{kato2}, we get
\begin{equation}\label{kato3}
\Pr\left(O_{M_s}-\sum_{i=1}^{M_s} \Pr \left(R_i = 1 \vert \mathcal{F}_{i-1} \right)\ge \sqrt{-0.5M_s{\ln\varepsilon_p}}\right)\le \varepsilon_p,
\end{equation}
With Eq.~\eqref{kato3}, we can construct the following estimator to estimate the upper bound of the observed value from its expected value
\begin{equation}\label{estimator3}
O_{M_s}^{\text{est}}=\text{KU}\left(\sum_{i=1}^{M_s} \Pr \left(R_i = 1 \vert \mathcal{F}_{i-1} \right)\right)=\sum_{i=1}^{M_s} \Pr \left(R_i = 1 \vert \mathcal{F}_{i-1} \right)+\sqrt{-0.5M_s{\ln\varepsilon_p}}.
\end{equation}

The failure budgets in this appendix concern only the virtual estimation experiment. They do not include a correctness failure probability. A separate $\varepsilon_{\mathrm{cor}}$ contribution to secrecy arises in the SCS application when transferring the estimate to counts inferred from the corrected string, as shown in Eq.~\eqref{rid_equal}; it is absent when the estimation inputs are valid independently of successful reconciliation.

\section{The Chernoff bound and inverse Chernoff bound}\label{chernoff}
The Chernoff bound establish the relationship between the expected values and their observed values for the independent random samples~\cite{chernoff1952measure}. Let $Y_1,Y_2,\dots,Y_n$ be $n$ independent random samples, detected with the value 1 or 0, and let $Y$ denote their sum satisfying $Y=\sum_{i=1}^nY_i$. $\Sigma$ is the expected value of $Y$. The Chernoff bound shows that
\begin{equation}\label{B1}
\Pr( Y \ge(1+\delta)\Sigma)\le \left[\frac{e^{\delta}}{(1+\delta)^{1+\delta}}\right]^{\Sigma}. 
\end{equation}
and 
\begin{equation}\label{B2}
\Pr( Y \le(1-\delta)\Sigma)\le \left[\frac{e^{-\delta}}{(1-\delta)^{1-\delta}}\right]^{\Sigma}. 
\end{equation}

With Eqs.~\eqref{B1} and \eqref{B2}, we can construct an estimator to get the lower or upper bound of the observed values according to the expected values,
\begin{equation}\label{estimator_c}
Y_U^{\text{est}}=\text{CU}(\Sigma)=[1+\delta_U(\Sigma)]\Sigma,
\end{equation}
where
\begin{equation}\label{B3}
\left[\frac{e^{\delta_U}}{(1+\delta_U)^{1+\delta_U}}\right]^{\Sigma}=\varepsilon_p,
\end{equation}
and
\begin{equation}\label{estimator_cl}
Y_L^{\text{est}}=\text{CL}(\Sigma)=[1-\delta_L(\Sigma)]\Sigma,
\end{equation}
where
\begin{equation}\label{B4}
\left[\frac{e^{-\delta_L}}{(1-\delta_L)^{1-\delta_L}}\right]^{\Sigma}=\varepsilon_p.
\end{equation}

With Eqs.~\eqref{B1} and \eqref{B2}, we can also construct an estimator to get the lower or upper bound of the expected values according to the observed values, 

\begin{equation}\label{estimator_cx}
\Sigma_U^{\text{est}}=\text{ICU}(Y)=\frac{Y}{1-\delta_U^\prime(Y)},
\end{equation}
where
\begin{equation}\label{B4y}
\left[\frac{e^{-\delta_U^\prime}}{(1-\delta_U^\prime)^{1-\delta_U^\prime}}\right]^{\frac{Y}{1-\delta_U^\prime}}=\varepsilon_p.
\end{equation}
and
\begin{equation}\label{estimator_clx}
\Sigma_L^{\text{est}}=\text{ICL}(Y)=\frac{Y}{1+\delta_L^\prime(Y)},
\end{equation}
where
\begin{equation}\label{B4x}
\left[\frac{e^{\delta_L^\prime}}{(1+\delta_L^\prime)^{1+\delta_L^\prime}}\right]^{\frac{Y}{1+\delta_L^\prime}}=\varepsilon_p.
\end{equation}

Eqs.~(\ref{estimator_cx}-\ref{B4x}) are also called as the inverse Chernoff bound.


\section{Confirm the validity of several commonly used concentration inequalities}\label{confirm}
In this section, we employ the variable-length QKD security framework to validate several concentration inequalities that are commonly used in QKD. We emphasize that this section is not intended to present a new technical contribution. In fact, many of the arguments discussed here have already appeared in the existing literature~\cite{tomamichel2012tight,zhang2017improved,shan2025improved,tupkary2025phase,mannalath2025sharp}, especially in Ref.~\cite{tupkary2025phase}. The main purpose of this section is instead to revisit several inequalities commonly used in QKD from the perspective of elementary probability calculations.

\subsection{The Serfling's inequality}

The finite-key effects in the qubit BB84 protocol have been extensively studied. The Serfling's inequality was employed in Ref.~\cite{tomamichel2012tight} to tightly bound the finite-key effects for the fixed-length qubit BB84 protocol. Through a straightforward generalization, the Serfling-based method presented in Ref.~\cite{tomamichel2012tight} can also be applied to the variable-length qubit BB84 protocol. 

Here we consider the case of variable-length qubit BB84 protocol, in which the final key length $\ell$ is determined by the observed number of clicks $n_X,n_Z$ and bit-flip error rate $e_{bx},e_{bz}$ in the $\mathbb{X}$ and $\mathbb{Z}$ basis respectively. To prove the variable-length qubit BB84 protocol is $\varepsilon_{\text{sec}}$-secret, as shown in Ref.~\cite{tupkary2025phase}, the key is to estimate the upper bound of the phase-flip error rate under a certain failure probability $\varepsilon$. For this purpose, we consider the following equivalent entanglement protocol. 

Alice prepare the states $\ket{\Phi_{AB}}^{\otimes{N}}$ and sends the subsystems $B$ to Bob through a quantum channel. After distribution stage, Alice and Bob share the states $\rho_{AB|\xi}$, where $\xi$ is the measurement results of Eve who can interact with the qubits sent to Bob in the distribution stage. Alice and Bob then randomly measure $n_X$ bits in $\rho_{AB|\xi}^{n}$ in $\mathbb{X}$ basis and publicly announced all their measurement results and obtains $e_{bx}$. The left $n_Z$ qubits are in state $\rho_{AB|\xi,\Omega}$, where $\Omega=\{n_X,n_Z,e_{bx}\}$ represents a group of observed values. Note the $\Omega$ defined here only contains the number of $n_X,n_Z,e_{bx}$, while in the proof of the SCS protocol $\tilde{\Omega}$ is the measurement results of all clicking windows, i.e., what the local subsystems measurement results for each time window are.

Let $\{\Omega_i\}$ be the set that contains all possible observations. Let $\{\xi_j\}$ be the set that contains all possible observations of Eve. Our goal is to prove the following inequality holds for the protocol
\begin{equation}
\Delta=\sum_{i,j} \Pr(\Omega_i,\xi_j) \Pr(e_{pz}\ge e_{pz}^{\text{est}}|\Omega_i,\xi_j)\le \varepsilon
\end{equation}
where $e_{pz}$ is a variable that represents the observed values of the bit-flip error rate if Alice and Bob measure the qubits in state $\rho_{AB|\xi,\Omega}$ in the $\mathbb{X}$ basis. 

The following two measurement procedures are clearly equivalent. 
In the first procedure, Alice and Bob randomly select $n_X$ out of the $n$ qubits in the state $\rho_{AB|\xi}$ and measure them in the $\mathbb{X}$ basis, observing $n_X e_{bx}$ errors; they then measure the remaining $n_Z = n - n_X$ qubits in the $\mathbb{X}$ basis and observe $n_Z e_{pz}$ errors. 
In the second procedure, Alice and Bob instead measure all $n$ qubits in the $\mathbb{X}$ basis, obtaining a total of $m$ errors, and subsequently randomly partition the $n$ measurement outcomes into two disjoint subsets—one containing $n_X$ bits with $n_X e_{bx}$ errors, and the other containing the remaining $n_Z$ bits with $n_Z e_{pz}$ errors. From Eve's perspective, these two procedures are statistically indistinguishable and yield identical joint distributions for the observed error counts.

We define $\vec{\xi}=\{\xi,n_X,n_Z\}$ and the set $\{\vec{\xi}_s\}$ contains all possible observed values. Let $\{m_k\}$ contains all possible values of total errors defined in above first procedure. Then we have
\begin{equation}
\Delta=\sum_{s}\Pr(\vec{\xi}_s) \sum_k\Pr(m_k|\vec{\xi}_s) \sum_{i} \Pr(e_{pz}\ge e_{pz,i}^{\text{est}},e_{bx,i}|\vec{\xi}_s,m_k).
\end{equation}

We also have
\begin{equation}
\Delta=\sum_{s}\Pr(\vec{\xi}_s) \sum_{i} \Pr(e_{pz}\ge e_{pz,i}^{\text{est}},e_{bx,i}|\vec{\xi}_s).
\end{equation}

The Serfling's inequity shows that if we define
\begin{equation}
e_{pz}^{\text{est}}=e_{bx}+\sqrt{-\frac{1}{2}\frac{(n_X+n_Z)(n_X+1)}{n_Zn_X^2}\ln\varepsilon_p},
\end{equation}
we have
\begin{equation}
\sum_{i} \Pr(e_{pz}\ge e_{pz,i}^{\text{est}},e_{bx,i}|\vec{\xi}_s,m_k)\le \varepsilon_p,
\end{equation}
which results
\begin{equation}
\Delta\le\sum_{s}\Pr(\vec{\xi}_s) \sum_k\Pr(m_k|\vec{\xi}_s) \varepsilon_p \le \varepsilon_p.
\end{equation}
{Thus the protocol-level phase-error estimation failure probability is bounded by the prescribed $\varepsilon_p$ in Serfling's inequality.}

\subsection{The inverse Chernoff bound}\label{inversec}
The inverse Chernoff bound was first introduced and proven to be legitimately applicable in the decoy-state parameter estimation of QKD in Ref.~\cite{zhang2017improved}. In this part, we shall use the security framework for variable-length QKD to prove its correctness.

It is standard practice in decoy-state analysis to invoke inverse Chernoff bounds when estimating lower bounds on the number (or yield) of single-photon (or the photons in other photon-number state) detections. Providing a complete and rigorous justification for the valid use of Chernoff bounds within the QKD protocol exceeds the present scope. In the following, we offer an elementary demonstration that elucidates the applicability of inverse Chernoff bounds to variable-length QKD protocols.

For concreteness, our proof focuses on estimating the lower bound on the number of single-photon detections in a three-intensity decoy-state variable-length BB84 protocol. In the protocol description below, we only highlight the steps relevant to the proof and omit irrelevant details.

In this protocol, Alice randomly selects weak coherent pulses from three different intensity settings: vacuum (denoted as \(o\)), decoy (denoted as \(x\)), and signal (denoted as \(y\)), with mean photon numbers \(0\), \(\mu_x\), and \(\mu_y\) (\(\mu_y > \mu_x > 0\)), respectively. The corresponding selection probabilities are \(p_o\), \(p_x\), and \(p_y\), satisfying \(p_o + p_x + p_y = 1\).

Alice sends a total of \(N\) pulses to Bob. After Bob announces which pulses yielded a detection (i.e., the sifted detection events), Alice learns the total counts originating from the three sources: \(n_o\), \(n_x\), and \(n_y\), respectively.

Our goal is to obtain a reliable lower bound on the number of detections caused by single-photon emissions from the signal source \(y\), using the observed counts \(n_o\), \(n_x\), and \(n_y\), under a prescribed failure probability.

We now consider the following equivalent entanglement-based formulation of the protocol.

In each time window, Alice prepared the following state
\begin{equation}
\ket{\Phi}=\sum_{\alpha}\sum_{n=0}\sqrt{p_{\alpha}p_{n|\alpha}}\ket{\alpha}_{L_1}\ket{t_n}_{L_2}\ket{n}_S,
\end{equation}
where $\alpha=o,x,y$, $L_1$ is the local memory storing the source information, $L_2$ is the local memory storing the photon-number information, $S$ is the system sent to Bob through a channel controlled by the eavesdropper, $\ket{n}$ represents a $n$-photon-number state, and 
\begin{align*}
&p_{0|o}=1,p_{n|o}=0\mbox{ for } n\ge 1,\\
& p_{n|x}=\frac{e^{-\mu_x}\mu_x^n}{n!},\\
& p_{n|y}=\frac{e^{-\mu_y}\mu_y^n}{n!}.
\end{align*} 

For the clicking windows, Alice perform the following measurement process to her local memories:

Step 1. Alice measures the subsystems $L_2$ to learn the information of the photon-number state of each clicking windows, and obtains $\tilde{\Omega}={ j_n | n \in \mathbb{N}_0 }$, where $j_n$ means Alice obtains $j_n$ windows of $n$-photon number state.

Step2. Alice then measures the subsystems $L_1$ to learn the information of the source choice of each clicking windows, and obtains $\Omega={n_o,n_x,n_y}$ and $n_{y1}$, where $n_{y1}$ is the number of the counts caused by the sing-photons of source $y$.

In the whole protocol, let $\{\xi_j\}$ be all possible measurement results of Eve, $\{\tilde{\Omega}_s\}$ be all possible measurement results in step 1, $\Omega_i$ be all possible measurement results in step 2. To prove the three-intensity decoy-state variable-length BB84 protocol is $\varepsilon_{\text{sec}}$-secret, we need to get the upper bound of 
\begin{equation}\label{D11}
\Delta = \sum_{i,j,s} 
\Pr(\xi_j, \tilde{\Omega}_s, \Omega_i) \;
\Pr\bigl( n_{y1} {<} n_{y1,i}^{\text{est}} \mid 
\xi_j, \tilde{\Omega}_s, \Omega_i \bigr),
\end{equation}
where $n_{y1,i}^{\text{est}}$ is a estimator construct by $\Omega_i$.

After step 1, the reduced density operator of subsystem $L_1$ (associated with the clicking window) is of the form
\begin{equation}\label{rhoa}
\rho_A=\frac{1}{\mathcal{N}}\bigotimes_{n=0}(\oprod{\psi_n}{\psi_n}_{L_1})^{\otimes j_n},
\end{equation}
where $\frac{1}{\mathcal{N}}$ is the normalization coefficient and 
\begin{align*}
&\ket{\psi_0}=\sqrt{{p_o}}\ket{o}+\sqrt{p_x e^{-\mu_x}}\ket{x}+\sqrt{p_y e^{-\mu_y}}\ket{y},\\
&\ket{\psi_n}=\sqrt{p_x e^{-\mu_x}\frac{\mu_x^n}{n!}}\ket{x}+\sqrt{p_y e^{-\mu_y}\frac{\mu_y^n}{n!}}\ket{y}, \mbox{ for } n\ge 1.
\end{align*}
Note after step 1, Alice has known the corresponding photon number of the pulse emitted in each window of $\rho_A$.

Let $F_u$ be the measurement results of the $u$-th window in step 2. Let $F_u^\prime$ be the measurement results of the $u$-th window in step 1. The states in Eq.~\eqref{rhoa} shows that the $\{F_u\}$ are independent random variables. For the measurement process in step 2, we define
\begin{equation}
\begin{split}
&\mean{n_o}=\sum_{u=1}^{M_s} \Pr(F_u=o)=p_oD_0j_0,\\
&\mean{n_x}=\sum_{u=1}^{M_s} \Pr(F_u=x)=\sum_{u=1}^{M_s} \sum_{n=0} \Pr(F_u=x)\Pr(F_u^\prime=t_n)=\sum_{n=0} p_x p_{n|x} D_n j_n,\\
 &\mean{n_y}=\sum_{u=1}^{M_s} \Pr(F_u=y)=\sum_{u=1}^{M_s} \sum_{n=0} \Pr(F_u=y)\Pr(F_u^\prime=t_n)=\sum_{n=0} p_y p_{n|y} D_n j_n,\\
 &\mean{n_{y1}}= \sum_{u=1}^{M_s} \Pr(F_u=y)\Pr(F_u^\prime=t_1)=p_y p_{1|y} D_1 j_1.
\end{split}
\end{equation} 
Note here $\Pr(F_u^\prime=t_n)$ equals $0$ or $1$ which determined by the measurement results in step 1 and
\begin{equation}
D_n=\frac{1}{p_o p_{n|o}+p_x p_{n|x}+p_y p_{n|y}}.
\end{equation}
The expected values defined above are fixed given the event $\tilde{\Omega}_s$ happened. 

{Conditioned on the preceding photon-number record, the source-choice outcomes in this virtual experiment are independent and their mean counts are fixed. These conditional properties allow the Chernoff argument below. They do not assert unconditional independence between the preceding record and the final observations. In a general correlated process, the compensator entering Kato\textquotesingle s inequality can instead depend on the observed history.}

Applying the standard decoy-state analysis, we have
\begin{equation}\label{eqmaind}
\mean{n_{y1}}\ge \frac{p_y\mu_y e^{-\mu_y}[e^{\mu_x}\mu_y^2 \mean{n_x}/p_x-e^{\mu_y}\mu_x^2 \mean{n_y}/p_y-(\mu_y^2-\mu_x^2)\mean{n_o}/p_o]}{\mu_x\mu_y(\mu_y-\mu_x)}.
\end{equation}
According to Eq~\eqref{eqmaind}, we can construct the following single-photon number counts estimator
\begin{equation}
\begin{split}
&\mean{n_o^{\text{est}}}=\text{ICU}(n_o),\mean{n_x^{\text{est}}}=\text{ICL}(n_x),\mean{n_y^{\text{est}}}=\text{ICU}(n_y),\\
&\mean{n_{y1}^{\text{est}}}= \frac{p_y\mu_y e^{-\mu_y}[e^{\mu_x}\mu_y^2 \mean{n_x^{\text{est}}}/p_x-e^{\mu_y}\mu_x^2 \mean{n_y^{\text{est}}}/p_y-(\mu_y^2-\mu_x^2)\mean{n_o^{\text{est}}}/p_o]}{\mu_x\mu_y(\mu_y-\mu_x)},\\
&n_{y1}^{\text{est}}=\text{CL}(\mean{n_{y1}^{\text{est}}}).
\end{split}
\end{equation}

To get the upper bound of $\Delta$ under the estimator above, we rewrite Eq.~\eqref{D11} into the form
\begin{equation}\label{D12}
\Delta = \sum_{j,s} 
\Pr(\xi_j, \tilde{\Omega}_s) \;
\sum_i \Pr\bigl( n_{y1} {<} n_{y1,i}^{\text{est}}, \Omega_i \mid 
\xi_j, \tilde{\Omega}_s \bigr).
\end{equation}
And we focus on the term $\Delta_{j,s}=\sum_i \Pr\bigl( n_{y1} {<} n_{y1,i}^{\text{est}}, \Omega_i \mid 
\xi_j, \tilde{\Omega}_s \bigr)$ in what follows.  The random process corresponding to the probability $\Delta_{j,s}$ is precisely the measurement process performed in step 2, and the associated random variables are mutually independent. Let $\{n_{y1,k}\}$ be the set containing all possible $n_{y1,k}$. We have
\begin{equation}\label{eqd13}
\begin{split}
\Delta_{j,s}&=\sum_{i,k} \Pr\bigl( n_{y1,k}, \Omega_i \mid \xi_j, \tilde{\Omega}_s \bigr)\Pr(n_{y1,k}{<} n_{y1,i}^{\text{est}} \mid \xi_j, \tilde{\Omega}_s),\\
&=\Pr\bigl( n_{y1} {<} n_{y1,i}^{\text{est}}, \mean{n_{y1}}\ge \mean{n_{y1}^{\text{est}}}  \mid \xi_j, \tilde{\Omega}_s \bigr) + \Pr\bigl( n_{y1} {<} n_{y1,i}^{\text{est}}, \mean{n_{y1}}< \mean{n_{y1}^{\text{est}}}  \mid \xi_j, \tilde{\Omega}_s \bigr),
\end{split}
\end{equation}
where
\begin{equation}
\begin{split}
&\Pr\bigl( n_{y1} {<} n_{y1,i}^{\mathrm{est}},\, 
\mean{n_{y1}} \ge \mean{n_{y1,i}^{\mathrm{est}}} \ \bigm|\ \xi_j, \tilde{\Omega}_s \bigr) 
\\[6pt]
&\quad= \sum_{i,k} \Pr( n_{y1,k}, \Omega_i \mid \xi_j, \tilde{\Omega}_s ) \;
      \Pr\bigl( n_{y1,k} {<} n_{y1,i}^{\mathrm{est}},\, 
      \mean{n_{y1}} \ge \mean{n_{y1,i}^{\mathrm{est}}} \ \bigm|\ \xi_j, \tilde{\Omega}_s \bigr)
\\[10pt]
&\Pr\bigl( n_{y1} {<} n_{y1,i}^{\mathrm{est}},\, 
\mean{n_{y1}} < \mean{n_{y1,i}^{\mathrm{est}}} \ \bigm|\ \xi_j, \tilde{\Omega}_s \bigr) 
\\[6pt]
&\quad= \sum_{i,k} \Pr( n_{y1,k}, \Omega_i \mid \xi_j, \tilde{\Omega}_s ) \;
      \Pr\bigl( n_{y1,k} {<} n_{y1,i}^{\mathrm{est}},\, 
      \mean{n_{y1}} < \mean{n_{y1,i}^{\mathrm{est}}} \ \bigm|\ \xi_j, \tilde{\Omega}_s \bigr)
\end{split}
\end{equation}

{For the first term in Eq.~\eqref{eqd13}, a zero lower count estimate gives an empty failure event. Otherwise the monotonicity of $\text{CL}$ and Eq.~\eqref{B2} give}
\begin{equation}
\begin{split}
&\Pr\bigl( n_{y1} {<} n_{y1,i}^{\text{est}}, \mean{n_{y1}}\ge \mean{n_{y1}^{\text{est}}}  \mid \xi_j, \tilde{\Omega}_s \bigr)\\
&=\Pr\bigl( n_{y1} {<} \text{CL}(\mean{n_{y1}^{\text{est}}}), \mean{n_{y1}}\ge \mean{n_{y1}^{\text{est}}}  \mid \xi_j, \tilde{\Omega}_s \bigr)\\
&\le \Pr\bigl( n_{y1} {<} \text{CL}(\mean{n_{y1}}), \mean{n_{y1}}\ge \mean{n_{y1}^{\text{est}}}  \mid \xi_j, \tilde{\Omega}_s \bigr)\\
&\le \Pr\bigl( n_{y1} {<} [1-\delta_L({\mean{n_{y1}}})]\mean{n_{y1}} \mid \xi_j, \tilde{\Omega}_s \bigr)\\
&\le \varepsilon_p.
\end{split}
\end{equation}

For the second term in Eq~\eqref{eqd13}, we have 
\begin{equation}
\Pr\bigl( n_{y1} {<} n_{y1,i}^{\mathrm{est}},\, 
\mean{n_{y1}} < \mean{n_{y1,i}^{\mathrm{est}}} \bigm|\ \xi_j, \tilde{\Omega}_s \bigr) \le \Pr\bigl(\mean{n_{y1}} < \mean{n_{y1,i}^{\mathrm{est}}} \ \bigm|\ \xi_j, \tilde{\Omega}_s \bigr) 
\end{equation}
If $\mean{n_o^{\text{est}}}\ge \mean{n_o}, \mean{n_y^{\text{est}}}\ge \mean{n_y},\mean{n_x^{\text{est}}}\le \mean{n_x}$, we have $\mean{n_{y1}}\ge \mean{n_{y1}^{\text{est}}}$. Applying the similar technique used in Eq.~\eqref{term2}, we have
\begin{equation}\label{eqd22}
 \Pr\bigl(\mean{n_{y1}} < \mean{n_{y1,i}^{\mathrm{est}}}  \bigm|\xi_j, \tilde{\Omega}_s \bigr)\le  \Pr\bigl(\mean{n_o^{\text{est}}}< \mean{n_o} \bigm| \xi_j, \tilde{\Omega}_s \bigr)+\Pr\bigl(\mean{n_y^{\text{est}}}< \mean{n_y} \bigm| \xi_j, \tilde{\Omega}_s \bigr) +\Pr\bigl(\mean{n_x^{\text{est}}}> \mean{n_x} \bigm| \xi_j, \tilde{\Omega}_s \bigr).
\end{equation}

{With the boundary convention in Appendix~\ref{chernoff}, $\text{ICU}$ is continuous and strictly increasing on $Y\ge0$. If $\mean{n_o}\le\text{ICU}(0)$, the event $\text{ICU}(n_o)<\mean{n_o}$ is empty. Otherwise, since $\mean{n_o}$ is fixed by the preceding record, there is a unique real threshold $n_o^e>0$ satisfying}
\begin{equation}\label{inverse_threshold}
\text{ICU}(n_o^e)=\frac{n_o^e}{1-\delta_U^\prime(n_o^e)}=\mean{n_o},
\end{equation}
where $\delta_U^\prime(n_o^e)$ satisfy
\begin{equation}
\varepsilon_p=\left[\frac{e^{-\delta_U^\prime}}{(1-\delta_U^\prime)^{1-\delta_U^\prime}}\right]^{\frac{n_o^e}{1-\delta_U^\prime}}=\left[\frac{e^{-\delta_U^\prime}}{(1-\delta_U^\prime)^{1-\delta_U^\prime}}\right]^{\mean{n_o}}
\end{equation}

{In this latter case, the first term in Eq.~\eqref{eqd22} obeys}
\begin{equation}\label{noest}
\begin{split}
&\Pr\bigl(\mean{n_o^{\text{est}}}< \mean{n_o} \bigm| \xi_j, \tilde{\Omega}_s \bigr)\\
&=\Pr\left(\text{ICU}(n_o)< \mean{n_o} \bigm| \xi_j, \tilde{\Omega}_s \right)\\
&=\Pr\left(\text{ICU}(n_o)< \text{ICU}(n_o^e) \bigm| \xi_j, \tilde{\Omega}_s \right)\\
&=\Pr\left(n_o< n_o^e \bigm| \xi_j, \tilde{\Omega}_s \right)\\
&=\Pr\left(n_o< \mean{n_o}[1-\delta_U^\prime(n_o^e)] \bigm| \xi_j, \tilde{\Omega}_s \right)\\
&\le \varepsilon_p.
\end{split}
\end{equation}

With the similar method, we can prove 
\begin{equation}
\Pr\bigl(\mean{n_y^{\text{est}}}< \mean{n_y} \bigm| \xi_j, \tilde{\Omega}_s \bigr)\le \varepsilon_p, \Pr\bigl(\mean{n_x^{\text{est}}}> \mean{n_x} \bigm| \xi_j, \tilde{\Omega}_s \bigr)\le \varepsilon_p.
\end{equation}

Finally, we have
\begin{equation}
\Delta=\sum_{j,s}\Pr(\xi_j, \tilde{\Omega}_s) \Delta_{j,s}\le \sum_{j,s}\Pr(\xi_j, \tilde{\Omega}_s)\cdot 4\varepsilon_p =4\varepsilon_p.
\end{equation}

{Thus the Chernoff bound and its inverse give valid averaged parameter-estimation bounds in this setting. By the union bound, the total failure probability is at most the sum of the failure probabilities allocated to the individual bounds, which is $4\varepsilon_p$ here.}

\subsection{The Kato's inequality}\label{kato_no}
In Sec.~\ref{rigorous}, we have rigorously shown that Kato’s inequality applies to variable-length QKD protocols. We now explain the restriction on parameters $a$ and $b$ when constructing the estimator in Eq.~\eqref{estimator1} to upper-bound the expectation from the observed value.

The QKD protocol typically involves multiple classes of observables (e.g., $n_\mathcal{O},n_\mathcal{B}$ in the SCS protocol), each with its own expectation to estimate. Within the same class, a single fixed pair $(a,b)$ must be prechosen for all observed samples of that class; it cannot be re-optimized for each realized sample. Different classes may use different pairs $(a_i,b_i)$. This sample-dependent optimization within a class is the incorrect approach. 


Compared with the Chernoff bound, the parameters $a,b$ in Kato’s inequality seem analogous to $\delta$. In the inverse Chernoff bound, an optimal $\delta$ can be chosen separately for each observed sample, and its validity was proven in the Sec.~\ref{inversec}. For the inverse Kato inequality, however, we must fix the pairs $(a_i,b_i)$ in advance. The reasons are as follows:

1. The failure probability  is ultimately evaluated using the original (non-inverted) form of the concentration inequality, Eqs.~(\ref{kato1},\ref{kato2},\ref{B1},\ref{B2}), even when we use the inverse form of the concentration inequality, i.e., to construct the estimator to upper-bound or lower-bound the expectation from the observed value.

2. The Chernoff bound admits a fictitious multi-step measurement such that the relevant expectation is fully determined by prior outcomes, making it deterministic when conditioned on earlier measurements and allowing the conditioning trick in Sec.~\ref{inversec}. In contrast, the expectation in Kato’s inequality necessarily depends on the final measurement outcome itself; we do not have, in general, an observable $\mathcal{X}_e$ such that $\mathrm{IKU}(\mathcal{X}_e) = \mean{\mathcal{X}_\Omega}$ for all possible $\Omega$, where \(\Omega\) denotes the collection of measurement outcomes across all windows in the final measurement step. Thus the same conditioning technique cannot be applied.

If we optimized $a$ and $b$ using the actual observation in Eq.~\eqref{abcal}, we would obtain data-dependent $a(\mathcal{X})$ and $b(\mathcal{X})$. This would turn the first equality of Eq.~\eqref{term2.1} into
\begin{equation}
\Pr\left(n_\mathcal{O}+\left[b(n_\mathcal{O})+a(n_\mathcal{O}) \left(\frac{2 n_\mathcal{O}}{M_s} - 1 \right) \right] \sqrt{M_s})< \mean{n_\mathcal{O}}|\xi_j\right)
\end{equation}
for which no provable failure-probability bound exists.

\section{{Using the actual reconciliation leakage through a Kraft bound}}\label{actual_leakage_general}
We give a sufficient criterion for inserting the realized reconciliation leakage into the key-length formula, allowing the communication length to depend arbitrarily on the raw strings. The proof combines a conditional Kraft bound with the QLHL and keeps the probabilities of all communication branches. Smooth entropy is handled by a single smoothing of the state before conditioning on the transcript. Cascade is a special case in which its reconstruction property permits counting only Alice's parity disclosures.

\subsection{Entropy before reconciliation}
Use $(\Omega_i,\xi_j,C_1=c)$ for a classical record available before the raw-bit measurement in an equivalent virtual protocol, and write
\[
\varrho^{i,j,c}_{ABQE}=\rho_{ABQE|\Omega_i,\xi_j,C_1=c},
\qquad p_{i,j,c}=\Pr(\Omega_i,\xi_j,C_1=c).
\]
The measurement produces $Z_A,Z_B$, and reconciliation retains Alice's reference string $Z_A$. We distinguish the pre-measurement quantum state $\varrho^{i,j,c}$ from its classical--quantum raw-key output $\rho^{i,j,c}_{Z_AQE}$. Writing $\mathcal Z$ for the raw-key $Z$-measurement channel with Bob's outcome discarded and $QE$ retained, we have $\rho^{i,j,c}_{Z_AQE}=\mathcal Z(\varrho^{i,j,c}_{ABQE})$. The tagging step is the one already included in the raw-key entropy argument. The record $C_1$ includes the reconciliation seeds and parameters, as well as any virtual information supplied in the proof. Independent random tapes can be sampled once their input sizes are known, before the raw-bit measurement, and retained in this record. An independent verification seed is likewise sampled in advance and retained in $C_1$, and reconciliation does not use it. Privacy amplification uses a fresh seed after its output length is determined.

Suppose the protocol's phase-error analysis and the smooth EUR establish the following bounds, with the full record $(\Omega_i,\xi_j,C_1=c)$ retained in the conditioning information:
\begin{equation}\label{actual_entropy_input}
\begin{split}
&q_{i,j,c}=\Pr(N_{\mathrm{ph}}\ge N_{\mathrm{ph},i}^{\mathrm{est}}|\Omega_i,\xi_j,C_1=c),\\
&H_{\min}^{\sqrt{q_{i,j,c}}}(Z_A|QE)_{\rho^{i,j,c}}\ge g_i,
\qquad q_{i,j,c}<1,\\
&\sum_{i,j,c}p_{i,j,c}q_{i,j,c}\le\varepsilon.
\end{split}
\end{equation}
Only branches eligible for a positive key require the entropy inequality. For the qubit setting with the tagging assumptions of Appendix~\ref{framework}, the established smooth max-entropy bound and EUR give $g_i=n_{\mathrm{un},i}[1-H(e_{\mathrm{ph},i})]$~\cite{tomamichel2011uncertainty,tupkary2025phase}. No conditioning on the subsequent reconciliation transcript or its length is introduced in Eq.~\eqref{actual_entropy_input}.

\subsection{A conditional Kraft bound for the complete transcript}
Let $T=t$ denote the complete reconciliation transcript, including both parties' messages and all data-dependent control, stopping, and abort information. Independent seeds already retained in $C_1$ are not counted again. Let $\lambda(t)$ be the nonnegative integer disclosure cost used in the real key-length calculation; its realized value is $\lambda_h$. For every fixed $(i,j,c)$, require
\begin{equation}\label{actual_kraft_condition}
\sum_{t\in\mathcal T_{i,j,c}}2^{-\lambda(t)}\le1,
\end{equation}
where $\mathcal T_{i,j,c}$ contains all possible transcripts on that branch, including aborted executions. Unlike a fixed-length condition, Eq.~\eqref{actual_kraft_condition} allows different transcripts on the same branch to have different costs. In particular, the raw-key state conditioned on $\lambda_h$ need not equal its unconditioned state.

One general way to meet this condition is to encode complete communication paths by the leaves of a binary communication tree~\cite{kushilevitz1997communication}, at each fixed value of the independent seeds. A leaf cannot be a proper prefix of another leaf. Kraft's inequality~\cite{kraft1949device,cover2006elements} therefore gives Eq.~\eqref{actual_kraft_condition} with $\lambda(t)$ equal to the actual encoded path length. Restricting to a compatible $(i,j,c)$ branch only removes leaves. This length includes every data-dependent choice that is not already specified by the preceding public record, including termination when it conveys additional information. Counting only payload bits is insufficient unless they determine the omitted messages. The full tree may be finite, or countable with almost-sure termination.

A shorter conditional description is also allowed: it suffices that the assigned costs obey Eq.~\eqref{actual_kraft_condition} given the record $(i,j,c)$, even if the corresponding description is not the encoding transmitted in the real protocol. The cost must still be computable from the real execution. This possibility will recover the Alice-only parity count for Cascade.

For the prescribed correctness parameter $\varepsilon_{\mathrm{cor}}$, choose $v=\lceil\log_2(2/\varepsilon_{\mathrm{cor}})\rceil$. Verification is accounted for separately by its $v$-bit value $V=a$. For each $t$ there are at most $2^v$ such values, so
\begin{equation}\label{actual_verification_kraft}
\sum_{t\in\mathcal T_{i,j,c}}\ \sum_{a\in\{0,1\}^v}
2^{-\lambda(t)-v}\le1.
\end{equation}
On every accepted run covered by the entropy bound, any additional public announcement must be determined by $(i,j,c,t,a)$ and the independent hash seeds, or be included in the transcript cost. The implemented key rule must agree with the following rule on these runs. If the estimation and tagging inputs are available independently of successful reconciliation, these conditions are required on all accepted runs. If those inputs are inferred from the corrected string and are justified only when $\Omega_g$ holds, the conditions are required on correctly reconciled accepted runs, and the remaining contribution is bounded separately. SCS belongs to the latter case by Appendix~\ref{verification_reduction}.

\begin{proposition}[Actual leakage under a Kraft bound]\label{actual_leakage_proposition}
Assume Eq.~\eqref{actual_entropy_input}, the conditional Kraft bound~\eqref{actual_kraft_condition}, and the accounting and independence conditions above. Use fresh two-universal hashing, with the following key-length rule on the runs covered by these conditions:
\begin{equation}\label{actual_general_keylength}
\ell_{i,t}=\max\left\{0,\left\lfloor
g_i-\lambda(t)-v-\log_2\frac1{4\varepsilon_{PA}^2}
\right\rfloor\right\},
\end{equation}
and output the empty key on abort. If the entropy inputs and the preceding accounting conditions apply on every accepted run, the variable-length secrecy distance is at most $\varepsilon_{PA}+2\sqrt\varepsilon$. If they are guaranteed only on correctly reconciled runs because the estimation inputs are inferred after error correction, the secrecy bound is $\varepsilon_{\mathrm{cor}}+\varepsilon_{PA}+2\sqrt\varepsilon$. In both cases the probability of accepting unequal keys is at most $\varepsilon_{\mathrm{cor}}$.
\end{proposition}

\subsection{Weighted QLHL on subnormalized communication branches}
We first prove the proposition without smoothing. Fix $(i,j,c)$ and abbreviate $\mathcal E=QE$, $\rho=\rho^{i,j,c}_{Z_A\mathcal E}$, and $g=g_i$. Suppose $H_{\min}(Z_A|\mathcal E)_\rho\ge g$. Let $G$ be a binary proof register selecting the accepted runs covered by the entropy input and the implemented key rule. When these are valid independently of reconciliation success, $G=1$ selects every accepted run, including incorrect ones. In the second case of the proposition, $G=1$ selects only correctly reconciled accepted runs. Before privacy amplification retain all executions, including aborts, in the classical transcript state $\sigma_{Z_ATVG\mathcal E}$. Since reconciliation retains $Z_A$ and does not act on $\mathcal E$,
\begin{equation}\label{actual_raw_marginal}
\tr_{TVG}\sigma_{Z_ATVG\mathcal E}=\rho_{Z_A\mathcal E}.
\end{equation}
Absent verification values on aborted runs can be assigned fixed dummy values. Eve's subsequent processing controlled by the public messages is covered by data processing.

For a transcript and verification value $(t,a)$, define the subnormalized state
\begin{equation}\label{actual_transcript_substates}
\begin{split}
&\tau^{t,a}_{Z_A\mathcal E}
=\langle t,a,1|\sigma_{Z_ATVG\mathcal E}|t,a,1\rangle_{TVG},\\
&p_{t,a}=\tr\tau^{t,a}=\Pr(T=t,V=a,G=1|i,j,c),\\
&\sum_{t,a}\tau^{t,a}\le\rho_{Z_A\mathcal E}.
\end{split}
\end{equation}
In particular, each $\tau^{t,a}\le\rho$. By the operator definition of conditional min-entropy~\cite[Definition~6.2]{tomamichel2015quantum}, a normalized $\eta_{\mathcal E}$ satisfying $\rho\le2^{-g}I_{Z_A}\otimes\eta_{\mathcal E}$ also bounds every substate. Consequently,
\begin{equation}\label{actual_substate_entropy}
H_{\min}(Z_A|\mathcal E)_{\tau^{t,a}}\ge g.
\end{equation}
This is a statement about subnormalized operators, not about their normalized conditional states.

Let $U$ be the fresh privacy-amplification seed, and let $\omega^{t,a}_{S_AU\mathcal E}$ be the corresponding output from $\tau^{t,a}$ with length $\ell_{i,t}$. For $p_{t,a}>0$, the normalized QLHL~\cite{tomamichel2011leftover}, together with
\[
H_{\min}(Z_A|\mathcal E)_{\tau^{t,a}/p_{t,a}}
=H_{\min}(Z_A|\mathcal E)_{\tau^{t,a}}+\log_2p_{t,a},
\]
gives, after multiplying the distance by $p_{t,a}$,
\begin{equation}\label{actual_subnormalized_qlhl}
\begin{split}
d_{t,a}&:=\frac12\left\|\omega^{t,a}_{S_AU\mathcal E}
-\rho^{\mathrm{unif},\ell_{i,t}}_{S_A}\otimes\omega^{t,a}_{U\mathcal E}\right\|_1\\
&\le\frac12\sqrt{p_{t,a}\,2^{\ell_{i,t}-H_{\min}(Z_A|\mathcal E)_{\tau^{t,a}}}}\\
&\le\frac12\sqrt{p_{t,a}\,2^{\ell_{i,t}-g}}
\le\varepsilon_{PA}\sqrt{p_{t,a}\,2^{-\lambda(t)-v}}.
\end{split}
\end{equation}
The same subnormalized QLHL is stated in Ref.~\cite[Lemma~18]{tomamichel2017largely}. The last inequality is needed only for positive output lengths; empty-key branches have $d_{t,a}=0$, as do zero-probability branches.

The classical information has not been removed from the output. The registers $T,V$, and the public length $L$ distinguish orthogonal blocks, with $L=\ell_{i,t}$ in each block. Explicitly, the selected contribution and its ideal version are
\begin{equation}\label{actual_classical_blocks}
\begin{split}
\omega&=\sum_{t,a}|t,a,\ell_{i,t}\rangle\langle t,a,\ell_{i,t}|_{TVL}
\otimes\omega^{t,a}_{S_AU\mathcal E},\\
\omega^{\mathrm{id}}&=\sum_{t,a}|t,a,\ell_{i,t}\rangle\langle t,a,\ell_{i,t}|_{TVL}
\otimes\rho^{\mathrm{unif},\ell_{i,t}}_{S_A}\otimes\omega^{t,a}_{U\mathcal E}.
\end{split}
\end{equation}
Thus the secrecy distance is the sum of the subnormalized block distances. Cauchy--Schwarz and Eq.~\eqref{actual_verification_kraft} give
\begin{equation}\label{actual_kraft_sum}
\begin{split}
D_{i,j,c}:=\sum_{t,a}d_{t,a}
&\le\varepsilon_{PA}\sum_{t,a}\sqrt{p_{t,a}\,2^{-\lambda(t)-v}}\\
&\le\varepsilon_{PA}
\sqrt{\left(\sum_{t,a}p_{t,a}\right)
\left(\sum_{t,a}2^{-\lambda(t)-v}\right)}
\le\varepsilon_{PA}.
\end{split}
\end{equation}
Thus the cost of publishing $T$ is controlled by the Kraft sum. No chain rule subtracting a branch-dependent dimension, and no independence between $\lambda$ and $Z_A\mathcal E$, is assumed. The same argument holds when $\sigma$ has trace at most one.

\subsection{Common smoothing and the phase-error failure probability}
Fix a branch $\alpha=(i,j,c)$ eligible for a positive key, put $q_\alpha=q_{i,j,c}$ and $\delta=\sqrt{q_\alpha}<1$, and abbreviate $\varrho=\varrho^\alpha_{AB\mathcal E}$, $\mathcal E=QE$, and $g=g_i$. In the qubit phase-error setting used above, the common smoothing can be constructed on this quantum state before the raw-key measurement. Let $\Pi_{\mathrm{good}}$ project onto the subspace $N_{\mathrm{ph}}<N_{\mathrm{ph},i}^{\mathrm{est}}$, and put $\overline\Pi=\Pi_{\mathrm{good}}\otimes I_{\mathcal E}$. Since $\tr(\overline\Pi\varrho)=1-q_\alpha$, the normalized projected state obeys
\begin{equation}\label{actual_quantum_smoothing}
\widetilde\varrho_{AB\mathcal E}
=\frac{\overline\Pi\varrho\overline\Pi}{1-q_\alpha},
\qquad P(\widetilde\varrho,\varrho)\le\sqrt{q_\alpha}=\delta.
\end{equation}
Here $P$ denotes purified distance. The projection selects the phase-error subspace without resolving the individual $X$ outcomes. It is a proof construction on the joint state, including all correlations with $\mathcal E$; the marginal $\widetilde\varrho_{\mathcal E}$ need not equal $\varrho_{\mathcal E}$.

Apply the same channel $\mathcal Z$ to both states. It acts trivially on $\mathcal E$, so data processing~\cite{tomamichel2015quantum} gives
\begin{equation}\label{actual_quantum_to_cq}
\begin{split}
\rho_{Z_A\mathcal E}&=\mathcal Z(\varrho),
\qquad \widetilde\rho_{Z_A\mathcal E}=\mathcal Z(\widetilde\varrho),\\
P(\widetilde\rho,\rho)&\le P(\widetilde\varrho,\varrho)\le\delta.
\end{split}
\end{equation}
For the alternative local $X$ measurements on the $n=n_{\mathrm{un},i}$ untagged pairs of the same projected quantum state, let $\widetilde\rho^X_{X_AX_B}$ denote the outcome state, with Bob's outcomes adjusted to the phase-error convention. Its support lies in the corresponding Hamming balls. For $e_{\mathrm{ph},i}\le1/2$, the non-smooth phase-error entropy bound and EUR~\cite{tomamichel2011uncertainty,tupkary2025phase} therefore yield, under the tagging assumptions stated above,
\begin{equation}\label{actual_projected_entropy}
\begin{split}
H_{\min}(Z_A|\mathcal E)_{\widetilde\rho}
&\ge n-H_{\max}(X_A|X_B)_{\widetilde\rho^X}\\
&\ge n[1-H(e_{\mathrm{ph},i})]=g_i.
\end{split}
\end{equation}
The $Z$ and $X$ measurements are alternative operations on $\widetilde\varrho$, rather than successive measurements. For SCS, the tagged values are fixed by $\omega_i$, as in Eq.~\eqref{phase_min_entropy}.

Thus this construction supplies a single classical--quantum state satisfying
\begin{equation}\label{actual_common_smoothing}
P(\widetilde\rho,\rho)\le\delta,
\qquad H_{\min}(Z_A|\mathcal E)_{\widetilde\rho}\ge g,
\qquad \tr\widetilde\rho\le1,
\end{equation}
which is the witness required by Eq.~\eqref{actual_entropy_input}. More generally, when that input bound is established by another argument, such a state exists by the definition of smooth min-entropy. The remaining proof uses this same state for every communication branch, with no additional smoothing error at the $Z$-measurement step. The extension property following from Uhlmann's theorem~\cite[Corollary~3.14]{tomamichel2015quantum} supplies an extension to the full transcript state with
\begin{equation}\label{actual_smoothing_extension}
\tr_{TVG}\widetilde\sigma_{Z_ATVG\mathcal E}=\widetilde\rho_{Z_A\mathcal E},
\qquad P(\widetilde\sigma,\sigma)\le\delta.
\end{equation}
The extension is taken on the same transcript alphabet $\mathcal T_{i,j,c}$. Dephasing its classical registers preserves the required marginal and cannot increase the distance. In particular, all smoothed substates come from this one extension and satisfy
\[
\sum_{t,a}\widetilde\tau^{t,a}\le\widetilde\rho,
\qquad \sum_{t,a}\widetilde p_{t,a}\le1.
\]
Equations~\eqref{actual_substate_entropy}--\eqref{actual_kraft_sum} therefore apply to the smoothed state and give $\widetilde D_{i,j,c}\le\varepsilon_{PA}$. Its transcript probabilities need not equal the real probabilities.

To transfer this result back to the real state, apply the same trace-nonincreasing map to $\sigma$ and $\widetilde\sigma$: restrict to $G=1$, choose the output length by Eq.~\eqref{actual_general_keylength}, append the independent hash seed, and perform privacy amplification while retaining $T,V,L,U$. Denote the resulting outputs by $\omega$ and $\widetilde\omega$. Their ideal versions replace the key in each length block by the uniform state of that length and preserve the corresponding side-information marginal. This replacement is a quantum channel controlled by the classical length register. Contractivity and the triangle inequality consequently imply
\begin{equation}\label{actual_smooth_weighted_qlhl}
\begin{split}
D_{i,j,c}
&=\frac12\|\omega-\omega^{\mathrm{id}}\|_1\\
&\le\frac12\|\omega-\widetilde\omega\|_1
+\frac12\|\widetilde\omega-\widetilde\omega^{\mathrm{id}}\|_1
+\frac12\|\widetilde\omega^{\mathrm{id}}-\omega^{\mathrm{id}}\|_1\\
&\le2\delta+\widetilde D_{i,j,c}
\le2\sqrt{q_{i,j,c}}+\varepsilon_{PA}.
\end{split}
\end{equation}
Each of the first and third distances is at most $\delta$, since the trace distance is bounded by the purified distance~\cite{tomamichel2015quantum}. This step compares the entire transcript states, including their different branch probabilities; it does not smooth each normalized transcript branch separately. If $q_{i,j,c}=1$, the same final bound follows directly from $D_{i,j,c}\le1$.

Restoring the original probabilities of the compatible records gives
\begin{equation}\label{actual_weighted_distance}
\sum_{i,j,c}p_{i,j,c}D_{i,j,c}
\le\varepsilon_{PA}+2\sum_{i,j,c}p_{i,j,c}\sqrt{q_{i,j,c}}
\le\varepsilon_{PA}+2\sqrt{\varepsilon}.
\end{equation}
The last step uses Cauchy--Schwarz and Eq.~\eqref{actual_entropy_input}. In the SCS application, the jointly defined pre-reconciliation records and phase-error event allow this average to be written, by the law of total probability, as
\begin{equation}\label{actual_phase_average}
\sum_{i,j,c}p_{i,j,c}q_{i,j,c}
=\sum_j\Pr(\xi_j)\sum_i
\Pr(N_{\mathrm{ph}}\ge N_{\mathrm{ph},i}^{\mathrm{est}},\Omega_i|\xi_j)
\le\varepsilon.
\end{equation}
Neither a phase-error measurement conditioned on the subsequent transcript nor commutation of that transcript with the phase-error observable is required. The phase-error analysis supplies the entropy before reconciliation, and the weighted QLHL controls the classical information subsequently disclosed.

If the estimation inputs and the implemented key rule are valid on every accepted run, $G=1$ already includes all such runs. Discarding virtual labels and grouping by the actual public output length in Eq.~\eqref{actual_weighted_distance} therefore gives
\begin{equation}\label{actual_general_security}
\begin{split}
\sum_{\ell\ge0}\frac{\Pr(L=\ell)}2
\left\|\rho_{S_A\tilde C QE|L=\ell}
-\rho^{\mathrm{unif},\ell}_{S_A}\otimes\rho_{\tilde C QE|L=\ell}\right\|_1
\le\varepsilon_{PA}+2\sqrt{\varepsilon}.
\end{split}
\end{equation}
Here $\tilde C$ retains all real public communication, seeds, the acceptance flag, and the length $L$; aborts produce the empty key. Incorrect reconciliation does not require a separate secrecy contribution in this case: Alice's reference key has already been bounded on every accepted run.

If the estimation inputs are instead justified only on $\Omega_g$, Eq.~\eqref{actual_weighted_distance} bounds the correctly reconciled contribution. Splitting the real accepted state into correct and incorrect contributions, as in Appendix~\ref{verification_reduction}, yields
\begin{equation}\label{actual_postec_security}
\mathcal D_{\mathrm{sec}}
\le\Pr(\Omega_g^c,\Omega_{EV})+\varepsilon_{PA}+2\sqrt\varepsilon
\le\varepsilon_{\mathrm{cor}}+\varepsilon_{PA}+2\sqrt\varepsilon.
\end{equation}
The independent verification hash with the prescribed $v$ bounds incorrect acceptance by $\varepsilon_{\mathrm{cor}}$ in both cases and hence establishes correctness separately. This proves both cases of the proposition. The extra term in Eq.~\eqref{actual_postec_security} concerns the validity of the reconstructed estimation inputs, not the use of a variable communication length.

\subsection{Cascade as a conditional encoding of the transcript}
The error-pattern-based treatment of Cascade's two-way communication follows Ref.~\cite{tupkary2023using}.
For SCS, the virtual window-type sequence $\omega_i$ determines the full error pattern $d(\omega_i)$. At fixed $R=r_c$ and $\Theta=\theta_c$, Appendix~\ref{scs_delta} proves that the Cascade queries, corrections, look-back choices, and stopping point depend only on this pattern. Thus all supported inputs on the same $(i,j,c)$ branch give the same number $m(i,c)$ of Alice parity disclosures. Writing the complete transcript as $T=(C_2,C_3)$, the reconstruction proved there gives
\begin{equation}\label{actual_cascade_alphabet}
\begin{split}
C_2&=H_{d(\omega_i),r_c,\theta_c}Z_A,
\qquad C_3=f(i,j,c,C_2),\\
|\mathcal T_{i,j,c}|&\le2^{\operatorname{rank}H_{d(\omega_i),r_c,\theta_c}}
\le2^{m(i,c)}.
\end{split}
\end{equation}
Consequently, Alice's actual parity count $\lambda(t)=m(i,c)$ satisfies the general condition:
\begin{equation}\label{actual_cascade_kraft}
\sum_{t\in\mathcal T_{i,j,c}}2^{-\lambda(t)}
=|\mathcal T_{i,j,c}|\,2^{-m(i,c)}\le1.
\end{equation}
Alice's parity string is a fixed-length description of the entire transcript given the compatible record, although the actual exchange also contains Bob's messages. This is why those messages do not require an additional leakage subtraction in this special case.

The SCS entropy and phase-error estimates supply Eq.~\eqref{actual_entropy_input} with $g_i=n_{\mathcal Z,i}[1-H(e_{\mathrm{ph},i})]$ and $\varepsilon=5\varepsilon_p$. The second case of Proposition~\ref{actual_leakage_proposition} therefore recovers Eq.~\eqref{keyrate1} with $\lambda_h$ equal to Alice's observed parity count and secrecy $\varepsilon_{\mathrm{cor}}+\varepsilon_{PA}+2\sqrt{5\varepsilon_p}$, because the real SCS counts are inferred after reconciliation. Retaining $Q$ also permits the joint variable-length key argument of Appendix~\ref{scs_delta}.

For a general reconciliation algorithm, constancy of $\lambda_h$ at fixed error pattern is not needed: the complete encoded communication path provides the more general Kraft bound. A smaller count, such as Alice's messages alone, requires its own conditional transcript bound. In either case, the argument uses the actual realized cost without asserting that conditioning on that cost leaves the raw-key state unchanged.

\end{document}